\documentstyle[epsfig,11pt]{mythesis}
\newcommand{\be}{\begin{equation}}
\newcommand{\ee}{\end{equation}}
\newcommand{\ba}{\begin{eqnarray}}
\newcommand{\ea}{\end{eqnarray}}
\newcommand{\pa}{\parallel}
\newcommand{\pe}{\perp}
\newcommand{\nn}{\nonumber}

\begin{document}



\thesistitle{\textbf Vertices and Vortices in High $T_c$ Superconductors}
\author{Mohammad H. Sharifzadeh Amin}
\previousdegrees{B.\ Sc. (Electrical Engineering) 
Shiraz University, Shiraz, Iran.
\\  M.\ Sc. (Physics) Sharif University of Technology, Tehran, Iran.}
\degreetitle{Doctor of Philosophy}
\department{Physics}
\signaturelines{6}
\copyrightyear{September 1999}
\titlepage

\authorizepage
\preface{Abstract}

This thesis is organized in two independent parts, in which I study
two different aspects of high $T_c$ superconductivity.

The first part begins with an introduction aimed to 
briefly introduce some relevant experimental and
theoretical works performed in recent years, that have helped us to think about
cuprates the way we do now. Afterwards, I introduce
Landau Fermi liquid theory in a standard text book way.
The question of validity of Fermi liquid theory in 2-d is then raised 
and investigated by searching for singularities in Landau's $f$-function.
I show that the interaction function between two quasiparticles 
whose momenta approach each other near a curved point of the Fermi 
surface, contains 
a 1-d singularity not strong enough to change the Fermi liquid behavior. 
On the other hand, inflection points provide 2-d singularities that have
to be taken seriously in Fermi liquid considerations.
I then introduce nearly antiferromagnetic Fermi liquid theory (NAFL), 
which is a phenomenological theory proposed to describe
high $T_c$ systems. I mainly focus on the 
self-consistency of the theory in calculations. I criticize the theory
on the basis of overlooking the vertex corrections in the strong coupling
calculations of the transition temperature $T_c$. I calculate the 
first vertex correction for an optimally doped system and show that
it is of the same order of magnitude as the bare vertex. Migdal's theorem
is therefore not valid and Eliashberg formalism is not applicable to
this situation. The same conclusion is obtained even after inclusion 
of the quasiparticle residue $Z$ to the calculation. The sign of the vertex
correction is then considered. I show that the positive sign of the
vertex correction for the optimally doped system requires a phase transition
of some sort as the doping is decreased.

Part II of the thesis is devoted to the vortex lattice properties of high
$T_c$ superconductors. I establish a method to study vortex lattice
properties of d-wave superconductors based on a generalization of the London
model. The method has the advantage of simplicity as well as having
very few free parameters (one at most) compared to other methods.
The generalized London free energy is obtained from an s-d mixing 
Ginsburg-Landau free energy and also from the microscopic theory
of Gorkov. The generalized London equation is found to be
analytic at high temperatures. At very low temperatures however
non-analyticities arise as a result of the 
nodes on the superconducting gap. I then present the results of our
calculations of some measurable quantities, such as the vortex lattice geometry
and the effective penetration depth (as defined in $\mu$SR experiments). 
Comparison between our results and different experimental data is then 
performed.
Especially our prediction for the magnetic field dependence of the effective
penetration depth at $T=0$, which is recently observed in $\mu$SR
experiments with excellent agreement, is discussed.

\eject
.
\vspace{6cm}

\begin{center}

{\bf To my parents}

\end{center}

    \tableofcontents
    \listoftables
    \addcontentsline{toc}{chapter}{List of Tables}
    \listoffigures
    \addcontentsline{toc}{chapter}{List of Figures}
\preface{Acknowledgement}

First, I would like to give my special thanks to my both supervisors
Ian Affleck and Philip Stamp for teaching me the basics of physics and 
research and also for all their support, encouragement and patience.
I especially would like to thank Philip Stamp for directing me to condensed
matter physics and teaching me superconductivity and basics of high $T_c$
superconductivity, and also Ian Affleck for teaching me the concepts of
condensed matter physics during several magnificent courses and informal 
discussions, and supporting me whenever I needed. I would like to 
thank both for being a friend and sometimes like a caring father to me.

I also would like to take the opportunity to thank Victor Barzykin, Alexander
Zagoskin and most of all Igor Herbut for very stimulating and encouraging
discussions and for explaining difficult materials to me in simple language.
I am also grateful to Walter Hardy, and Doug Bonn and Rob Keifl for providing
me up to date experimental information, Jeff Sonier for several 
useful discussions about $\mu$SR experiments and providing me his data, 
Marc Julien for teaching me the basics of NMR experiments and especially
Saeid Kamal for helping me understand microwave experiments as well as
being a supporting friend for me during these five years.

I also would like to thank all my friends, specially Shahrokh Parvizy and
Ali Sheikholeslam for encouraging me to choose physics as my career,
all my friends in Canada especially Siavash Jahromi, Mahan Movassaghi, 
Ali Mirabedini and Nima Ahmadvand for all good times I had with them
during my years of Ph.D., and Andrea and Jeff for lending me their 
laptop with which I wrote most parts of this thesis.  

Outside the physics world, I would like to thank my parents and my sister
Parvaneh for their loving support and encouragement and believing in me
and providing me all the supports I needed all through the years of my study.
I also want to thank my uncle Ali Joukar and his wife Roya Sadeghi for making
Vancouver my home by all their love and support.

Finally I would like to give my especial thank to my gorgeous wife Roxana
Pardis for her love and support, for always being with me during difficult times 
of life, giving me lessons of living and making me believe in myself.


\part{Fermi Liquids, Non-Fermi Liquids and Nearly Antiferromagnetic Fermi Liquids}

\chapter{Introduction}

\begin{figure}[t]
\epsfysize 8cm
\epsfbox[-90 180 600 600]{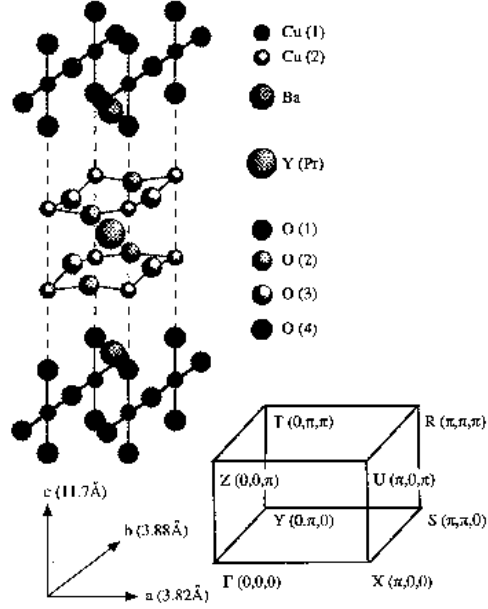}
\caption{Crystal structure of YBCO}
\label{htc}
\end{figure}

It has been more than a decade since the discovery of the first high 
temperature superconductor \cite{bednorz}, and still
no consensus about the microscopic theory relevant to 
cuprates exists. New theories have been emerging
one after another and disputes about the older theories are still unsettled.
A general feature of all cuprates is their strong anisotropy which makes
them effectively 2-d electronic systems. Fig. \ref{htc} shows the crystal 
structure of the YBCO compound which is one of important high $T_c$ materials.
At the center there are two CuO$_2$ planes separated with Y atoms in between. 
Between these two planes and another two planes above or below them, there are
oxygen chains (that play crucial role in the doping of the system) and 
also Ba atoms. It is generally agreed that the CuO$_2$ planes play an
important role in determining the characteristics of cuprates at all dopings.
Fig. \ref{htc} shows the crystal structure of YBCO material
In the early days of high Tc's, Anderson \cite{anderson} proposed an 
effective Hamiltonian
which describes a single band 2D Hubbard model with an infinite on site 
repulsion $U$. Soon after, Zhang and Rice 
\cite{zhang} tried to justify the Anderson's suggested model 
(known as the t-J model) for cuprates. 
They proposed that the holes which primarily reside
on oxygen sites would delocalize onto four O sites around Cu$^{+2}$ ions.
The combination of a delocalized O hole and a Cu$^{+2}$ ion forms 
a singlet which would hop between Cu$^{+2}$ sites 
in the  same way as a hole does in a simple Cu$^{+2}$ square lattice.

In the absence of doping, CuO$_2$ planes are half filled and
form a quasi-2D antiferromagnet with an insulating N\'eel ground state.
Numerical simulations,
experiments, and scaling theory, all suggest that above the N\'eel 
temperature, the antiferromagnetic spin
correlations are well described by the
isotropic 2D Heisenberg model \cite{Manousakis}
with exchange coupling $J \simeq 1550 K$ between nearest neighbors.
However, according
to the Hohenberg-Mermin-Wagner theorem \cite{MW}, classical fluctuations
destroy N\'eel ordering at an arbitrarily small nonzero temperature.
The small ($\sim 10^{-5}J$), but finite correlations between
the planes are believed to be responsible for the observed long-range order
\cite{Manousakis}. Chakravarty, Halperin, and Nelson
\cite{CHN} on the other hand, have employed quantum non-linear
$\sigma$ model for the long-wavelength action
of the $2D$, $s=1/2$ Heisenberg model.
A scaling analysis of this model 
describes correctly both experimental and numerical data
\cite{Manousakis}.

\begin{figure}[t]
\epsfysize 5cm
\epsfbox[-100 320 600 580]{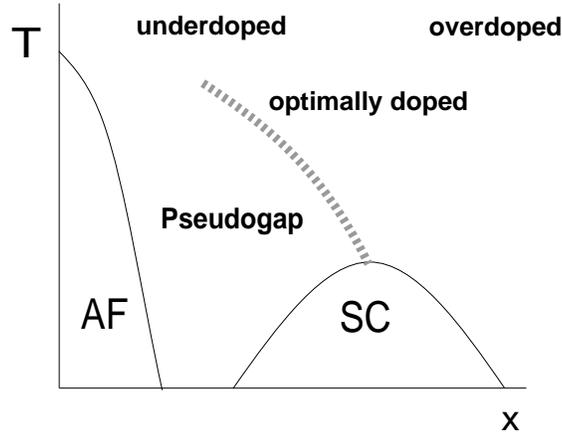}
\caption[Phase diagram of high temperature superconductors. ]
{Phase diagram of high temperature superconductors. $x$ represents
doping, AF antiferromagnetically ordered phase and SC superconducting phase.
Upper line is the crossover to pseudogap state. \cite{timusk0} }
\label{pd}
\end{figure}

Introducing a small amount of holes into the CuO$_2$ planes, 
the N\'eel temperature falls off rapidly from its undoped value.
Increasing the doping level even further, the system moves into a state
with  a metallic normal state and a transition to a superconducting
ground state at low temperatures (cf. Fig. \ref{pd}). 
The superconducting transition 
temperature increases initially as doping is increased. After
reaching its maximum at optimal doping, it starts to decrease and
the superconductor enters a so called overdoped regime (Fig. \ref{pd}). 
All over the metallic phase,
the antiferromagnetic correlation is short ranged with $\xi$
between one and (up to) fifteen lattice spacings, depending on the 
material, doping and temperature.

Experiments evidently indicate that the electronic properties
of the underdoped and overdoped materials are different
\cite{marshall,Mason,fukuzumi,ando,berthier1}; especially the pseudogap
phase which exits only for underdoped compounds \cite{timusk0} (Fig. \ref{pd}).
Among experiments, inelastic neutron scattering (INS)
and nuclear magnetic resonance (NMR) experiments 
provide important information about the magnetic properties of the system.
INS directly probes the spin response 
function, $\chi''({\vec q}, \omega)$. 
In La$_{1.85}$Sr$_{0.15}$CuO$_4$ the INS
experiments \cite{Mason} display four sharp magnetic peaks located at
$\delta q \simeq \pi/4$ away from the antiferromagnetic wave vector.
These incommensurate peaks were also observed recently in 
YBa$_2$Cu$_3$O$_{6.6}$.  \cite{dai}
Like many experiments, INS has
some inherent problems. For example it is difficult
to determine experimentally the absolute value of intensity.
It is also not easy to separate the phonon
spectrum from the magnetic excitations.

NMR experiments \cite{berthier2,slichter} on the other hand, 
are much more exact, and can detect 
much lower intensities. However, NMR is an indirect probe and 
its data analysis involves using a theoretical Hamiltonian
connecting nuclear and electron spins, the hyperfine Hamiltonian.
Moreover, since NMR experiments are done at very low frequencies
($\sim$ nuclear spin Zeeman splitting),
no information about $\omega$-dependence of
$\chi(\omega, T)$ can be obtained. Early measurements
of the Magnetic Hyperfine Shift (Knight shift) at the oxygen sites
showed that most of the total spin polarization was carried by the 
copper sites \cite{takigawa,horvatic}. 
Alloul {\it et al.} \cite{alloul} reported that in  
YBa$_2$Cu$_3$O$_{6+x}$, the knight shift of Yttrium scales linearly with
the macroscopic susceptibility. This was interpreted as a proof of the 
existence of a single spin fluid in agreement with the t-J model, because
yttrium is coupled to the magnetic susceptibility of the CuO$_2$ 
planes mainly through the polarization of the oxygen orbitals.
Further measurements of knight shift showed
the same T-dependence on all nuclear sites \cite{berthier2}. 
Similar result was obtained for all other cuprates indicating that the single
spin fluid nature is a common feature of all these materials.
Assuming a single fluid picture, Shastry \cite{shastry} and also 
Mila and Rice \cite{mila} (SMR) proposed a hyperfine 
Hamiltonian in which the only source of coupling is the localized 
electronic spins at the copper sites.

While the magnetic hyperfine shift (Knight shift) gives a measure of 
the uniform magnetic susceptibility,  the Nuclear
Spin Lattice Relaxation Rate ($1/T_1$)  provides information about the
imaginary part of the dynamic spin susceptibility $\chi({\vec q}, \omega)$.
An important experiment by Takigawa \cite{takigawa} on 
YBa$_2$Cu$_3$O$_7$, in the early days, 
displayed a significant difference 
between the temperature dependence of  
$1/T_1$ for the CuO$_2$ plane copper and oxygen 
sites. The relaxation rate for oxygen
sites (and also $^{89}$Y sites) obeys Korringa relation 
$^{17}T_1T\approx Const.$
while the relaxation rate for copper sites is much shorter than for
$^{17}$O and $^{89}$Y sites and shows non-Korringa behavior.
Using SMR Hamiltonian, Moriya \cite{moriya} showed that 
\be
(T_1 T)^{-1} \propto \lim_{\omega \rightarrow 0} \sum_{\vec q}
F({\vec q}){\chi^{\prime\prime}({\vec q}, \omega)\over \omega}
\label{T1}
\ee
where $\chi^{\prime\prime}({\vec q}, \omega)$ is the imaginary part of the
magnetic susceptibility and $F({\vec q})$ is a form 
factor which depends on the site and the direction of the magnetic field.
For $^{17}$O sites, $F({\vec q})$ is peaked at ${\vec q}=0$ and vanishes at
the antiferromagnetic (AF) wave vector ${\vec Q}=(\pi/a, \pi/a)$. 
For $^{63}$Cu sites on the other hand, it is peaked at ${\vec q=\vec Q}$. Thus different nuclei
probe different regions of the  momentum space of $\chi^{\prime\prime}({\vec q},0)$.
The significant difference between
the $^{63}$Cu and $^{17}$O relaxation rates could then be understood as 
a result of an enhanced susceptibility at ${\vec q=\vec Q}$. 

Another important quantity measured by NMR experiments is the nuclear
spin-spin relaxation rate ($1/T_{2G}$) which is the rate of relaxation of
the nuclear spins among themselves and is usually much larger than 1/$T_1$.
In most solids $T_{2G}$ is temperature independent and is related to
spin-spin dipolar coupling and therefore doesn't give any information
about electron susceptibility. In cuprates however, because of the
antiferromagnetic enhancement of the spin susceptibility, indirect
coupling of spins via nonlocal spin susceptibility dominates the 
nuclear spin-spin coupling. Thus $T_{2G}$ can provide important
information about the antiferromagnetic exchange between electron spins
as was pointed out first by Pennington {\it et al.} \cite{pennington}.
Measurements of $1/T_{2G}$ in different materials with different dopings
have found $1/T_{2G}$ to increase with decreasing temperature, with sometimes
a flattening or small decrease close to $T_c$ \cite{berthier2}. This is
considered as a proof for the existence of AF correlations in all cuprates.

To have a quantitative description, Millis, Monien and Pines (MMP) 
\cite{MMP} proposed a phenomenological 
form for the magnetic susceptibility $\chi ({\vec q},\omega)$ 
which has a peak near the 
Antiferromagnetic (AF) wave vector ${\vec Q}=(\pi/a, \pi/a)$
\be
\chi ({\vec q},\omega)={\chi_Q\over {1+\xi^2({\vec q- \vec Q})^2-i\omega /\omega_{sf}}}
+{\chi_0\over 1-i\omega/\Gamma_0}
\label{chi}
\ee
where $\chi_0$ is the measured bulk spin susceptibility which in general is 
temperature dependent, $\chi_Q$ is the staggered susceptibility which is some
orders of magnitude greater than the bulk susceptibility and  
$\Gamma_0$ is an energy scale of order of 
quasiparticle bandwidth or Fermi energy. The first term in Eq.(\ref{chi})
produces a peak at AF wave vector ${\vec Q}$ and the second term is the
ordinary Fermi liquid susceptibility. For the $^{17}$O sites, the effect of 
the first term is suppressed by the form factor $F(\vec q)$ (cf. Eq. (\ref{T1}))
and the second term (the Fermi
liquid term) is the only term that remains. To the lowest order,
Eq.(\ref{chi}) gives $(^{17}T_1 T)^{-1} \propto \chi_0(T)$ which is consistent with 
the Korringa relation. For copper sites however, the contribution of the
first term is dominant near ${\vec Q}$ and one can safely neglect the second term. 
Assuming large correlation length $\xi$, the form factor can be treated as a
constant and the integration can be simply carried out to get 
\be
(^{63}T_1 T)^{-1} \propto {\chi_Q \over \omega_{sf}\xi^2}={\alpha \over \omega_{sf}}
\ee 
where $\alpha$ is defined by $\chi_Q=\alpha \xi^2$ and assumed to 
be temperature independent. Thus the spin lattice relaxation rate provides  
information about the temperature dependence of $\omega_{sf}$ in the MMP model
\be
^{63}T_1 T \propto \omega_{sf}/\alpha
\ee
This is completely different from the $^{17}$O case and has non-Korringa behavior.

In the early days of high $T_c$'s, there existed some speculations 
that antiferromagnetic correlation was responsible for Cooper pairing 
in high $T_c$ materials \cite{bealmonod,bc,scalap1,scalap2,moriya2} (the idea of
d-wave pairing already existed in the context of heavy fermions 
\cite{vg,sigrist}). Inspired by paramagnon theory, Pines and collaborators 
\cite{MBP,pines1} proposed a (phenomenological)
nearly antiferromagnetic Fermi liquid (NAFL) theory to describe pairing
process as well as normal state properties of high $T_c$ superconductors.
The theory assumes that the low lying excitations of a high $T_c$ 
system in its normal state, are Fermi liquid quasiparticles 
interacting with each other via
antiferromagnetic spin-fluctuations. The susceptibility (\ref{chi}) 
can be thought of as the
propagator of bosonic particles that describe spin fluctuations. 
One of the first predictions of NAFL theory was d$_{x^2-y^2}$-wave
pairing which was then confirmed by various experiments \cite{Hardy,van}.
Transition temperature was calculated in weak \cite{MBP} and strong 
\cite{MP1,MP2} coupling limits and a $T_c$ as high as 90K was obtained
using several fitting parameters. 

NAFL theory, as we mentioned above, is constructed based on the 
assumption that the excitations of the system are Fermi liquid quasiparticles. 
Almost none of the normal state properties of cuprates agree with Fermi
liquid theory (at least in low to moderate doping regimes). 
The proximity to Mott-insulating antiferromagnetic phase might
suggest that non-Fermi liquid properties are actually signatures of
non-Fermi liquid excitations. NAFL theory on the other hand believes
that the anomalous behavior is a result of strong interaction between 
the Fermi liquid quasiparticles due to antiferromagnetic spin fluctuations.

Orthogonal to the above philosophy was 
Anderson's proposal in 1990 \cite{anderson1,anderson2} that 2-dimensional
Fermi liquids are unstable to the formation of a 2-d Luttinger liquid.
Since Anderson,
theorists have been trying to understand if and when ``non-Fermi liquid"
(NFL) can appear in itinerant Fermi systems. There is strong evidence
\cite{willett} for NFL behavior in the $\nu=1/2$ quantum Hall liquid;
theoretical
understanding of this assumes a singular gauge interaction between the
quasiparticles \cite{halperin}.
The high $T_c$ problem is more complex because the
lattice introduces an on-site constraint for strong Hubbard repulsion- which
means the lattice problem is fundamentally different from a continuum model.
Perturbative investigations of the nature and existence of singular interactions
in the low-density regime \cite{engelbrecht,fukuyama1,stamp1}
have not shown NFL behavior, even in a finite
box \cite{stamp1} (although they do reveal weak singular structure in the
quasiparticle interactions \cite{fukuyama1,stamp1}). However, these results
do not rule out non-perturbative singular behavior in the higher density
lattice problem- no consensus yet exists on whether these exist in 2 dimensions.
We stress that this question is quite different
from that of the existence of instabilities to ordered phases
(e.g., Superconducting or SDW instabilities),
which can be understood using standard methods \cite{shankar}.

As discussed by various authors \cite{anderson1,anderson2,stamp1}, NFL
behavior is typically expected if the interaction between quasiparticles with
momenta ${\vec p}$ and ${\vec p^\prime}$ is singular when $\vec Q=\vec p-\vec{p^\prime}
\rightarrow 0$. It is nevertheless essential to satisfy all relevant Ward
identities and crossing symmetry in any calculation, given the way singular
behavior in one channel can influence that in the others.
An approach which reveals the structure in the irreducible interactions
as a function of ${\vec Q}$ and which satisfies the above requirements,
involves calculating the irreducible 4-point vertex
$f^{\sigma\sigma^\prime}_{p p^\prime}$ perturbatively \cite{abrikosov},
and then using this to determine the full scattering function and
quasiparticle dispersion. If $f^{\sigma\sigma^\prime}_{pp'}$ is not
singular, then it is nothing but Landau's {\it f}-function. 

In the next chapter, after an overview on Landau Fermi liquid theory, 
we perform a calculation of
$f^{\sigma\sigma'}_{pp'}$ along the line mentioned above for a regular
Fermi surface and also near inflection points of the Fermi surface. 
Our calculation for a curved
Fermi surface reproduces the previous results \cite{stamp1} existing in
the literature. Near inflection points however, we find
that $f^{\sigma\sigma'}_{pp'}$ is singular in ${\vec Q}$ already
at 2nd order in a short-ranged repulsive interaction.  
We stress that the NFL behavior found previously for Fermi surfaces having
finite curvature everywhere appeared only at infinite order, and
was too weak to destroy Fermi liquid behavior \cite{fukuyama1,stamp1}.
Calculating the imaginary part of the self energy to second order in
the singular interaction, we find a frequency
dependence different from the expected Fermi liquid form. 

Besides the controversy about the Fermi liquid starting point, the validity
of Eliashberg formalism used in NAFL theory to calculate $T_c$ is under
question. Migdal's theorem is an essential ingredient of Eliashberg
formalism which states that the vertex corrections to the bare coupling
between electrons and the bosonic medium (here spin fluctuations) are small.
In the case of phonons, the small ratio of Debye frequency and Fermi energy
($\omega_D/E_{_F}$) makes the vertex corrections
small \cite{migdal}; despite the large bare coupling. In the case of
spin-fermion coupling on the other hand, 
the only relevant energy scales are $E_{_F}$
and the antiferromagnetic exchange coupling $J$. The ratio $J/E_{_F}$
however, is of order one \cite{schrieffer1}. Thus it is not clear why
vertex corrections should be small in the spin fermion interaction.

Calculations of the vertex corrections close to the antiferromagnetic ordered
state \cite{schrieffer1,altshuler,chubu2} showed that the vertex corrections
are actually the same order as the bare vertex and therefore are not small. 
Schrieffer \cite{schrieffer1} showed that in the ordered state, the dressed
vertex should vanish at the antiferromagnetic wave vector. 
The vertex corrections therefore should be equal to the bare vertex but 
with opposite sign. The magnitude of the vertex correction
at higher dopings is more important because the superconducting transition is
suppressed at low doping.

In chapter 3, we first give a brief introduction to paramagnon theory, which
is a theory for Fermi liquid systems near a ferromagnetic instability. We then
turn to the nearly antiferromagnetic Fermi liquid theory pointing out the
similarities and differences between NAFL and paramagnon theory.
The self-consistency of NAFL theory is then discussed. 
We calculate the first
correction to the bare spin-fermion vertex. We find that the vertex correction
is actually big and of the same order as the bare vertex. The self-consistency
of the calculation including quasiparticle renormalization is discussed
afterwards.

The sign of the vertex corrections is another subject of dispute. As we mentioned
above, the sign of the vertex corrections is negative near the ordered state.
At optimal doping the sign of our vertex correction is also negative in
disagreement with some other results \cite{CMM}. We talk about the
sign of the vertex corrections at the end of the 3rd chapter. Our discussion is
based on the flow of the renormalized vertex versus change in doping.
We argue that the positive sign of the vertex corrections at the optimal
doping requires a phase transition at some intermediate doping. In fact, for 
other reasons, this phase transition has been suggested by Chubukov to be 
a Lifshitz transition in the topology of the Fermi surface \cite{chubu3}.
However, there is no experimental evidence for this or any other phase 
transitions at intermediate doping at the time of writing \cite{arpes}.

\chapter{Fermi Liquid Theory}

\section{Landau Theory of Fermi Liquids}

Systems of strongly interacting electrons are very difficult, if not 
impossible, to solve. Perturbative methods fail to work as a 
result of the strong coupling between particles. Non-perturbative methods,
also, are not quite well developed for systems of more than one dimension.
It was Landau's ingenuity that realized the low energy excitations of a
strongly interacting fermion system could still be fermionic quasiparticles
similar to non-interacting electrons. Proposed in 1957 \cite{landau1,landau2},
Landau's Fermi liquid theory has been extremely successful in describing 
low temperature properties of liquid $^3$He \cite{leggett0,baym0} as well as 
interacting electrons in metals \cite{pines0}. It especially
explains the success of non-interacting electron theory for low temperature 
properties of metals.

In his theory, Landau assumes that if one adiabatically 
turns on the interaction between electrons,
the non-interacting ground state evolves into an interacting
ground state with a one to one correspondence between the bare particle
states and the quasiparticles in the interacting state. A necessary condition 
for this to happen is that no bound states should form during this
process. A system that possesses this property is called ``Normal Fermi 
Liquid". Superconductors therefore do not belong to this category. 
In Landau's picture, the number of quasiparticles, N, is the same as 
the number of bare electrons. They carry the same charge as electrons
and obey Fermi statistics. One has to be careful not to go too far with this 
picture. For example, the ground state energy is not the same as 
the sum of the quasiparticles energies. As we will see later, the 
quasiparticle states are meaningful only
near the Fermi surface.  A better way, is to define the 
quasiparticles as excitations of the system above its ground state and 
also quasiholes as absence of particles in the ground state.
More precisely, a quasiparticle state with momentum 
$\vec p$ ($>p_{_F}$, Fermi momentum) is defined by adiabatic evolution 
of a non-interacting Fermi sea with an extra electron at momentum 
$\vec p$ (and likewise a quasihole state). 

Yet another assumption is necessary to establish the Landau Fermi 
liquid theory. The assumption is that the quasiparticles interact with 
each other via an interaction function $f^{\sigma \sigma'}_{\vec p \vec{p'}}$.
(For simplicity from now on we suppress the spin indices and denote 
the $f$-function by $f_{pp'}$. We put them back when necessary later.) 
The scattering of quasiparticles due to this interaction causes the
quasiparticles to decay. If the decay rate is large, then the quasiparticle 
lifetime, $\tau_p$ may be too short. The quasiparticle states therefore will
be ill-defined. In other words, the quasiparticle states are not 
exact eigenstates of the interacting Hamiltonian, i.e. the energy
levels have finite width ($\sim 1/\tau_p$).  In order to have distinguished
energy levels, the width should be smaller than the energies, or
\be
{1 \over \tau_p} \ll (\epsilon_p - E_F)
\ee
The key point here
is that the scattering of these quasiparticles occurs at a vanishingly small 
rate as the quasiparticles get close to the Fermi surface. The reason behind 
this lies in the Pauli exclusion principle which restricts the phase space
available for quasiparticles to scatter to. At $T=0$ this phase
space is proportional to $(p-p_{_F})^2$ where $p$ is the momentum of
the scattered particle \cite{mattuck}. The scattering therefore, is very small 
when $p \approx p_{_F}$. At higher temperatures, the scattering rate 
increases like  $T^2$. As a result, the quasiparticle picture
is meaningful only at low temperatures and for quasiparticles 
with momenta close to the Fermi surface. Thus Landau's picture, with
N quasiparticles filling up the states up to the Fermi surface, should 
be considered only as a formal assumption. Still this picture is useful 
especially in the context of Luttinger's theorem that allows to state that
for a spherical Fermi surface, the value of $K_F$ is the same in the interacting
and non-interacting system.
 
Let us apply a weak perturbation to the system to take it away from
its ground state. The effect of this perturbation is to change the occupation
number by $\delta n_{\vec p}$. Landau postulated that the change in the
energy of the system is given by the expansion
\be
\delta E=\sum_{\vec p} \epsilon^0_{\vec p} \delta n^0_{\vec p}
+ {1 \over 2V} \sum_{\vec p, \vec{p'}} f_{pp'}
\delta n^0_{\vec p} \delta n^0_{\vec{p'}}
\label{Enn}
\ee
where $\epsilon^0_{\vec p}$ is the quasiparticle energy and $V$ is the
volume of the system. In general, one can expand to higher powers
of $\delta n_{\vec p}$. However, expansion to the second order is enough for 
most practical purposes. The quasiparticle energy affected by other 
quasiparticles is given by
\be
\epsilon_{\vec p} = {\delta E \over \delta n^0_{\vec p}} 
=\epsilon^0_{\vec p} + {1 \over V}
\sum_{\vec{p'}} f_{pp'} \delta n^0_{\vec{p'}}
\ee
Landau's $f$-function is therefore  the second derivative of the energy
with respect to $\delta n_{\vec p}$ and $\delta n_{\vec{p'}}$
\be
f_{pp'}={\delta^2 E \over \delta n^0_{\vec p} \delta n^0_{\vec{p'}} }
\ee
Since there is a one to one correspondence between quasiparticle
states and non-interacting electron states, the quasiparticles should 
obey Fermi statistics (a more rigorous proof involves maximization of
the entropy. \cite{pines0})
\be
n^0_{\vec p}={1 \over e^{\beta (\epsilon_{\vec p} - \mu)} +1 }
\ee
where $\mu$ is the chemical potential; which can be shown to be equal
to the Fermi energy $E_F$ at T=0 \cite{pines01}.
For a SU(2) symmetric system, one can write
(putting back spin indices again)
\be
f^{\sigma \sigma'}_{\vec p \vec{p'}}=f_{\vec p \vec{p'}} + 
4\vec \sigma \cdot \vec{\sigma'}  \phi_{\vec p \vec{p'}}
\ee
These functions are usually expressed in terms of spherical 
harmonics 
\ba
f_l=\left({2l+1 \over 4\pi}\right) \int d\Omega P_l(\cos \theta) 
\left[ f_{\vec p \vec{p'}} \right]_{_{|\vec p|=|\vec p'|=p_{_F}}} \nn \\
\phi_l=\left({2l+1 \over 4\pi}\right) \int d\Omega P_l(\cos \theta) 
\left[ \phi_{\vec p \vec{p'}} \right]_{_{|\vec p|=|\vec p'|=p_{_F}}} 
\ea
$f_l$ and $\phi_l$ are known as Landau parameters.
In terms of these parameters, the quasiparticle's effective mass is given by 
(for all derivations and proofs see Ref. \cite{negele})
\be
{1 \over m} = {1 \over m^*} + {p_{_F} \over 3 \pi^2} f_1
\ee 
Observable quantities are usually given in terms of dimensionless
Landau parameters $F_l=N(0) f_l$ and $Z_l=N(0) \phi_l$, 
with the density of states at  the Fermi surface 
\be
N(0) = {1 \over V} \sum_{\vec k,\sigma} \delta (\epsilon^0_k - \mu)
= {m^* p_{_F} \over \pi^2}
\ee
In terms of $F_l$'s and $Z_l$'s, we have \cite{negele}
\ba
&&{\rm Effective \ Mass:} \qquad\qquad\quad
{m^* \over m}=\left( 1 + {F_1 \over 3} \right) \nn \\
&&{\rm Specific \ Heat:} \qquad\qquad\qquad c_V = {1 \over 3} m^* p_{_F} k_B^2  T \nn \\
&&{\rm Compressibility:} \qquad\qquad\ \ {1 \over \kappa}={n p_{_F}^2 \over 3m^*} (1+F_0)\nn \\
&&{\rm Sound \ Velocity:} \qquad\qquad\ \ \ c_1^2 = {p_{_F}^2 \over 3mm^*} (1+F_0) \nn \\
&&{\rm Magnetic \ Susceptibility:} \quad \ \chi_{_M}={\gamma^2 p_{_F} m^* \over
4\pi^2 (1+Z_0)} \nn
\ea
where $n=N/V$ is the density of electrons and $\gamma=e \hbar/mc$ 
is the electron's gyromagnetic ratio.

\section{Microscopic Foundation of Fermi Liquid Theory}

As emphasized in the previous section, the basic ingredient  of Landau's
Fermi liquid theory is the function $f_{pp'}$; which describes 
the interaction of low energy quasiparticles. The simple form of the 
interaction term in (\ref{Enn}), essential in Fermi liquid  
calculations, is far from obvious. 
Renormalization group theory \cite{shankar} provides a plausible derivation
for a low energy effective Hamiltonian in agreement with Fermi liquid theory. 
As we will see later however, renormalization group might miss 
some essential singularities in the quasiparticle interaction by 
putting all the momenta on the Fermi surface. Moreover, although renormalization group is a conserving approximation \cite{baym1,baym2}, 
it does not satisfy the Pauli principle \cite{stamp2,tremblay,chitov}. 
Another way, is to find a relation between $f_{pp'}$ and
the dressed two-particle interaction or 4-point vertex function in many-particle
theory. What we do in this section is to first derive  
$f_{pp'}$ for a low density electron gas by variation of the
ground state energy using second order perturbation theory.
We then discuss the connection between 
many-particle theory and Fermi liquid theory. In particular, we find a 
relation between Landau's $f$-function and 4-point vertex function in
many-particle theory.

\subsection{Calculation of ${\bf f_{pp'}}$ for a Dilute Fermi Gas}

In a dilute Fermi gas, one can calculate $f_{pp'}$
using Abrikosov-Khalatnikov's formalism \cite{abrikosov}
(for more detailed derivation, see also Ref. \cite{abrikosov0}, or
Ref. \cite{stamp1,beydaghyan} for derivation in 2-d). In the dilute limit 
one can assume short range interaction between electrons. The interaction Hamiltonian
therefore can be approximated by
\be
H_{\rm int}= {U \over V} \sum_{\vec q,\vec p,\vec{p'}}
c^\dagger_{\vec  p+ \vec q \uparrow}
c^\dagger_{\vec{p^\prime} - \vec q \downarrow}
c_{\vec{p^\prime}\downarrow}
c_{\vec  p\uparrow}
\ee
where $U$ is the coupling constant and $V$ is the volume of the system.
We do our expansion in powers of the s-wave {\it scattering amplitude}, 
$a$, which is given to the first order in $U$ by
\be
U={4 \pi a \over m}
\label{a0}
\ee
Applying perturbation theory to $H_{\rm int}$, the first order correction
to the ground state energy is given by
\be
E^{(1)}={U \over V} \sum_{1,2} n_1 n_2 Q_{12} = {\pi a N^2\over m V}
\label{E1}
\ee
Here the index (1) represents both momentum and spin, 
$n_i$ is the occupation
number and $Q_{ij}$ requires spins to be antiparallel
\be
Q_{ij} = {1 \over 4} (1- \vec \sigma_i \cdot \vec \sigma_j)
\ee
where $\sigma$'s are the Pauli matrices. To second order in perturbation 
theory we have
\be
E^{(2)}= \sum_{m \ne 0} {|(H_{\rm int})_{m0}|^2 \over E^{(0)} - E^{(0)}_m}
={U^2 \over V^2} \sum_{1,2,3,4} \delta_{_{p_{_1}+p_{_2},p_{_3}+p_{_4}}} 
{n_1 n_2 (1-n_3)(1-n_4) Q_{12}Q_{34} \over
\epsilon_1 + \epsilon_2 - \epsilon_3 - \epsilon_4 }
\label{E2}
\ee
Eq. (\ref{E2}) is divergent at large momenta. To get away from this divergence 
we have to expand everything in terms of $a$ instead of $U$. First notice that
(\ref{a0}) is not exact. To second order in $U$,  (\ref{a0}) can be written as
\be
{4 \pi a \over m} = \bar{U} = U + {U^2 \over V} \sum_{1,2,3,4}
 \delta_{_{p_{_1}+p_{_2},p_{_3}+p_{_4}}}
{Q_{34} \over \epsilon_1 + \epsilon_2 - \epsilon_3 - \epsilon_4 } 
\label{Ua}
\ee
Here, $\bar{U}$ is the renormalized coupling.
Inverting (\ref{Ua}),  $U$ can be written in terms of $a$
\be
U = {4 \pi a \over m} - {(4 \pi a )^2 \over m^2V} \sum_{1,2,3,4} 
 \delta_{_{p_{_1}+p_{_2},p_{_3}+p_{_4}}} {Q_{34} \over
\epsilon_1 + \epsilon_2 - \epsilon_3 - \epsilon_4 }
\label{ar}
\ee
substituting this into (\ref{E1}), we get a term second order in $a$ which should 
be combined with (\ref{E2}) to get the second order correction to the ground 
state energy. Combining all, after some manipulations, we get
\be
E^{(2)}= - {32 \pi^2 a^2 \over m^2 V^2} \sum_{1,2,3,4} 
 \delta_{_{p_{_1}+p_{_2},p_{_3}+p_{_4}}}{ n_1 n_2 n_3 Q_{12}Q_{34} \over
\epsilon_1 + \epsilon_2 - \epsilon_3 - \epsilon_4 }
\ee
This now converges at large momenta, since all momenta are limited
because of the $n_i$'s.
Taking the derivatives with respect to $n_{p\sigma}$ and $n_{p'\sigma'}$, we get
\ba
f^{\sigma \sigma'}_{pp'} &=& {4\pi a \over m}\delta_{\sigma,-\sigma'}
- {(4 \pi a)^2 \over m^2 V} \sum_{\vec k} n_{\vec k}
[ {2 \delta_{\sigma,-\sigma'}
\over \epsilon_{\vec p}+\epsilon_{\vec{p'}}
-\epsilon_{\vec k}-\epsilon_{\vec p + \vec{p'} - \vec k}} \nn \\
&& + {1 \over \epsilon_{\vec p}-\epsilon_{\vec{p'}}+
\epsilon_{\vec k}-\epsilon_{\vec k+\vec p - \vec{p'}}} +
{1 \over \epsilon_{\vec{p'}}-\epsilon_{\vec p}+
\epsilon_{\vec k}-\epsilon_{\vec k+\vec{p'}-\vec p}} ]
\ea
The first term in the bracket is the ``Cooper" channel and the last two terms 
belong to the ``crossed" channels. The denominators of the last two terms 
vanish when $\vec p = \vec{p'}$.  The singularities however get canceled 
between the two terms except for special cases (we will return to 
this point later).

\subsection{Fermi Liquid Theory and Many-Particle Theory}

In many-particle theory \cite{mattuck,negele,abrikosov0,mahan}, 
the excitations of the system above its ground
state are given by the poles of the single particle Green's function
$G(\omega,\vec k)$ \cite{mattuck}. In the absence of any phase 
transition,  $G(\omega,\vec k)$ has a form very similar to 
non-interacting Green's function. One can write \cite{mattuck}
\be
G(\omega,\vec k) = {Z_k \over \omega - \epsilon_{k} + i\tau^{-1}_k}
+ \Phi(\omega,\vec k)
\label{Grn}
\ee
where $Z_k$ and $\tau_k$ are quasiparticle residue and quasiparticle
lifetime respectively. The first term in the right hand side of (\ref{Grn})
is similar to the non-interacting electron Green's function
\be
G^{(0)}(\omega,\vec k) = {1 \over \omega - \epsilon_{k} + i\delta_k}
\label{G0}
\ee
where $\delta_k = \delta \ {\rm sgn}(\epsilon_{k}-\mu)$ with 
$0< \delta \ll 1$.  The second 
term, $\Phi(\omega,\vec k)$, represents the incoherent
part of the Green's function. 
For most calculations, the first term in (\ref{Grn}) is sufficient to give
correct physical behavior up to the desired accuracy. This explains why Fermi
liquid theory, which is very similar to non-interacting electron theory,
works so well for complex systems like metals. In any
calculation however, one has to make sure that the incoherent
part of the Green's function does not make significant contribution.

In terms of quasiparticle self-energy $\Sigma_{\vec k} (\omega)$, 
$Z_k$ and $\tau_k$ are give by \cite{mahan}
\ba
&&\tau^{-1}_k = - Z_k \ {\rm Im} \Sigma_{\vec k} (\epsilon_k) 
\label{tauk} \\
&& Z_k =  \left(1- {\partial \over  \partial \omega}{\rm Re} 
\Sigma_{\vec k} (\omega) \right)_{\omega=\epsilon_k}^{-1}
\label{Zk}
\ea
We can see from (\ref{G0}) that the quasiparticle lifetime in 
non-interacting systems is infinite ($\tau_k \rightarrow \infty$).
In interacting systems however, $\tau_k$ is finite. In fact, one can show that
\cite{mattuck} for a 3-d Fermi liquid $\tau_k^{-1} \propto 
(\epsilon_k - \mu)^2$, and therefore the lifetime diverges as the particle
gets close the Fermi surface. The closer to the Fermi surface, the more
stable the quasiparticle state is. For states with momenta far away from 
the Fermi surface, the quasiparticle concept is not meaningful as 
emphasized before. 

For small $\omega$, the self energy behaves like
\ba
{\rm Im} \Sigma_{p_{_F}} (\omega) \sim \omega^2 \qquad {\rm and} \qquad
{\rm Re} \Sigma_{p_{_F}} (\omega) \sim \omega
 \ea
Substituting into (\ref{Zk}) we find $Z$ ($\equiv Z_{(k=p_{_F})}$) a nonzero 
positive constant 
(always $<1$). $Z$ measures the amount of the spectral weight 
accumulated in the quasiparticle peak \cite{mahan}. 
It has other meanings as well. 
It can be shown \cite{luttinger1,pines01} that the occupation 
number $n_k$ is always discontinuous at the Fermi surface 
with a jump exactly equal to $Z$. Thus the Fermi surface concept 
survives even after turning on the interaction. The existence of the Fermi 
surface, in fact, is equivalent to the validity of Fermi liquid theory.
When $Z = 0$, as it is in 1-d Luttinger liquids, the occupation 
number doesn't have a jump at $p_{_F}$. Consequently, there is 
no well defined Fermi surface; although higher order derivatives of 
$n_k$ still are singular right at the position of the Fermi surface. 
Fermi liquid theory therefore breaks down in such cases.

The dressed interaction between two quasiparticles
is given by 4-point vertex function $\Gamma^{(2)} (P_1,P_2;P_3,P_4)$ 
which is related to the 2-particle Green's function $G^{(2)}$ by 
\ba
&&G^{(2)}(P_1,P_2;P_3,P_4)= (2\pi)^8 G(P_1)G(P_2)
\left[\delta(P_1-P_3)\delta(P_2-P_4) - \delta(P_1-P_4)\delta(P_2-P_3)\right]
\nn \\ && + \ i (2\pi)^4 \delta (P_1+P_2-P_3-P_4) \
G(P_1)G(P_2)G(P_3)G(P_4)\ \Gamma^{(2)} (P_1,P_2;P_3,P_4) 
\ea
with $P=(\nu, \vec p)$. Since in Fermi liquid theory, interaction 
between two quasiparticles is described by $f_{pp'}$, it is
natural to expect connection between $f_{pp'}$ and $\Gamma^{(2)}$.
To see the connection, let us define $\Gamma (P,P'; K)$ by
\be
\Gamma (P,P'; K) = \Gamma^{(2)} (P,P';P+K,P'-K)
\ee
where $K=(\omega, \vec k)$.
It has been shown that the following integral equation 
holds under certain assumptions \cite{negele}
\be
\Gamma (P,P'; K) = \Gamma^{(\omega)} (P,P') +
{Z^2 p_{_F}^2 \over (2\pi)^3} \int d\Omega \Gamma^{(\omega)} (P,Q)
{\hat{q} \cdot \vec k \over \omega - \hat{q} \cdot \vec k}
\Gamma (Q,P'; K)
\label{Gamma0}
\ee
where $Z$ is the quasiparticle residue, $Q=(\mu, p_{_F}\hat{q})$,
with $\mu$ being the chemical potential, and the function 
$\Gamma^{(\omega)} $ is defined by
\be
\Gamma^{(\omega)} (P,P')= \lim_{\omega \rightarrow 0} 
\lim_{\vec k \rightarrow 0} \Gamma (P,P',K)
\label{gmo}
\ee
The order of the limits plays important role here (we will see 
this shortly). 
The integration in (\ref{Gamma0}) is over all directions of $\hat{q}$.
The function $\Gamma^{(\omega)}$ is closely related to the Landau's  
$f$-function \cite{abrikosov0}. More precisely, at low energies 
putting all the momenta on the Fermi surface we have
\be
f_{pp'} = Z^2 \Gamma^{(\omega)} (\hat{p},\hat{p}')
\ee
The notation means that $P=(\mu, p_{_F}\hat{p})$ and 
$P'=(\mu, p_{_F}\hat{p}')$. 
$ f_{pp'} $ does not have direct physical meaning by itself \cite{abrikosov0}.
However, one can change the order of limits in (\ref{gmo}) to define
\be
\Gamma^{(k)} (P,P')=  \lim_{\vec k \rightarrow 0}  
\lim_{\omega \rightarrow 0} \Gamma (P,P',K)
\label{gmk}
\ee 
$Z^2\Gamma^{(k)}$ is actually the physical forward scattering 
interaction between two quasiparticles \cite{abrikosov0}. 
Taking the limit $\vec k \to 0$ and $\omega/k \to 0$ in (\ref{Gamma0}), one
can find the relation between $Z^2\Gamma^{(k)}$ and $f_{pp'}$ 
\be
Z^2 \Gamma^{(k)} (\hat{p},\hat{p}') = f_{pp'} -
{p_{_F}^2 \over (2\pi)^3} \int d\Omega \ f_{pq} \
Z^2 \Gamma^{(k)} (\hat{q},\hat{p}')
\ee
The quantity $Z^2 \Gamma^{(k)}$ is usually called scattering amplitude
or T-matrix \cite{baym0,stamp1}. More precisely the T-matrix 
$t_{pp'}(\vec q, \omega)$ is defined by
\be
t_{pp'}(\vec q, \omega) = Z_p Z_{p'} \Gamma (P,P';Q)
\ee
A modified version of (\ref{Gamma0}) gives $t_{pp'}(\vec q, \omega)$
in terms of $f_{pp'}$ \cite{baym0}
\begin{eqnarray}
t_{pp'}(\vec q,\omega)= f_{pp'}
+\sum_{\vec k} f_{pk} \ {{\vec q \cdot \vec v_{k}} \over
{\vec q \cdot \vec v_{k}} - \omega + i\delta} \ n^{(0)}_{\vec k}
\ t_{kp'}(\vec q,\omega). 
\label{B-S}
\ea
This equation is usually known as Bethe-Salpeter equation. One can again
write
\be
t^{\sigma \sigma'}_{\vec p \vec{p'}}=t^s_{\vec p \vec{p'}} + 
4\vec \sigma \cdot \vec{\sigma'}  t^a_{\vec p \vec{p'}}
\ee
In the limit $q \to 0$, $\omega/q \to 0$, and for $p$ and $p'$ on the Fermi
surface, Eq. (\ref{B-S}) gives
\ba
t^s_l={F_l \over 1+ F_l/(2l+1)} \nn \\
t^a_l={Z_l \over 1+ Z_l/(2l+1)}
\label{tZ}
\ea
where $t^s_l$ and $t^a_l$ are spherical harmonics.

\section{Fermi Liquid Theory in 2-d}

In the last section, we introduced Landau's Fermi liquid theory as a
powerful tool to describe fermionic systems in 3-d. It is well known, on
the other hand, that 1-d electron systems are not Fermi liquids but
Luttinger liquids \cite{1-d}. The question for 2-d systems is more subtle and has
been a subject of controversy for almost a decade. Anderson in 1990
\cite{anderson1,anderson2} suggested that 2-d interacting electron systems 
are also Luttinger liquids and not Fermi liquids. He conjectured that  
Landau's quasiparticle interaction function would become singular in 2-d.   
The singularity he suggested had the form
\be
f_{pp'} \sim {\vec p \cdot (\vec p - \vec{p'}) \over |\vec p - \vec{p'}|^2}
\label{andf}
\ee
This singular interaction will result in vanishing quasiparticle residue
and break down of Fermi liquid theory \cite{stamp1}.
One immediately notices that the numerator of (\ref{andf})
vanishes (as $p \to p'$) when both particles are located on the Fermi surface.
Thus the singularity can be detected only when the particles are
away from the Fermi surface. Methods such as Renormalization 
Group, which put their initial momenta on the Fermi surface, are 
unable to detect this singularity. Since Anderson, there have been 
several attempts to prove or disprove Fermi liquid theory in 2-d 
\cite{stamp1,engelbrecht,fabrizio,fukuyama1,fukuyama2};
all of them in agreement with Fermi liquid theory (at least for systems
with isotropic Fermi surface). On the other hand, it is well known that
Van Hove singularities \cite{polchinski} or nested Fermi surfaces 
\cite{VR,shankar} do not preserve Fermi liquid theory. Therefore the geometry 
of the Fermi surface plays an important role in the Fermi liquid
behavior of the system. In this section, we shall study the effect of local 
geometry of the Fermi surface on the interaction function $f_{pp'}$. 
We especially find a singularity at points where the Fermi surface has
zero curvature. This singularity is shown, in the next section, to change
the Fermi liquid behavior of the quasiparticles near the inflection points.

\subsection{Calculation of ${\bf f_{pp'}}$ for a Curved Fermi Surface}

\begin{figure}[t]
\epsfysize 5cm
\epsfbox[-40 320 600 550]{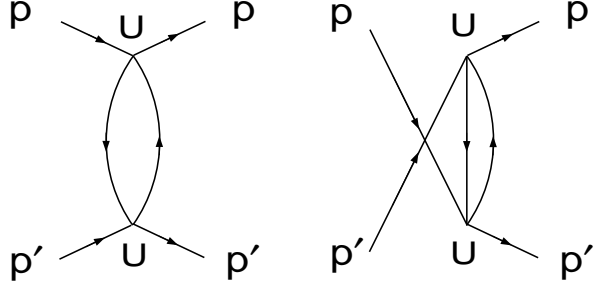}
\caption{Crossed channel diagrams contributing in $I_{pp'}$.}
\label{crossed}
\end{figure}

Let us start with a 2-d Hubbard Hamiltonian
\be
H=\sum_{\vec  p,\sigma} \epsilon_{\vec  p}c^\dagger_{\vec  p\sigma}c_{\vec  p\sigma}
+ {U \over \Omega} \sum_{\vec  q, \vec p, \vec{p'}}c^\dagger_{\vec  p+ \vec q \uparrow}
c^\dagger_{\vec{p^\prime} - \vec q \downarrow}c_{\vec{p^\prime}\downarrow}
c_{\vec  p\uparrow} 
\ee
where $\epsilon_{\vec  p}$ defines the ``bare" band structure and 
$\Omega$ is the area of the sample. We wish to find
$f^{\sigma\sigma^\prime}_{p p^\prime}$ up to second order in small coupling
$U$, or more precisely, the renormalized coupling $\bar{U}$, the
same way as we did in Sec. (2.2.1). Note that by small coupling
we mean $U/t \ll 1$ (this is not necessarily a low-density limit, 
but it is a weak-interaction limit). 
Following the method we described in Sec. (2.2.1), we find
\be
f^{\sigma\sigma^\prime}_{p p^\prime} = (\hbar^2 g_{pp^\prime}/m)
\delta_{\sigma,-\sigma'} -(\hbar^2 g_{pp^\prime}/m)^2
(I_{pp^\prime} + C^{\sigma\sigma'}_{pp'})
\ee
where the dimensionless coupling $g_{pp^\prime}$ is given by
\cite{abrikosov,stamp1}
\be 
g_{pp^\prime}={m \bar{U}\over \hbar^2}={m \over \hbar^2}
\left[U+{U^2 \over 2} \sum_{\vec q} {1 \over \epsilon_{\vec p}+\epsilon_{p\prime}-\epsilon_{\vec{p'}+\vec q}-\epsilon_{\vec p-\vec q}} \right] \\
\ee
we also identify the ``Cooper" channel and ``crossed" channel by
(cf. Fig. \ref{crossed})
\ba
C^{\sigma\sigma^\prime}_{pp'} &=& \sum_{\vec k} \left(
{2\delta_{\sigma,-\sigma'} \ n_{\vec k}  \over \epsilon_{\vec p}+ 
\epsilon_{\vec{p'}} 
- \epsilon_{\vec k}-\epsilon_{\vec p + \vec{p'} -\vec k}} \right) \\
I_{pp^\prime} &=& \sum_{\vec k} 
\left( {n_{\vec k} \over \epsilon_{\vec p}-\epsilon_{\vec{p'}}+\epsilon_{\vec k}
-\epsilon_{\vec k+\vec Q}} + (\vec p \leftrightarrow \vec{p'}) \right) 
\label{Ip0}
\ea
The Cooper term, $C^{\sigma\sigma^\prime}_{pp'}$, 
is singular when $\vec{\bar p}= (\vec p+ \vec{p'})/2 
\rightarrow 0$, but regular near ${\vec Q}\equiv \vec p - \vec{p'}=0$ 
(as is $g_{pp^\prime}$). We thus drop 
$C^{\sigma\sigma^\prime}_{p p^\prime}$ and approximate 
$g_{pp^\prime} \approx g_0$. 
$I_{pp'}$, on the other hand, may become singular when 
${\vec Q} \rightarrow 0$.
This can be seen by the fact that the denominators in (\ref{Ip0}) vanish at
${\vec Q}=0$. We therefore have to examine $I_{pp'}$ more carefully.
Let us define
\be
{\vec v}_{\vec k}={\partial \epsilon_{\vec k}\over \partial {\vec k}} \hspace{1cm}
{\rm and} \hspace{1cm}
(M^{-1}_{\vec k})_{ij}={\partial^2 \epsilon_{\vec k} \over
\partial k_i \partial k_j}
\ee
Expanding the denominators to second order in $\vec Q$, we have 
\ba
\label{Ip}
&I_{pp^\prime}&=\int {d^2 k \over (2\pi)^2} 
\left( {n_{\vec k} \over ({\vec v_{\bar p} - \vec v_k)} \cdot  {\vec Q} - {1 \over 2} {\vec Q} \cdot {\cal M}^{-1}_{\vec k} \cdot {\vec Q}}-{n_{\vec k} \over ({\vec v_{\bar p} - \vec v_k)} \cdot {\vec Q} + {1 \over 2} {\vec Q} \cdot 
{\cal M}^{-1}_{\vec k} \cdot {\vec Q}} \right) \nn \\
&&= {1 \over (2\pi)^2} \int d^2 k {n_{\vec k} (2\Delta_{\vec k,Q}/Q) \over 
[({\vec v_{\bar p} - \vec v_k)} \cdot \widehat{Q}]^2-\Delta_{\vec k,Q}^2} 
\propto  I_{\vec{\bar p}}({\vec Q}) 
\ea
where
\be
I_{\vec{\bar p}}({\vec Q}) \equiv \int d^2 k {n_{\vec k} \over 
[({\vec v_{\bar p} - \vec v_k)} \cdot \widehat{Q}]^2-\Delta_{\vec k,Q}^2} 
\ee
The "gap" $\Delta_{\vec k,\vec Q}$ is defined by
\be
\Delta_{\vec k,\vec Q}={{\vec Q}\cdot {\cal M}^{-1}_{\vec k} \cdot {\vec Q} \over
2 |{\vec Q}|} \propto |{\vec Q}|
\ee
and $\widehat{ Q}\equiv {\vec Q}/|{\vec Q}|$. 
For a quadratic dispersion relation $\epsilon_{\vec k}=k^2/2m$, 
${\cal M}^{-1}_{\vec k}$ ($=1/m$) is independent of 
${\vec k}$. In more general cases however, it has 
${\vec k}$-dependence but this does not affect our calculation (we will
consider the ${\vec k}$-dependence of ${\cal M}^{-1}_{\vec k}$ when 
appropriate later). Thus for the moment, we 
can safely neglect the ${\vec k}$-dependence and write 
$\Delta_{\vec k,\vec Q}=\Delta_{\vec Q}$.

\begin{figure}[t]
\epsfysize 5cm
\epsfbox[-40 350 600 580]{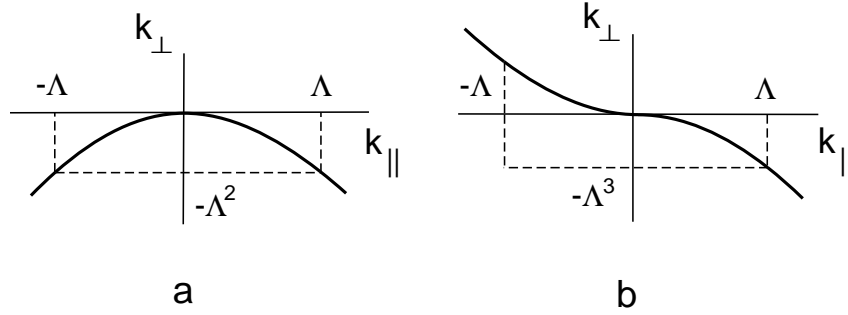}
\caption[Fermi surface near a curved point and an inflection point.]
{\small The two situations considered in the text. In (a) we assume
the Fermi surface has finite curvature at 
${\vec k= \vec k}_0$; in (b) the
curvature is zero when ${\vec k} = {\vec k}_0$ 
(inflection point). Hatched lines show the integration
regions discussed in the text.}
\label{fl1}
\end{figure}

Our goal is now to understand the behavior of $I_{\vec{\bar p}}(\vec Q)$ 
as $\vec Q \rightarrow 0$.
Infrared singularity occurs when $(\vec v_{\bar p} - \vec v_k)
\cdot \widehat{ Q}=0$. This condition
is usually satisfied on a line in $k$-space
inside the Fermi surface (corresponding to a
straight line in ${\vec v}$-space). Let's confine our 
integration to a region within a cutoff $\Lambda$ around this line, where
$|{\vec Q}| \ll \Lambda \ll p_{_F}$, and $p_{_F}$ is the Fermi momentum. The
integration inside the Fermi surface then has the form  
\be
\int_{-\Lambda}^{\Lambda}{du \over u^2-\Delta_{\vec Q}^2}
={1 \over \Delta_{\vec Q}}\ln ({|\Lambda-\Delta_{\vec Q}| \over |\Lambda+\Delta_{\vec Q}|})
\sim {1 \over \Delta_{\vec Q}}(-{2 \Delta_{\vec Q} \over \Lambda})=-{2 \over \Lambda} 
\ee
Thus the integration inside the Fermi surface contains no singularity
 as $\Delta_{\vec Q} \rightarrow 0$. Singularity, if any, should therefore 
come from the integration
near the Fermi surface. Let ${\vec k}_0$ be a point on the Fermi surface,
 satisfying
$({\vec v_{\bar p} - \vec v_{k_0})} \cdot \widehat{ Q} = 0$.
We first consider points where Fermi surface has finite curvature 
(Fig. \ref{fl1} (a)). Writing
${\vec k}={\vec k}_0+k_\pe\hat{e}_\pe +k_\pa\hat{e}_\pa$, 
where $\hat{e}_\pa$ ($\hat{e}_\pe$) are unit vectors
parallel (perpendicular) to the Fermi surface at $\vec k_0$, 
we then define $\epsilon_{\vec k}$ by the expansion
\be
\epsilon_{\vec k} = E_{_F}+v_{k_0}k_\pe+a k_\pe^2
+b k_\pa^2 + c k_\pe k_\pa 
\ee
 Near ${\vec k}_0$ therefore, ${\vec v_k}$ can be written as 
\be
{\vec v_k}=v_{k_0} \hat{e}_\pe + {\vec \gamma_{\pa}}k_\pa+{\vec 
\gamma_{\pe}}k_\pe 
\ee
where 
\ba
\vec \gamma_{\pe}=2 a \hat{e}_\pe + c \hat{e}_\pa  \qquad , \qquad 
\vec \gamma_{\pa}= c \hat{e}_\pe + 2 b \hat{e}_\pa.
\ea
Fermi surface also is defined by
\be
v_{k_0}k_\pe+a k_\pe^2+b k_\pa^2 + c k_\pe k_\pa=0 
\label{FS} 
\ee
Since the Fermi surface has finite curvature at $\vec k_0$, $\vec \gamma_{\pa}$
and $b$ are non-zero. Rescaling the integral in (\ref{Ip}) so that
$v_{k_0}=b=1$, we have a Fermi surface near $k_0$ obeying 
$k_\pa^2 \approx - k_\pe$ (cf. (\ref{FS})). It is not difficult to see that
singularities in $I_{pp^{\prime}}$, for $\vec p, \vec{p'}$ close to 
${\vec k_0}$, arise only when 
\be
\vec \gamma_\pa \cdot \widehat{ Q} \rightarrow 0
\label{gma}
\ee
In that case:
\be
I_{p p^{\prime}}\sim \int_{-\Lambda^2}^0 dk_\pe
\int_{-\sqrt{-k_\pe}}^{\sqrt{-k_\pe}} {dk_\pa \over k_\pe^2-\Delta_{\vec k, \vec Q}^2}
\sim {\pi \over \vert Q \vert^{1/2}} 
\label{Ip2}
\ee
To get (\ref{Ip2}) we have made use of the following integral 
\be
\int du {\sqrt{u} \over u^2 - \Delta^2}= {1 \over \sqrt{\Delta}}
({1 \over 2} \ln {|\sqrt{u}-\sqrt{\Delta}| \over |\sqrt{u}+\sqrt{\Delta}|}
+ \tan^{-1}\sqrt{u \over \Delta})
\ee
This singularity is just that found already for a circular Fermi surface
\cite{fukuyama1,stamp1}. It only occurs if ${\vec p}$ and
$\vec{p'}$ approach each other along the direction orthogonal to
$\vec \gamma_\pa$. For a circular Fermi surface, this direction is
orthogonal to the Fermi surface. For non-circular Fermi surfaces (with
no inflection point) on the other hand, $\vec \gamma_\pa \cdot
\widehat{Q} =0$ for directions which are not necessarily normal to 
the Fermi surface; but the singularity is still along lines.
This case was also studied away from the Fermi surface by Fukuyama and
Ogata \cite{fukuyama2}. A closer look at the
singular region in $k$-space for the case of a circular Fermi surface shows
that it is bounded by circular arcs of opposite curvature which touch at the
Fermi surface \cite{stamp1}.
 
This singularity does not change the Fermi liquid behavior \cite{stamp1}.
A simple way to see this, is to look at the quasiparticle energy
\be
\epsilon_{\vec p}=\epsilon^0_{\vec p} + \sum_{\vec p'} f_{\vec pp'}
\delta n_{\vec{p'}}
\ee
Singularity in $f_{pp'}$ is summed over in this equation.
In our case here, the $|Q|^{1/2}$ singularity is always along a line. 
Integration over this term removes the singularity and gives
Fermi liquid behavior as one approaches the Fermi surface
\cite{fukuyama1,stamp1}.

\subsection{Calculation of ${\bf f_{pp'}}$ near an Inflection Point}

At points where the Fermi surface has zero curvature,  
$\vec \gamma_\pa \rightarrow 0$
and consequently, condition (\ref{gma}) is always satisfied. 
It is therefore natural to expect a singularity in any direction of $\widehat{Q}$;
which may now cause break down of Fermi liquid theory.  
We wish to study this singularity near an inflection point. We must therefore
extend (\ref{FS}) to higher order, and also include the $k$-dependence
of ${\cal M}^{-1}_{\vec k}$ (which now vanishes in certain directions).
Expanding about an inflection point at ${\vec k= \vec k}_0$, we write 
\be
\epsilon_{\vec k}=v_{k_0} k_{\perp} + a k_{\perp}^2 + g k_{\parallel}^3 
\ee
(cf. Fig.1(b)). Substituting directly into (\ref{Ip0}) we get
\be
I_{\bar p}(\vec Q)= {1 \over 3gQ_\pa}\int_{-\Lambda}^{\Lambda}dk_{\parallel}
\int_{-g\Lambda^3/v_{k_0}}^{-gk_\pa^3/v_{k_0}}  \left( {dk_{\perp} \over
c_{_{\hat{ Q}}}k_{\pe}+ k_\pa^2 + Q_\pa k_\pa + T(\vec \kappa,\vec Q)} \right)
+ ({\vec Q} \rightarrow -{\vec Q})
\label{IpQ1}
\ee
We have defined ${\vec \kappa}=\vec{\bar p}-{\vec  k}_0$ (i.e., the
center of mass momentum relative to ${\vec  k}_0$), 
$c_{_Q}=2aQ_\pe/3gQ_\pa$ and
\be
T(\vec \kappa,\vec Q)=\kappa_\pa^2 - {Q_\pa^2 \over 4} + c_{_Q}
(\kappa_\pe + {Q_\pe \over 2})
\ee
Integration over $\kappa_\pe$ is straightforward. One finds
\be
I_{\bar p}(\vec Q)= {1 \over 2aQ_\pe}\int_{-\Lambda}^{\Lambda}dk_\pa \ln 
\left|{k_\pa^2 + Q_\pa k_\pa - (c_{_Q}g/v_{k_0}) k_\pa^3 
+ T(\vec \kappa,\vec Q) \over k_\pa^2 + Q_\pa k_\pa - (c_{_Q}g/v_{k_0})
\Lambda^3 + T(\vec \kappa,\vec Q)} \right|  + (\vec Q \rightarrow -\vec Q)
\ee
The $k_\pa^3$ term in the numerator is small compared to the other terms
unless if $Q_\pa \rightarrow 0$ (or $c_{_Q} \rightarrow \infty$).
We therefore neglect this term except when $Q_\pa \ll Q_\pe$ 
(we will come back to this case later). To calculate the above integrals we 
should use 
\ba
&&\int du \ln |u^2+\beta u + \gamma|=(u+\beta/2)\ln |u^2+\beta u+ \gamma| - 2u
\nn \\ &&+ 2 \sqrt{|W|} 
\left[\theta (-W){1\over 2}\ln \left( {u+ \beta/2 + \sqrt{|W|} \over u+ \beta/2 - \sqrt{|W|}}\right) + \theta (W)\tan^{-1} \left({u+ \beta/2 \over \sqrt{|W|}}\right) \right]
\ea
where $W=\gamma - \beta^2/4$. However , it turns out that all the terms in
the above integral are regular as $\vec Q \rightarrow 0$ except for 
the ``$\tan^{-1}$" term. Keeping only the singular terms we find
\be
I_{\bar p}(\vec Q) \sim \theta (W_{\vec \kappa,\vec Q})
\left( {\sqrt{W_{\vec \kappa,\vec Q}} \over 2Q_\pe} \right) 
\left[ \tan^{-1} \left( {\Lambda + Q_\pa/2 \over 
\sqrt{W_{\vec \kappa,\vec Q}}}\right) + \tan^{-1} \left( {\Lambda - Q_\pa/2 \over 
\sqrt{W_{\vec \kappa,\vec Q}}}\right) \right]
+ (\vec Q \rightarrow -\vec Q)
\ee
where 
\be
W_{\vec \kappa,\vec Q}=T(\vec \kappa, \vec Q)+ {Q_\pa^2 \over 4}
=-\kappa_\pa^2 - c_{_Q} (\kappa_\pe + Q_\pe/2)
\ee
In the limit where $\kappa, Q \ll \Lambda$ we approximately get
\be
I_{\vec{\bar p}}({\vec Q}) = {\pi \over a Q_\pe}
[|W_+|^{1/2} \theta(W_+)  - |W_-|^{1/2} \theta(W_-)]  + ...
\label{WkQ}
\ee
where (...) represents non-singular terms and 
$W_\pm \equiv W_{\vec \kappa,\pm \vec Q}$.

This behavior of $I_{pp'}$ is rather complicated. We 
can get some feeling for it if we first take the 
case where ${\vec  p} = {\vec  k}_0$, i.e., where one quasiparticle is at the 
inflection point (see Fig. \ref{fl2}). Then for most regions in $k$-space,
(\ref{WkQ}) gives
\be
I_{\vec  k_0, \vec k_0+ \vec Q} \sim 
{\theta(4cQ_\pe^2-Q_\pa^3) \theta(Q_\pa)\over \vert Q_\pa \vert^{1/2}} 
\label{butter}
\ee
This is the same inverse square root as found for the circular Fermi 
surface (cf. (\ref{Ip2})), but now extending over large ``butterfly wing" 
regions in $k$-space.) Fig. \ref{fl2} shows different regions in $k$-space with 
different singular behavior. The singularity in the left wing of the butterfly 
region comes from switching ${\vec  p}$ and $\vec{p^\prime}$, i.e. putting 
$\vec{p^\prime}$ at the inflection point. 

\begin{figure}[t]
\epsfysize 5cm
\epsfbox[-40 320 600 550]{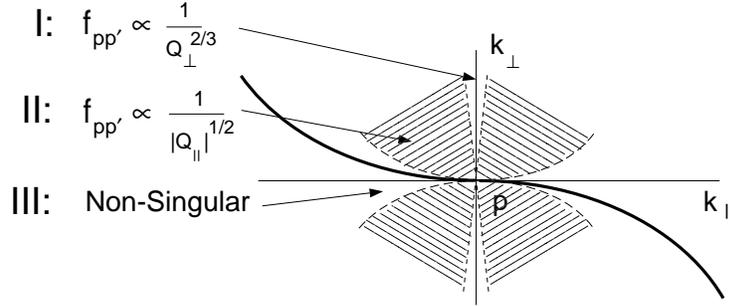}
\caption[Singular and non-singular regions near an inflection point.]
{\small Three different regions in $k$-space around the inflection
points with different singular behavior. 
We have put ${\vec p=\vec k}_0$ therefore
the regions shown are actually the regions where $\vec{p'}$ can be.
Between I and II, there is a crossover whereas between II and III there is
a sharp edge.}
\label{fl2}
\end{figure}

When $Q_\pa \rightarrow 0$, the above approximation does not hold anymore.
In that case (\ref{IpQ1}) becomes
\ba
I_{\bar p}(\vec Q)&=&{1 \over 2aQ_\pe} \int_{-\Lambda}^\Lambda dk_\pe 
\int_{-(v_{k_0} k_\pe/g)^{1/3}}^{(v_{k_0} \Lambda/g)^{1/3}} dk_\pa 
\left[ {1 \over k_\pe - p_\pe + k_{0\pe}} + (\vec p \leftrightarrow \vec{p'}) \right]
\nn \\
&=&{\zeta (v_{k_0}/g)^{1/3} \over 2aQ_\pe} \left[ (\kappa_\pe + Q_\pe/2)^{1/3} - (\kappa_\pe - Q_\pe/2)^{1/3} \right] + ...
\ea
where $\zeta \approx \int_{-\infty}^\infty dz/(z^3 -1)= -\pi/\sqrt{3}$.
If we again put $\vec p = \vec k_0$, we get 
\be
I_{\bar p}(\vec Q) \sim Q_\pe^{-2/3} \qquad \quad (Q_\pa \ll Q_\pe)
\label{IQpe}
\ee
The singularity here is stronger than what we found before 
($\sim Q_\pa^{-1/2}$). 
There is a crossover between these two different behaviors 
(cf. Fig. \ref{fl2}). 

It is very important to note also the non-singular region (region 
III in Fig. \ref{fl2}) which lies close to the Fermi surface as ${Q}$ 
approaches the origin.
There is no singularity in the interaction between 2 quasiparticles if they 
are both on the Fermi surface. Thus any  
calculation (perturbative or renormalization group) 
of the scattering between quasiparticles confined
to the Fermi surface will 
miss all the singular behavior in (\ref{butter}).

In the more general case where neither particle is at $k_0$, 
the behavior of $I_{pp'}$ is not so easily displayed. As the
quasiparticles approach each other one finds the simple behavior 
\be
I_{\vec k_0, \vec k_0+ \vec Q} \propto 
{\theta(-\kappa_\pa^2 - c_{_Q} \kappa_\pe) \over 
[-\kappa_\pa^2 - c_{_Q} \kappa_\pe]^{1/2}} \qquad \quad (Q \ll \kappa)
\ee

\section{Calculation of Self Energy}

Fermi or non-Fermi liquid nature of the system can be checked by calculating
the self energy $\Sigma_{\vec p}(\omega)$. We already have seen how the
quasiparticle lifetime and quasiparticle residue are related to the 
real and imaginery parts of the self energy (cf. (\ref{tauk}) and (\ref{Zk})).
The imaginary part of $\Sigma_{\vec p}(\omega)$ is specially important 
because it can be calculated easily using the Fermi golden rule. 
We saw that in a 3-d Fermi liquid, at low energies, we have
\be
{\rm Im} \Sigma_{p_{_F}}(\omega) \sim \omega^2
\label{sgm3d}
\ee
In this section we first see how this relation is different for a 2-d
Fermi liquid with non-singular interaction.
We then find the effect of our singularities on 
${\rm Im} \Sigma_{p_{_F}}(\omega)$ and show the deviation from Fermi liquid
behavior.

\subsection{Non-Singular Interaction}

Let us first consider the non-singular case. Landau's $f$-function 
is therefore regular and given by the cylindrical harmonics 
$f_l$ the same way as before. For the present calculation, it is enough 
to consider the isotropic harmonic and write
\be
f_{pp'}=f_0
\ee
Having Landau's $f$-function, we can calculate the 
quasiparticle T-matrix using the Bethe-Salpeter equation (\ref{B-S}).
In the isotropic channel, which we consider here, the T-matrix is 
simply given by \cite{baym0}
\be 
t_{pp'}(\vec q,\omega)=t_0(\eta)={f_0 \over 1 + f_0 \chi^0({\vec  q},\omega)}
\label{tf0}
\ee
where $\chi^0({\vec  q},\omega)$ is the 2-dimensional Lindhard function
\cite{stamp1}
\be
\chi^0({\vec  q},\omega)=N(0)\left[1-\eta\left({\theta(\eta^2-1)\over\sqrt{\eta^2-1}}
+ i{\theta(1-\eta^2)\over\sqrt{1-\eta^2}} \right) \right],
\label{lindhard}
\ee
$\eta=\omega / qv_{_F}$ and density of states at the Fermi surface 
$N(0)=p_{_F}/2\pi v_{_F}$.
All the physical quantities can then be found from the T-matrix. What we
wish to do here however, is to calculate the quasiparticle self energy 
and compare it with the 3-d case discussed before (calculation we
present here is based on Ref. \cite{stamp1}).

It is always easier to calculate the imaginary part of the self energy
which is equivalent to the scattering rate and is given by Fermi golden rule
\cite{baym0}
\be
{\rm Im}\Sigma_p(\omega) = \pi \sum_{\vec p',\vec q} 
|t_{pp'}(\vec q,\omega)|^2
n^0_{\vec{p'}}(1-n^0_{\vec p-\vec q})(1-n^0_{\vec{p'}+\vec q}) 
\delta( \omega+\epsilon_{\vec{p'}}-\epsilon_{\vec p-\vec q}-
\epsilon_{\vec{p'}+\vec q})
\label{slf0}
\ee
We calculate the energy on shell (i.e. $\omega=\epsilon_{\vec p}$).
Assuming $q \ll p_{_F}$ we have
\be
\epsilon_{\vec p-\vec q}=\epsilon_{\vec p} - v_{_F} q \mu = \omega - v_{_F} q \mu
\ee
where $\mu=\hat{p} \cdot \hat{q}$.
The $\delta$-function in (\ref{slf0}) can be written as
\ba
\delta( \omega+\epsilon_{\vec{p'}}-\epsilon_{\vec p-\vec q}-
\epsilon_{\vec{p'}+\vec q}) &=& \int_{-\infty}^{\infty} d \nu
\delta (\nu - \omega + \epsilon_{\vec p-\vec q})
\delta (\nu - \epsilon_{\vec{p'}} - \epsilon_{\vec{p'}+\vec q})) \nn \\
&=& \int_{-\infty}^{\infty} d \nu 
\delta (\nu - qv_{_F} \mu) \delta (\nu - qv_{_F} \mu')
\ea
with $\mu'=\hat{p'} \cdot \hat{q}=\cos \theta'$. On the other hand (at $T=0$)
\be
n^0_{\vec{p'}}(1-n^0_{\vec{p'}+\vec q})=\theta(p_{_F}-p') \theta(p'+q \mu' -p_{_F})
\label{n0}
\ee
Eq. (\ref{n0}) requires $p' < p_{_F}$ and $\mu' > 0$.
Therefore
\be
\sum_{vec{p'}} n^0_{\vec{p'}}(1-n^0_{\vec{p'}+\vec q})=
\int_{-\pi/2}^{\pi/2} {d\theta' \over 2\pi} 
\int_{p_{_F}-q\mu'}^{p_{_F}} {p'dp' \over 2\pi}= {p_{_F} q \over (2\pi)^2}
\int_{-\pi/2}^{\pi/2} \mu' d\theta'
\ee
Substituting all back into (\ref{slf0}) we find
\ba 
{\rm Im} \Sigma_p(\omega) = {p_{_F} \over 2\pi  v_{_F}}  \sum_{\vec q} 
\int {\nu d\nu \over \sqrt{v_{_F}^2 q^2 - \nu^2}} |t(q,\nu)|^2 
\theta(\omega - qv_{_F} \mu) \theta(qv_{_F} - \nu) \delta(\nu - qv_{_F} \mu)
\ea
One can write
\be
\sum_{\vec q}= \int {d^2q \over (2\pi)^2}=\int {qdq \over (2\pi)^2} 
\int {d \mu \over \sqrt{1-\mu^2}}
\ee
Because of the $\delta$-function, the $\mu$-integral will require
$\mu=\nu/qv_{_F} \equiv \eta$. As a result
\be
{\rm Im} \Sigma_p(\omega) = {p_{_F} \over (2\pi)^3  v_{_F}}  \int qdq 
\int \nu d\nu {|t(\vec q,\nu)|_{\mu=\eta}^2 \over q^2 v_{_F}^2 - \nu^2}
\theta(\omega - \nu) \theta(qv_{_F} - \nu)
\label{sgm0}
\ee

In the isotropic case, $t(\vec q,\nu)$ is only a function of $\eta$ 
(cf. (\ref{tf0})).
For $\eta < 1$, which is actually the case here, we have 
\be
t(\eta)=  {f_0 \over 1+f_0(1-i\eta / \sqrt{1-\eta^2})N(0)}
={1 \over N(0)}{A_0 \over 1+iA_0 \eta / \sqrt{1-\eta^2}}
\label{t-eta}
\ee
where $A_0=F_0/(1+F_0)$ with $F_0=N(0)f_0$  dimensionless 
Landau parameter. 
In this case (\ref{sgm0}) becomes
\ba
{\rm Im} \Sigma_p(\omega) &=&
{p_{_F} \over (2\pi v_{_F})^3 } \int_0^\Lambda {dq \over q}
\int \nu d\nu {|t(\eta)|^2 \over 1- \eta^2}
\theta(\omega - \nu) \theta(qv_{_F} - \nu)  \nn \\
&=& {p_{_F} \over (2\pi v_{_F})^3 } \int_0^\omega \nu d\nu
\int_{\nu/\omega_c}^1 {d\eta \over \eta}{|t(\eta)|^2 \over 1- \eta^2}
\ea
where $\omega_c= \Lambda v_{_F}$ is the energy cutoff. 
Substituting $t(\eta)$ from (\ref{t-eta}) we get 
\be
{\rm Im} \Sigma_p(\omega) = K
\int_0^\omega \nu d\nu \int_{\nu/\omega_c}^1 {d\eta \over \eta}
{1 \over 1+(A_0 - 1)\eta^2} 
\approx  K \int_0^\omega \nu d\nu \int_{\nu/\omega_c}^1 
{d\eta \over \eta} 
\label{sgm3}
\ee
where $K={A_0^2 / 2\pi v_{_F} p_{_F} }$. In the last step
we have dropped the $\eta^2$ term in the denominator.
We therefore simply get
\be
{\rm Im} \Sigma_p(\omega) \sim \int_0^\omega d\nu \nu \ln 
|{\omega_c \over \nu}| \sim \omega^2(1+ 2\ln 
|{\omega_c \over \omega}| )
\ee
The leading $\omega$ dependence is therefore
\be
{\rm Im} \Sigma_p(\omega) \sim \omega^2 \ln \omega 
\label{sgm2d}
\ee
Although this is different from the 3-d case (\ref{sgm3d}), it 
does not change Fermi liquid behavior \cite{stamp1}.
Now let us see how our singular interaction will change this result.

\subsection{Singular Interaction}

We now calculate the self energy for the singular interaction we found
near inflection points.
As in some previous studies of various kinds of singular interaction
\cite{stamp1,fukuyama1,halperin,varma,stern}, it is convenient to begin 
with that part of $\Sigma_{\vec  p}(\omega)$ which is 2nd-order in the 
singular interaction. Let us write  
\be
f_{pp^\prime} = f_{pp^\prime}^{FLT} + \lambda_{pp^\prime}
\ee
where $f_{pp^\prime}^{FLT}$ is the non-singular
part of $f_{pp^\prime}$ and $\lambda_{pp^\prime}$ is the singular part. 
We calculate $\Sigma_{\vec  p}(\omega)$ to second order in 
$\lambda_{pp^\prime}$, but to all orders in $f_{pp'}^{FLT}$. Since we are
not particularly interested in the angular dependence of the regular Fermi 
liquid contribution, we will simply let $f_{pp^\prime}^{FLT} \to f_0$, a 
constant as before. 
We do our calculation for $\vec p = \vec k_0$, i.e. at the inflection point. 
The singular term $\lambda_{pp^\prime}$ therefore is given by (\ref{butter})
and (\ref{IQpe}). Since (\ref{IQpe}) is only valid in a region very close to the
normal axis, we take $\lambda_{pp^\prime}$ everywere to be
\be
\lambda_{pp^\prime} \sim {1 \over \sqrt{|Q_\pa|}}
\ee

\begin{figure}[t]
\epsfysize 5cm
\epsfbox[-40 320 600 550]{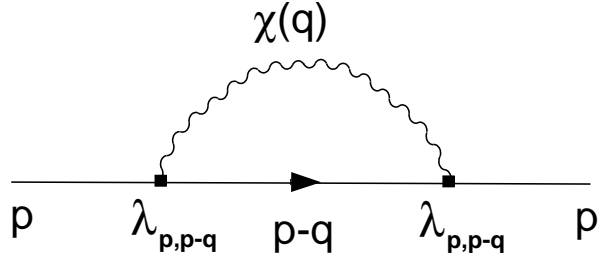}
\caption[Self energy to second order in $\lambda_{p,p-q}$]
{\small Self energy to second order in $\lambda_{p,p-q}$.}
\label{fl3}
\end{figure}

We again use (\ref{slf0}) as our starting point (note that (\ref{slf0}) 
is valid for $\omega > 0$, as can be immidiately seen by
the form of the occupation numbers. A full version is given
in Ref. \cite{baym0}). In order to calculate to
2nd order in $\lambda_{pp'}$,
$|t_{pp'}|^2$ in (\ref{slf0}) should be replaced by \cite{stamp1}
\be
|t_{pp'}(\vec q, \omega)|^2 \sim {\lambda_{pp'}^2 \over 
|1+f_0 \chi^0(\vec q, \omega)|^2}
\ee
where $\chi^0(\vec q, \omega)$ is the generalization of (\ref{lindhard}) for
a non-circular Fermi surface. This is equivalent to calculating the 
diagram presented in Fig. \ref{fl3} \cite{stamp1},
where $\chi(\vec q,\nu)$ is the usual fluctuation propagator 
\be 
\chi(\vec q,\nu)={\chi^0(\vec q,\nu) \over 1 + f_0 \chi^0(\vec q,\nu)}
\ee
Since we are going to calculate everything near the inflection point, we
can assume that $\chi^0(\vec q, \omega)$ behaves regularly at the 
inflection point and treat it as a constant. At $T=0$ we have
\be
{\rm Im}\Sigma_{k_0}(\omega) \sim \sum_{\vec p',\vec q} {1 \over Q_\pa}
\theta(-\epsilon_{\vec{p'}}) \theta(\epsilon_{\vec p- \vec q})
\theta(\epsilon_{\vec{p'}+ \vec q})
\delta( \omega+\epsilon_{\vec{p'}}-\epsilon_{\vec p-\vec q}-
\epsilon_{\vec{p'}+\vec q})
\ee
We expand $\epsilon_{\vec k}$ near the inflection point $\vec k_0$ by
\be
\epsilon_{\vec k}= v_{k_0} \kappa_\pe + g \kappa_\pa^3
\label{ek0}
\ee
where $\vec \kappa = \vec k - \vec k_0$. From now on, for simplisity, 
we take $v_{k_0}=g=1$.Thus 
\ba
&&\epsilon_{\vec{p'}} = -Q_\pe - Q_\pa^3 \nn \\
&&\epsilon_{\vec p-\vec q}= -q_\pe - q_\pa^3 \nn \\
&&\epsilon_{\vec{p'}+\vec q}=q_\pe - Q_\pe + (q_\pa-Q_\pa)^3
\ea
The self energy now becomes
\ba
{\rm Im}\Sigma_{k_0}(\omega) &\sim& \int {dQ_\pa \over Q_\pa}
\int dq_\pa \int_{-Q_\pa^3}^\Lambda dQ_\pe
\int_{-\Lambda}^{-q_\pa^3}dq_\pe
\theta(\omega - Q_\pe -Q_\pa^3 + q_\pe + q_\pa^3) \nn \\
&& \times \delta(\omega -3Q_\pa^2q_\pa + 3Q_\pa q_\pa^2) 
\ea
Performing the $q_\pe$ integral, we get
\ba
{\rm Im}\Sigma_{k_0}(\omega) &\sim& \int {dQ_\pa \over Q_\pa} 
\int dq_\pa \int_{-Q_\pa^3}^\Lambda dQ_\pe
(\omega - Q_\pe - Q_\pa^3) \theta(\omega - Q_\pe - Q_\pa^3) \nn \\
&& \times \delta(\omega -3Q_\pa^2q_\pa + 3Q_\pa q_\pa^2) 
\ea
Integrating over $Q_\pe$ is also trivial
\be
{\rm Im}\Sigma_{k_0}(\omega) \sim {1 \over 2}\omega^2 
\int_{-\Lambda}^\Lambda {dQ_\pa \over Q_\pa}
\int_{-\Lambda}^\Lambda dq_\pa
\delta(\omega -3Q_\pa^2q_\pa + 3Q_\pa q_\pa^2)
\label{1}
\ee
We now change the variables to $Q_\pa \rightarrow \omega^{1/3} x$
and $q_\pa \rightarrow \omega^{1/3} y$. Integration limits also change to
$\pm \Lambda/\omega^{1/3}$. For small $\omega$ ($\ll \Lambda^3$, or 
more precisely $\ll g\Lambda^3$), the
integration limits can be replaced by $\pm \infty $. However, one has to 
make sure that the integral does not diverge as the limits
go to infinity.  We finally get
\be
{\rm Im}\Sigma_{k_0}(\omega) \sim \omega^{4/3} 
\int_{-\infty}^{\infty} {dx \over x}
\int_{-\infty}^{\infty} dy \delta(1 -3x^2 y+ 3x y^2)
\ee
The integral is regular at integration limits and also at $x=0$. Therefore
we can confidently write
\be
{\rm Im}\Sigma_{k_0}(\omega) \sim \omega^{4/3} 
\label{4/3}
\ee
This is clearly different from the Fermi liquid form (\ref{sgm3d})
or the 2-d version of it (\ref{sgm2d}). This is also different from the result
obtianed by Fukuyama and Ogata \cite{fukuyama2} which they find the power 
to be (5/3) instead of (4/3) for inflection points. 
The reason for this descripency is
that they miss the singular interaction in their calculation. Notice that
if we drop the $1/Q_\pa$ factor (which comes from the singular interaction)
in the integrand of (\ref{1}), we get ${\rm Im} \Sigma \sim \omega^{5/3}$
in agreement with Ref. \cite{fukuyama2}.

We can repeat this calculation for a curved Fermi surface using
\be
\epsilon_k= \kappa_\pe + \kappa_\pa^2
\ee
instead of (\ref{ek0}). Eq. (\ref{1}) then becomes
\ba
{\rm Im}\Sigma_{k_0}(\omega) &\sim& \omega^2 
\int_{-\Lambda}^\Lambda dQ_\pa
\int_{-\Lambda}^\Lambda dq_\pa
\delta(\omega +2Q_\pa q_\pa - 2q_\pa^2) \nn \\
&\sim& \omega^2 
\int_{-\Lambda}^\Lambda {dQ_\pa \over |Q_\pa|} \nn \\
&\sim& \omega^2 \ln \omega \ 
\label{o2}
\ea
which agrees with (\ref{sgm2d}). Although (\ref{4/3}) is different from
(\ref{o2}), the difference is not enough to break down Fermi liquid theory
(in the sense that $Z \ne 0$ and Fermi surface exists) \cite{fukuyama2}.
On the other hand, higher order calculations can change this result 
significantly. For example simple estimate indicates that the 4th order
calculation gives {\it marginal} Fermi liquid behavior 
${\rm Im} \Sigma \sim \omega$. Systematic higher order calculation therefore
is necessary to acheive the final conclusion \cite{amin6}.

\section{Summary}

Our calculation of $f_{pp'}$ for a 2-d electron system reveals 
a 1-d singularity 
near curved points of the Fermi surface as $\vec p \to \vec{p'}$.
This is in agreement with previous results and is shown to preserve Fermi
liquid behavior \cite{stamp1}. 
Near inflection points of the Fermi surface,
on the other hand, we find a 2-d singularity again with 
$|\vec p - \vec{p'}|^{-1/2}$ behavior. 
To second order in the singular term, the imaginary
part of the self energy shows an $\omega^{4/3}$-dependence in disagreement
with the ordinary Fermi liquid result (i.e. $\omega^2 \ln \omega$). 
This however does not invalidate Fermi liquid theory \cite{fukuyama2}. 
Inclusion of Higher order diagrams is necessary 
for any rigorous conclusion. 

We should point out that our result is different from Ref. \cite{fukuyama2}
in a sense that they do not incorporate singular interaction 
at the inflection points in their calculation of the self energy.
They report $\omega^{5/3}$-dependence for the imaginary self energy
near an inflection point, different from what we find ($\omega^{4/3}$).
Ref. \cite{guinea} has also investigated the effect of the inflection points.
They however consider the scattering between two quasiparticles located
on two inflection points opposite to each other with respect to the
center of the Brillouin zone. This is clearly different from our calculation.
Employing renormalization group technique they find that inflection
points would increase the lower critical dimension for instability from Fermi 
liquid fixed point from d=1 to d=3/2. Thus 2-d systems will still be stable
under such an effect.


\chapter{Nearly Antiferromagnetic Fermi Liquid Theory}

At zero doping, all cuprates are Mott insulators with 
antiferromagnetically ordered spins. On the other hand, the physical 
behavior of overdoped samples is well described by Fermi liquid theory. 
However, the most interesting region in the phase space of high 
$T_c$'s is none of the above phases, but the intermediate doping regime. 
One can think about the intermediate doping as a continuation
of the metallic phase or the extension of the insulating phase.
Depending on which way to choose, one may end up with two different
schools of thought based on Fermi liquid or non-Fermi liquid
theories. At this time not only no consensus about the
older theories of high $T_c$ has been achieved, but also different new theories 
such as, SO(5) theory \cite{so5}, theory of stripes \cite{stripe}, 
and more recently theory of nodal liquids \cite{nodal}, etc. have emerged.
In this chapter we shall focus on one of the older theories which is based on
the Fermi liquid assumption, and because of its simplicity in calculations,
has attracted considerable attention especially among experimentalists. 

Nearly antiferromagnetic Fermi liquid theory (NAFL) is a phenomenological
theory motivated by nearly ferromagnetic spin fluctuation theory or
paramagnon theory, proposed in 1960's to describe metals near their 
ferromagnetic instability as well as liquid $^3$He. Both theories are based
on the assumption that the excitations of the electronic system are Fermi
liquid quasiparticles, interacting among each other via the damped spin waves.
Besides the similarities, there are fundamental differences between these
two theories which 
have made NAFL not as widely accepted as paramagnon theory. In this 
chapter, after a brief review of paramagnon theory, we introduce NAFL
emphasising on the similarities and differences between the two theories.  

\section{Paramagnon Theory}

Fermi liquid theory is primarily constructed to describe properties
of low energy electron-hole excitations \cite{mahan}. There also exist 
other excitations in a Fermi liquid, which are inherently
different from electron-hole excitations (although they can be views as
collective resonances of them \cite{mahan}), such as
density oscillations and damped spin waves. These excitations are
usually known as {\it plasmons} and {\it paramagnons} respectively.
On the other hand, although
thermodynamic and static measurements such as specific heat and spin
susceptibility can be expressed in terms of the Fermi liquid parameters,
the information contained in Landau parameters is not enough to
describe some other measurements, e.g.  transport properties and superconductivity (or superfluidity in $^3$He). 
In particular, the interaction between the 
quasiparticles plays important role for these properties. We saw in the
last chapter that the effective interaction between quasiparticles can
be obtained from the Landau interaction function via Bethe-Salpeter equation
(\ref{B-S}). This is however, not always the easiest route to take.
The other way is to assume a model Hamiltonian for interaction between
quasiparticles and apply field theory techniques to calculate the physical
properties. The reliability of these calculations should then be judged by
experiments. One of the theories that follow this path is paramagnon
theory which we describe briefly here.

Magnetic susceptibility of a Fermi liquid is given by \cite{negele}
\be
\chi_{_M}= {\gamma^2 p_{_F} m^* \over 4\pi^2 (1+Z_0)} 
\label{chi0}
\ee
If the Landau parameter $Z_0 \to  -1$,
the denominator of (\ref{chi0}) vanishes and therefore magnetic susceptibility 
diverges. The system in that case undergoes a phase transition to a
ferromagnetically ordered state. 
When the denominator in (\ref{chi0}) is not zero but
very small, the magnetic susceptibility is still large and the neighboring
spins still have the tendency to align, although the long range order 
no longer exists. On the other hand, the interaction between the 
quasiparticles is given by  the T-matrix which can be found by Bethe-Salpeter
equation (\ref{B-S}). In particular (\ref{tZ}) gives
\be
t^a_0=Z_0/(1+Z_0)
\ee
which also diverges as $Z_0 \to -1$.
Thus strong magnetic susceptibility results in strong interaction between
quasiparticles; which tends to align the neighboring spins.
The dynamic susceptibility of a Fermi liquid
was suggested by Leggett \cite{leggett0} to be
\be
\chi (\vec q, \omega) ={\chi^{(0)}(\vec q,\omega) \over 1+ Z_0\chi^{(0)}(\vec q,\omega)/N(0)}
\label{chi1}
\ee
where $\chi^{(0)}$ is the susceptibility of the non-interacting 
electron system (particle-hole bubble). In 3-d isotropic systems, $\chi^{(0)}$ 
is given by Lindhard function \cite{mahan}.

Eq. (\ref{chi1}) was proposed by Leggett using deductive reasoning. 
An attempt to derive it was first made by paramagnon theory \cite{doniach}. 
The theory assumes that the (effective) interaction occurs only between 
quasiparticles with the opposite spins.
The model Hamiltonian for this interaction was first proposed by Berk and 
Schrieffer \cite{berk} as
\be
H_{\rm int}= I \int d^3r \ n_{\uparrow}(\vec r) n_{\downarrow}(\vec r)
= {I \over V} \sum_{\vec k,\vec k',\vec q}
c^\dagger_{\vec k + \vec q}c_{\vec k}c^\dagger_{\vec k' - \vec q}c_{\vec k'}
\label{HBS}
\ee
where $c_{\vec k}^\dag$ ($c_{\vec k}$) is an operator that creates
(annihilates) a quasiparticle with momentum $\vec k$ and $I>0$ is 
the coupling constant.
Performing Eliashberg type calculations, Berk and 
Schrieffer showed that in nearly ferromagnetic metals, such as Pd, 
superconductivity is suppressed as a result
of ferromagnetically enhenced coupling between the quasiparticles.
This model was then adopted by Doniach and Engelsberg \cite{doniach} 
and Rice \cite{rice} for liquid $^3$He. 

Spin susceptibility can be calculated from (\ref{HBS}) using Kubo 
formula \cite{mahan} and RPA approximation \cite{doniach}
\be
\chi (\vec q, \omega) ={\chi^{(0)}(\vec q,\omega) \over 1-I \chi^{(0)} (\vec q, \omega) }
\label{chi2}
\ee
Comparing (\ref{chi1}) and (\ref{chi2}) we find
\be
Z_0=-N(0) I
\label{ZI}
\ee
Thus the interaction $I$ is closely related to Fermi liquid interaction
function $f_{pp'}$.
In low energy and momentum, one can expand (\ref{chi2}) 
to get
\be
\chi (\vec q, \omega) ={N(0) \over 1-\bar{I} + (\bar{I}/12)\bar{q}^2  
-i(\pi/4) \bar{I} \omega/\bar{q} }
\label{chi3}
\ee
where $\bar{I}=I N(0)$ and $\bar{q}=q/p_{_F}$. 
The susceptibility $\chi(\vec q, \omega) 
$ in (\ref{chi3}) is peaked at $q=0$, reflecting the proximity to ferromagnetism.
Near the antiferromagnetic instability, one expects the susceptibility
to be peaked near $\vec q = (\pm \pi/a, \pm \pi/a,\pm \pi/a)$ 
where $a$ is the lattice spacing.

$\chi(\vec q, \omega)$ can be viewed as the propagator
of a bosonic field that mediates the interaction between 
the fermionic quasiparticles. The corresponding bosons are called 
paramagnons which are critically damped spin waves. 
Eq. (\ref{chi3}) can be written as
\be
\chi (\vec q, \omega) ={\chi_0 \over 1 + \xi^2 q^2 - 
i \omega/\omega_q}
\label{chiFM}
\ee
where $\chi_0=SN(0)$ is the enhanced static susceptibility, 
$\xi=\sqrt{S\bar{I}/12 p_{_F}^2}$ is the ferromagnetic correlation 
length and  $\omega_q= 4q/\pi \bar{I}p_{_F}$ is an energy scale that 
characterizes the damping of the bosonic particles. Here,
$S=1/(1-\bar{I})$ is the Stoner enhancement factor which measures
the proximity to the ferromagnetic instability.

Although paramagnon theory gives the right form for the spin susceptibility,
it has some internal inconsistencies \cite{stamp2}. For example if we
calculate the density-density correlation, instead of (\ref{ZI}) we get 
\cite{mahan}
\be
F_0=N(0) I
\label{FI}
\ee
(\ref{ZI}) and (\ref{FI}) can be satisfied simultaneously only if $F_0=-Z_0$. 
This is obviously not always the case. For example
for liquid $^3$He \cite{mahan}, $F_0=10$ but $Z_0=-0.67$.
Thus paramagnon theory cannot be the right quantitative theory for the bare
quasiparticles in liquid $^3$He. Even without comparison with experiments
the theory has internal inconsistencies. For example RPA approximation does
not satisfy Pauli principle \cite{tremblay}. Another problem is the absence
of Migdal's theorem in the electron-spin fluctuation interaction, 
first raised by Hertz {\it et al.} \cite{hertz}. 
Vertex corrections was shown to be of the same order as the bare interaction.
This invalidates the Migdals theorem and therefore any Eliashberg type
calculation. We will study this point in the context of NAFL theory later.
Besides quantitative inconsistency, paramagnon theory gives some qualitatively 
right results. We will not discuss the applications of paramagnon theory in
this thesis. Instead, we proceed to the antiferromagnetic version of the 
theory which is relevant to high $T_c$ superconductivity.
Interested reader should refer to Ref. \cite{stamp2} 
for more information about paramagnon theory.

\section{Antiferromagnetic Spin Fluctuations}

It is more or less well established that high $T_c$ materials are effectively  
2-dimensional electron systems close to the antiferromagnetic instability. 
As we showed in the last chapter, a two-dimensional electron 
system with a smoothly curved Fermi surface can be a Fermi liquid (at least
in weak interaction regime). On the other hand, NMR experiments
\cite{berthier2} and inelastic neutron scattering (INS) \cite{hayden} 
indicate enhanced staggered susceptibility, even at optimal doping. 
One therefore might be tempted to generalize paramagnon theory 
to antiferromagnetic systems to describe high $T_c$ superconductors.
It is well known that for a nested Fermi surface, the RPA susceptibility
(\ref{chi2}) diverges at the wave vector $\vec q=\vec Q =
(\pm \pi/2a, \pm \pi/2a)$ resulting in spin density wave instability 
\cite{shankar}.  For a general Fermi surface however, microscopic 
derivation of the spin susceptibility fails to succeed because the 
incoherent parts of the quasiparticle spectral function
play important role in the calculation of the real part of the
particle-hole bubble \cite{chubu1}. 
The low frequency part of the imaginary part of the particle-hole bubble, 
on the other hand, is given merely by the coherent parts and can be easily
calculated. We will discuss this issue when we talk about vertex corrections
later. Based on NMR and NIS results, a phenomenological susceptibility 
analogous to  (\ref{chiFM}) was suggested by Millis, Monien and Pines 
\cite{MMP} for high $T_c$ systems
\be
\chi (\vec q, \omega) ={\chi_Q \over 1 + \xi^2 (\vec q-\vec Q)^2 - 
i \omega/\omega_{sf}}
\label{chiAF}
\ee
where $\chi_Q \gg \chi_0$ is now the static staggered susceptibility, $\xi$
is the antiferromagnetic correlation length and $\omega_{sf}$ specifies
low frequency relaxation behavior. 
As in paramagnon theory, quasiparticles are again assumed to interact
with each other via the spin fluctuations. The Hamiltonian that describes
this interaction was written as \cite{MBP}
\be
{\cal H}= \sum_{\vec p,\sigma}\epsilon_{\vec p} 
\psi_{\vec p \sigma}^\dag \psi_{\vec p \sigma}+{\bar{g}\over 2} 
\sum_{\vec q,\vec k,\alpha,\beta}\psi_{\vec k+\vec q,\alpha}^\dag
\psi_{\vec k, \beta} \ \sigma_{\alpha\beta} \cdot \vec S(-\vec q)
\label{Hsf}
\ee
where $\bar{g}$ is spin-fermion coupling constant and
\be
\epsilon_{\vec p}=-2t[\cos(p_xa)+\cos(p_ya)]-4t' \cos(p_xa)\cos(p_ya)
-\mu
\label{epsilon}
\ee
is the quasiparticle dispersion relation \cite{MP1,14}, with $t=0.25$eV and 
$t'=-0.45 t$. $\vec S(\vec r)$ is the spin fluctuation density operator with 
a bosonic propagator given by (\ref{chiAF}). 

\begin{figure}
\epsfysize 8cm
\epsfbox[-250 150 130 700]{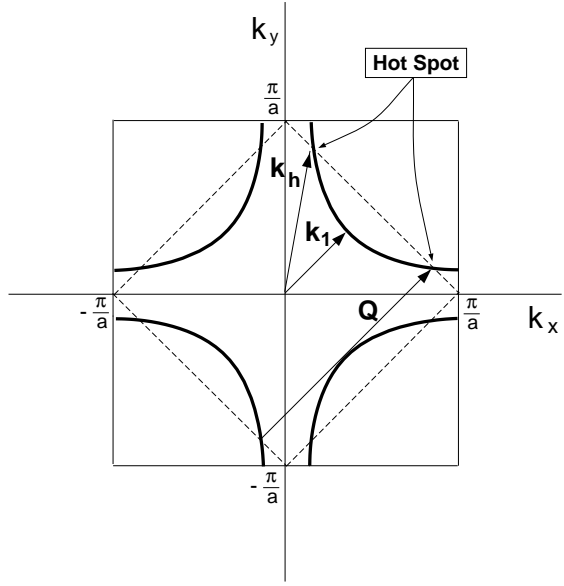}
\caption[Fermi surface in the first
Brillouin zone, with the value for $t$ and $t'$ given in the text.]
{\small Fermi surface in the first
Brillouin zone, with the value for $t$ and $t'$ given in the text; and we
assume n=0.75. Calculations are presented here for the wave vectors 
$k_1$ and $k_h$.} 
\label{fg1}
\end{figure}

The presence of $t^{\prime}$ allows ``hot spots" \cite{hlubina}
on the Fermi surface 
(Fig. \ref{fg1}) which are points that can be connected by $\vec Q$; 
this leads to singular behavior when $\xi$ in (\ref{chiAF}) diverges 
\cite{chubu2,hlubina}. Quasiparticles near the hot spots are highly scattered and 
play important role in NAFL calculations. Full analysis of the scattering
rate \cite{hlubina,SP} indicates linear $T$ dependence at the hot spots, 
while $T$ or $T^2$ dependence at other points, depending 
on the temperature and the distance
from the hot spots. This is used to explain the linear $T$ dependence of
the normal state resistivity of cuprates and the crossover between
linear $T$ behavior to $T^2$ behavior
\cite{hlubina,SP}.  Some other properties of the cuprates such as 
superconducting $T_c$ \cite{MBP,MP1,MP2}, Hall conductivity \cite{SP},
magnetic and scaling properties \cite{sokol,barzykin}, evolution of 
Fermi surface \cite{CM,chubu3}, Raman scattering \cite{morr1}, etc.,
have been calculated using NAFL theory. It is not our intention in this thesis
to describe these calculations in detail. Instead, we would like to study the
self-consistency of the theory. We proceed by briefly discussing how
d-wave superconductivity naturally arises from NAFL theory. We then
discuss the self-consistency of NAFL and calculations made 
by this theory.

\section{d-wave Superconductivity}

In NAFL theory, antiferromagnetic spin fluctuations play the same role in
the pairing process as phonons do in conventional superconductors. One can
therefore apply BCS theory, replacing phonons with magnons. There is 
however a principle difference here, that is the effective interaction potential
\be
V_{\rm eff} = g^2 \chi (\vec q)
\ee
is always positive (repulsive). Here $g=\sqrt{3/4} \bar{g}$ (the factor
comes from the consideration of the Pauli matrices).
BCS gap equation for this interaction is
\be
\Delta_k= -g^2 \int d^2k' \Delta_{k'} \chi(k - k') {\tanh (\epsilon_{k'}/2T)
\over 2 \epsilon_{k'} }
\label{BCS}
\ee
Since $\chi$ is positive, for an s-wave gap $\Delta_k$, 
both sides of (\ref{BCS}) appear with different
signs and therefore s-wave paring is not
possible. However, since $\chi(\vec q)$ is peaked at $\vec q=\vec Q$,
d$_{x^2-y^2}$ pairing is possible because in that case 
$\Delta_{k+Q}=-\Delta_k$ and this changes the negative sign in (\ref{BCS}) 
to positive. Therefore d$_{x^2-y^2}$ pairing is a natural consequence
of NAFL theory. Experimental observation of d-wave pairing is 
considered as one of the triumphs of NAFL theory.

In order to get high $T_c$, spin fermion coupling should be strong. At the
beginning it was thought that short quasiparticle lifetime 
due to the strong 
scattering would reduce $T_c$ dramatically \cite{pines1}. Weak coupling 
\cite{MBP} and also strong coupling (Eliashberg) \cite{MP1,MP2}
calculations of Urbana group however, have found a $T_c$ as 
high as 90K, even including the lifetime effect \cite{MP2}. 
Weak coupling calculation in Ref. \cite{MBP} gives a
$T_c$ which is well approximated by
\be
T_c= \alpha \left( {\Gamma \over \pi^2} \right) e^{-1/\lambda}
\label{TcMBP}
\ee
where $\Gamma \approx 0.04 eV$ is an energy scale,
\be
\lambda = \eta g^2 \chi_0 N(0)
\ee
is the dimensionless coupling constant, and $\alpha$ and $\eta$ are
dimensionless constants of $O(1)$.
Strong coupling calculation of $T_c$ was also done by Monthoux and
Pines (MP) \cite{MP1,MP2} using Eliashberg formalism. For $t'=0$, a 
relation similar to (\ref{TcMBP}) was again obtained
\be
T_c= 0.636 \left( {\Gamma \over \pi^2} \right) e^{-1/0.402 N(0) g}
\label{Tc}
\ee
where
\be
N(0)= - {2 \over \pi} \sum_p {\rm Im} G_{_R} (\vec p,0)
\label{NG}
\ee
is the tunneling density of state calculated from one particle 
retarded Green's function at $T_c$ \cite{MP1}.
It is interesting that $T_c$ is only sensitive to $g$ and not other parameters.
Fitting to the measured value for $T_c$, one can determine $g$.
For $T_c=90K$, one gets $g \approx 0.8$ \cite{MP1}.

\section{Self-Consistency of NAFL}

Besides the similarities, there is one fundamental difference between NAFL 
theory and paramagnon theory: charge properties of
ferromagnets are still metallic while antiferromagnets are insulators. 
The reason behind this difference as pointed out by Anderson 
\cite{andrsn,anderson3}
is that ferromagnetism is a natural consequence of Hund's rule:
exchange interaction favors spins to be parallel in order to keep electrons
as far as possible and make the repulsion energy minimum. 
Antiferromagnetism on the other hand, is a result of superexchange 
interaction which is inherently a non-perturbative effect and is a property
of Mott insulators. Going out of the antiferromagnetically ordered state
involves a metal-insulator transition which does not have
a counterpart in ferromagnetic case (there is a slight difference between
strong coupling and weak coupling here. In weak coupling the N\'{e}el 
ordering and metal-insulator transition - which is not of the Mott type -
happen at the same time whereas in strong coupling the system is already
an insulator above the N\'{e}el temperature). It therefore
completely makes sense to talk about a nearly ferromagnetic Fermi liquid
but one should be very careful to talk about a nearly antiferromagnetic 
Fermi liquid. On the other hand, almost none of the properties of high $T_c$
superconductors resembles of a Fermi liquid, even outside the 
antiferromagnetically ordered (insulating) region of the phase diagram 
(except for very high dopings). Thus it is hard to believe that the electron 
system turns into a Fermi liquid state from a highly non-Fermi liquid 
insulating state by slightly adding doping. 

Another problem, which is more computational than fundamental, is
consideration of the vertex corrections (note that in pseudogap problem
the strong frequency dependence of the vertex correction may make a
difference between having or not having a pseudogap \cite{tremblay}. 
I that sense vertex
correction is a fundamental problem). In order to calculate $T_c$ in strong
coupling limit, Eliashberg formalism \cite{mahan} is commonly used. A necessary
condition for the applicability of Eliashberg formalism is Migdal's theorem 
\cite{migdal}; which states that the corrections to the bare spin-fermion vertex
are negligible compared to the bare vertex. In case of electron-phonon 
interaction, the vertex corrections are of order of $(\omega_D/E_{_F}) \sim
\sqrt{m/M} \ll 1$, where $m$ and $M$ are the masses of electron and nucleus
respectively. Spin fluctuations however are 
different from phonons in three aspects. First, phonons are external
excitations while spin fluctuations are excitations of the electronic system
itself. Secondly, unlike phonons, spin fluctuations are not propagating 
modes. Thirdly, the ratio $J/E_{_F}$, with $J$ being the exchange
coupling responsible for antiferromagnetism, is of order of one, 
so there is no small parameter like $(\omega_D/E_{_F})$ in 
spin-fluctuation interaction \cite{schrieffer1}.
     
Early estimates reported small vertex corrections 
\cite{millis1}. Calculations of Ref. \cite{altshuler} and \cite{chubu2} 
on the other hand, indicated that close to the antiferromagnetic instability point,
the first vertex correction contains a logarithmically divergent term. 
However, sum of a series of divergent diagrams results in a power law
behavior for the dressed coupling \cite{chubu2,CMM}
\be
g_{\rm tot} \sim g ({\xi \over a})^\beta
\label{gtot}
\ee
where $\beta$ is a small positive number 
(we should mention that in the original publication \cite{chubu2},
$\beta$ was reported to be negative but in later
publications $\beta$ was claimed to be positive \cite{CMM}). 
Thus near the antiferromagnetic 
instability ($\xi \to \infty$), $g_{\rm tot}$ become very large.
In other words, vertex corrections tend to increase the spin-fermion coupling.

On the other hand, a work by 
Schrieffer \cite{schrieffer1} showed that right outside
the antiferromagnetically ordered phase, the dressed spin-fermion vertex
is given by
\be
{\Lambda_{\vec q}\over \Lambda_\circ} \propto
[ (\vec q - \vec Q)^2 + \xi^{-2} ]^{1/2}
\label{schr}
\ee
Thus very close to the antiferromagnetic instability ($\xi \to \infty$),
the full vertex vanishes as $\vec q \to \vec Q$. The vanishing of the
dressed vertex is a consequence of Adler principle: in the ordered state,
magnons are Goldstone modes and their interaction with other degrees 
of freedom should vanish at the ordering momentum to preserve the
Goldstone modes to all orders in perturbation theory \cite{CM}.
The above statement means that the vertex corrections should 
be equal and opposite sign to the bare vertex to make the dressed 
vertex vanish.
Consequently, Migdal's theorem can not exist near the antiferromagnetic 
instability. However, superconductivity does not happen at 
very low dopings either. The relevant question therefore is the validity
of Migdal's theorem at higher dopings, outside the ordered state
where superconducting transition occurs at high temperatures.
 
The first attempt to calculate vertex corrections at optimal doping 
was made by us \cite{amin0}. Using the 
same parameters as used in NAFL theory,  we showed that the first correction 
to the bare vertex is actually the same order as the bare vertex \cite{amin0}.
The sign of the correction is also controversial because it determines 
whether vertex corrections increase or suppress the effect of the spin 
fluctuation coupling. Our finding had a sign opposite to the bare vertex. 
The effect of the correction is therefore to reduce the spin-fermion 
coupling, in agreement with Schrieffer's result (\ref{schr}).
In the next section we describe our calculation in detail.
We then discuss the effect of inclusion of the quasiparticle renormalization
in our calculation. 

To conclude this section we mention one experimental consideration.
The susceptibility (\ref{chiAF}) is adopted by NAFL theory only 
on phenomenological basis and by fitting to experiments. Consequently,
NAFL theory is incapable of explaining the incommensurate peaks observed in
inelastic neutron scattering
(INS) experiments on ${\rm La_{2-x}Sr_xCuO_4}$ \cite{ins} and
recently on ${\rm YBa_2Cu_3O_{6+x}}$ \cite{dai}. The peaks are 
observed in the magnetic response $\chi''(\vec q,\omega)$  
at $\vec Q_i=(1\pm \delta,1)\pi$ and $\vec Q_i=(1,1\pm \delta)\pi$.
It should be pointed out that some other theories such as the theory of stripes
have natural explanations for the incommensurate peaks 
\cite{stripe,tranquada}.

\section{Calculation of Vertex Correction}

Let us first assume that the quasiparticle residue is 
$Z \approx 1$. We will discuss the effect of the quasiparticle 
renormalization later. The first vertex 
correction is shown in Fig. \ref{fg2} and can be written as
\be 
\delta\Lambda(3) = {{\bar g}^2\over 4}
\int \Pi(1-2) \ G(1-3) \ G(3-2)
\ee
where numbers are representing the coordinates and $\Pi$ is defined by
\ba
\Pi &=& {{\bar g}\over 2}\sum_{j=1}^3 
\sigma^j \sigma^i \sigma^j <S^j_1 S^j_2>
={{\bar g}\over 2}\sum_{j=1}^3  
(2\delta^{ij} - \sigma^i \sigma^j)
\sigma^j  <S^j_1 S^j_2> \nn \\
&=&{{\bar g}\over 2} \sigma^i \left( 2<S^i_1 S^i_2> - \sum_{j=1}^3 
<S^j_1 S^j_2> \right)
= - \Lambda_\circ \ \chi(1-2)
\ea
where $\Lambda_\circ = ({\bar g}/ 2) \sigma^i$
is the bare vertex. In the last step we assumed three equivalent 
modes of spin fluctuation, reflecting the rotational symmetry of
the system. In momentum representation the first vertex
correction can be calculated for general external momenta. Here
we concentrate on the interaction between a Fermi surface electron and a
spin fluctuation with ${\vec q}={\vec Q}$. We have
\be
{\delta\Lambda_{\vec k}\over \Lambda_\circ} = -{{\bar g}^2\over 4}V
\sum_{{\vec Q^\prime}}\int {d^2q\over (2\pi)^2}{d{\tilde \omega}\over 2\pi}
\chi ({\vec Q^{\prime}+q},{\tilde \omega})G({\vec k+\vec Q'+\vec q},
{\tilde \omega})G({\vec k+ \vec{Q''}+\vec q},{\tilde \omega})
\label{eq.4}
\ee
where $\delta\Lambda_{\vec k}=\delta\Lambda({\vec k},\epsilon=0;{\vec q=\vec Q},
\Omega=0)$ and ${\vec{Q''}}={\vec Q}+{\vec Q^{\prime}}$; the sum over 
${\vec Q^{\prime}}$ takes care of Umklapp processes.
In the reduced Brillouin zone ${\vec Q^{\prime\prime}}=(0,0)$.
Note that the overall sign of the graph in Fig. \ref{fg2} is 
{\it negative}, i.e., Eq. (\ref{eq.4}) is negative.

\begin{figure}
\epsfysize 6cm
\epsfbox[-40 260 468 500]{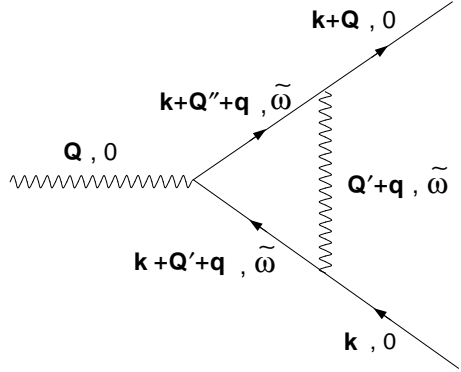}
\caption[First correction $\delta\Lambda_{\vec k}$ to the bare vertex.]
{\small First correction $\delta\Lambda_{\vec k}$ to the bare vertex, 
for an incoming fermion with momentum ${\vec k}$ and energy 0 (relative to the
Fermi energy), interacting with a fluctuation of wave-vector ${\vec Q}$ and
zero energy.
\label{fg2}}
\end{figure}

We concentrate on $\delta\Lambda_{\vec k}$, with ${\vec k}$ being a point 
on the Fermi surface, since it gives an indication of the size 
of vertex corrections. This is because in general 
$|\delta\Lambda({\vec k,\epsilon;q},\Omega)|$ exceeds   $|\delta\Lambda_{\vec k}|$ 
(in fact it diverges along a surface in k-space, for given values of 
${\vec q},\Omega$ and $\epsilon$); thus one cannot argue, even after integrating 
over one or more of its arguments, that 
$|\delta\Lambda({\vec k,\epsilon;\vec q},\Omega)|$ will lead to corrections smaller 
than one would get from just using $\delta\Lambda_{\vec k}$ (our argument here 
parallels Migdal's \cite{migdal}). More generally one finds that if 
$\Omega\ll{\rm qv_F},\Delta$ , where $\Delta$ is the spin gap, then 
$\delta\Lambda_{\vec k}$ is a good approximation to 
$\delta\Lambda({\vec k,\epsilon;\vec q},\Omega)$. Writing (\ref{eq.4}) as

\begin{eqnarray}
{\delta\Lambda_{\vec k}\over \Lambda_\circ} &=&-{{\bar g}^2\over 4}V
\sum_{{\vec Q^\prime}}\int {d^2q\over (2\pi)^2}{d\omega\over \pi}
{\chi^{\prime\prime} ( {\vec Q'+ \vec q},\omega)\over 
\epsilon_{\vec k+\vec Q^{\prime}+\vec q}-
\epsilon_{\vec k+\vec Q^{\prime\prime}+\vec q}} \nn \\
&&\times \left( {f_{\vec k+\vec Q'+\vec q}\over 
\epsilon_{\vec k+\vec Q'+\vec q}-\omega}+ 
{1-f_{\vec k+\vec Q'+\vec q}
\over \epsilon_{\vec k+\vec Q'+\vec q}+\omega} 
- {f_{\vec k+\vec Q''+\vec q}\over \epsilon_{\vec k+\vec Q''+\vec q}
-\omega}-{1-f_{\vec k+\vec Q'+\vec q}\over 
\epsilon_{\vec k+\vec Q'+\vec q}+\omega} \right) \nn \\
&=& -g^2{\chi_Q\over {12\pi^3\omega_{sf}}} ({\omega_{sf}\over \mu})^2
{\cal I}_{\vec k}
\label{vtx}
\end{eqnarray}
where  $g^2={3\over 4}{\bar g}^2$ and ${\cal I}_{\vec k}$ is dimensionless; we
define $g$ to correspond directly with the coupling constant $g$ used in
Monthoux and Pines \cite{MP1,MP2}.
At zero temperature one has
\be
{\cal I}_{\vec k} = \int_{-2\pi}^{2\pi}d{\bar q}_x \int_{-2\pi}^{2\pi}d{\bar q}_y 
{1\over {{\bar \epsilon}_1-{\bar \epsilon}_2}}[{\rm Sgn}({\bar \epsilon}_1)
G_1({\vec {\bar q}},{\bar \epsilon}_1)- {\rm Sgn}({\bar \epsilon}_2)G_1
({\vec {\bar q}},{\bar \epsilon}_2)
-G_2({\vec {\bar q}},{\bar \epsilon}_1)+G_2({\vec{\bar q}},{\bar \epsilon}_2)]
\label{eq.6}
\ee
where ${\bar q_x} = q_xa$, etc., ${\bar \epsilon}_1$ and ${\bar \epsilon}_2$
are 
\be
{\bar \epsilon}_1={\epsilon_{\vec k+\vec Q'+\vec q}\over \mu} \quad , \quad
{\bar \epsilon}_2={\epsilon_{\vec k+Q''+\vec q}\over \mu} 
={\epsilon_{\vec k+\vec q}\over \mu}
\ee
and $G_1$ and $G_2$ are defined by
\ba
G_1({\vec q},{\bar \epsilon})={\pi X\over 2({\bar \epsilon}^2+X^2)}
\quad , \quad
G_2({\vec q},{\bar \epsilon})={{\bar \epsilon} \ln (X/|{\bar \epsilon}|)\over 
{\bar \epsilon}^2+X^2}
\label{eq.8}
\ea
with $X=[1+({\xi\over a})^2({\bar q}_x^2+{\bar q}_y^2)](\omega_{sf}/|\mu|)$.

${\cal I}_{\vec k}$ can be investigated both analytically and numerically. 
Here we calculate it for two different points in k-space, both on the 
Fermi surface (see Fig. \ref{fg1}); ${\vec k_1}$ makes a $45^o$ angle with 
$k_x$, and ${\vec k_h}$ is a ``hot spot" wave vector. 

In order to obtain a value for $\delta\Lambda_{\vec k}$, we need values for 
the spin fluctuation energy $\omega_{sf}$, the correlation length $\xi$, 
the susceptibility $\chi_Q$, the coupling constant $g$, and the chemical 
potential $\mu$ (which is determined by the electron filling factor n). 
There are different values reported in the references \cite {MMP,MBP,pines1}. 
We have evaluated $\delta\Lambda_{\vec k}$ for ${\vec k}={\vec k_1},{\vec k_h}$, 
in two ways, viz. (a) by assuming various published values for the 
different parameters, and (b) by making the simple assumption that 
one is very close to
an AFM instability, and then, in the spirit of RPA, imposing the 
condition  $|({\bar g}/2)\chi^{(0)} ({\vec Q},0)| \approx 1$ where 
$\chi^{(0)} ({\vec q},\omega)$ is the electron-hole bubble. Numerical 
calculation gives  $|\chi^{(0)} ({\vec Q},0)|= 2.6 ({\rm eV})^{-1}$ and 
thereby $g=0.67$ eV, assuming n=0.75 and the band structure in (\ref{epsilon}). 
Since $\omega_{sf}/|\mu| \ll 1$, this value of $g$ should be a very 
good guess, within a naive RPA scheme. 

\begin{table}
\begin{center}
\begin{tabular}{|c|c|c|c|c|c|c|}  \hline
& $g$ (eV) & $\omega_{sf}$ (meV) & ${\cal I}_{\vec k_1}$ & $\delta
\Lambda_{\vec k_1}/\Lambda_\circ$ & ${\cal I}_{\vec k_h}$ &
$\delta\Lambda_{\vec k_h}/\Lambda_\circ$\\ \hline\hline

MPI  & 1.36 & 7.7 & 78.6 & -1.81 & 105.6 & -2.43 \\ \hline
MPII  & 0.64 & 14 & 49.6 & -0.46 & 73.4 & -0.68 \\ \hline
``RPA" & 0.67 & 7.7 & 78.6 & -0.44 & 105.6 & -0.59 \\ \cline{3-7}
       &      & 14 & 49.6 & -0.50 & 73.4 & -0.74 \\ \hline
\end{tabular}
\end{center}
\caption[Calculated values of the vertex correction $\delta\Lambda_{\vec k}$
for two different wave-vectors $\vec k_1$ and $\vec k_h$ on the Fermi surface.]
{\small Calculated values of the vertex correction $\delta\Lambda_{\vec k}$ for two
different wave-vectors $\vec k_1$ and $\vec k_h$ on the Fermi surface
(columns 4 and 6 in the table). Ref. \cite{MP1} and \cite{MP2}
give different values for $g$, and different values for $\omega_{sf}$.
From the values for these two models one calculates ${\cal I}_{\vec k}$
in equation (\ref{eq.6}), and thence $\delta\Lambda_{\vec k}/\Lambda_\circ$.
The third model is the naive ``RPA" model described in the text, for
which $g$ is determined; we have calculated  ${\cal I}_{\vec k}$ and
$\delta\Lambda_{\vec k}/\Lambda_\circ$ for two values of $\omega_{sf}$
given in MPI and MPII respectively.}
\label{table1}
\end{table}

The results are summarized in Table \ref{table1}. We use two different values 
for $\omega_{sf}$ ; these are the two different values quoted by 
Monthoux and Pines {\it et al}. \cite{MP1,MP2}. 
We also use two different values for 
the coupling constant $g$, quoted from Ref. \cite{MP1} and \cite{MP2}. 
We use values of $\xi=2.5a$, $\chi_Q = 80$ states/eV (from \cite{MP1,MP2}) 
and $n=0.75$ , appropriate to  ${\rm YBa_2Cu_3O_7}$ (again quoted from 
\cite{MP1,MP2}). This value of $n$ corresponds, with the band structure in 
(\ref{Hsf}), to a value of $|\mu|\sim 1.46t \equiv 0.365$ eV.   

We see that even the values for the vertex correction calculated from the 
simple RPA model (b) are not small; as in the standard discussion of Migdal's 
theorem, the importance of vertex corrections appears in the ratio 
$|\delta\Lambda_{\vec k}|/\Lambda_\circ$. If one takes values of $g$ from 
the literature \cite{MP1,MP2}, this ratio is quite unreasonably large (as large as 2.43 
for the hot spots in the model used by Ref. \cite{MP1}). 
Thus vertex corrections are clearly 
very important. The values we quote for $|\delta\Lambda_{\vec k}|/\Lambda_\circ$ 
are considerably larger than previous estimates \cite{MMP,millis1,22}.  
The reason for this difference with previous work can be tracked back to 
the factor ${\cal I}_{\vec k}$, which is impossible to guess from purely 
dimensional arguments. In fact if we drop the factor ${\cal I}_{\vec k}$ 
from $\delta\Lambda_{\vec k}$, we get an order of magnitude estimate for 
$\delta\Lambda_{\vec k}$ given by
\be 
{|\delta\Lambda_{\vec k}|\over \Lambda_\circ} \sim O[g^2{\chi_Q \over 
4\pi^3|\mu|}{\omega_{sf} \over |\mu|}] \ll 1\nonumber
\label{eq.9}
\ee
which is broadly in agreement with previous estimates (see e.g. Millis \cite{millis1})
; $\chi_Q$, $\omega_{sf}$ and $g$ must be redefined to conform with the 
parametrizations in this paper).

In fact however  ${\cal I}_{\vec k}$ is surprisingly large, and also shows 
a significant variation around the Fermi surface, with a maximum at the hot 
spots, and a minimum at intermediate wave-vector like ${\vec k_1}$.  
We should emphasize here that analytic calculations of $\delta\Lambda_{\vec k}$ 
have to be approximated rather carefully in order to give reasonable 
agreement with the numerical results in Table \ref{table1}. Approximations 
such as those of Hertz {\it et al.} \cite{hertz} (see also \cite{22}), 
which try to separate off a rapidly-varying (in ${\vec q}$-space) contribution 
from $\chi'' ({\vec q},\omega)$, give quantitatively incorrect results 
(including a completely unphysical $\ln [{(\vec k-\vec k_h)}a]$ divergence as 
one approaches the hot spot). 

One might suppose that ${\cal I}_{\vec k}$ is large simply because of the
band structure (i.e., because of van Hove singularities, or the hot spots).
If this were true one could argue that the quasiparticle weight ought to be
renormalized down near these singular points in the Brillouin zone, and that
this would considerably reduce the vertex correction. In fact however we find 
this is not the case; this can be checked analytically by suppressing the 
regions immediately around the hot spots in the integral for 
${\cal I}_{\vec k}$, or by simply redoing the numerical calculation for
a slightly different band structure. We find that suppressing the hot spots 
entirely, reduces the vertex correction by a factor which is everywhere less
than 2 (and which differs very little from unity when ${\vec k}$
is far from a hot spot).

\subsection{The Effect of Quasiparticle Renormalization $Z$}

Our calculation of the vertex correction was criticized by Chubukov, Monthoux
and Morr (CMM) \cite{CMM} on the basis of inappropriate quasiparticle
renormalization factor $Z$. As we mentioned earlier the value we used was 
$Z \approx 1$ which is the value used in NAFL
calculations of $T_c$ \cite{MBP,MP1,MP2}. CMM's criticism was based on
the fact that self consistent consideration of spin fluctuation damping
gives spin fluctuation frequency $\omega_{sf}$ in terms of other parameters.
More precisely, one can write \cite{CM} 
$\omega_{sf}=c_{sw}^2/2\xi^2 \gamma$,
where $c_{sw}$ is the spin wave velocity, $\xi$ is the correlation rate
and $\gamma$ is the spin damping rate.
In fact two points of views can be taken here. One can assume the spin
environment as independent degrees of freedom. In that case $\omega_{sf}$
will be an independent parameter. On the other hand if one considers a one
band model then assuming that the damping of the spin excitations is 
dominantly due to the spin-fermion interaction (and not other sources such
as spin-spin exchange, impurity, etc.), the damping rate $\gamma$ can
be obtained from the imaginary part of the particle-hole bubble at transfer
momentum $\vec Q$. The result of the calculation given in \cite{CMM} is 
\be
\omega_{sf}={\pi \over 4} |\sin \phi_0| {v^2 \over g^2 Z^2 \chi_Q}
\label{osf}
\ee
where $v$ is the Fermi velocity at the hot spots and $\phi_0$ is the angle
between normals to the Fermi surface at  the hot spots. CMM claim
that if one uses the parameters reported in \cite{MP2} and at the
same time requires $Z=1$, one finds $\omega_{sf} \approx 1.06 meV$ which
is one order of  magnitude smaller than $\omega_{sf}=14 meV$
used in Ref. \cite{MP2} and our calculations. CMM then conclude that
this is a sign of inconsistency in our calculations. 
Using the same parameters and solving (\ref{osf}) 
for $Z$, with $\omega_{sf}=14 meV$ one finds $Z=0.28$.
Substituting this value of $Z$ in the calculations, the vertex correction
will be reduced by a factor of $Z^2=0.08$. This way the magnitude
of the correction will be one order of magnitude smaller than what we 
obtained, and therefore the vertex corrections will be small. 

We do not dispute the fact that all the calculations should be 
self-consistent. What we argue actually is that the self-consistency should be
considered everywhere throughout the calculation. We also agree
that the Hamiltonian (\ref{Hsf}) is valid for renormalized quasiparticles 
and therefore it is necessary to include the renormalization factor
$Z$ in all calculations. All the parameters we used in our 
calculation  (including $Z$) were based on the reported values by
Monthoux and Pines \cite{MP1,MP2}. We emphasize again that the coupling
constant $g$ in Ref. \cite{MP1,MP2} is obtained by fitting the calculated
$T_c$ to the measured value ($\approx $ 90K). If  $Z$,
as claimed in Ref. \cite{CMM} is much smaller than one, then this value 
should be also considered in $T_c$ calculations and would affect 
the magnitude of $g$ which gives a reasonable $T_c$. 

In a strong 
coupling calculation, this means that all the fermionic lines have to be
replaced by renormalized Green's functions. This multiplies the
coupling constant $g$ by a factor $Z$ which exactly compensates the 
reduction of the vertex correction discussed in \cite{CMM}. 
This is more clear by looking at (\ref{NG}). One realizes
that a renormalized $G_{_R}$ is going to reduce $N(0)$ by a factor
of $Z$. This factor multiplies $g$ in the exponent in (\ref{Tc}).
Keeping all the other parameters fixed the coupling constant should be
increased to preserve the value of $T_c$. The new coupling constant is 
now given by
\be
g_{_Z} = {g \over Z}
\label{gZ}
\ee
where $g$ is the (old) coupling obtained using $Z=1$ 
(i.e. the values reported in Ref. \cite{MP1} and \cite{MP2}). 
The same statement holds for weak coupling calculation (\ref{TcMBP})
keeping in mind that $\lambda \propto g^2 N(0)^2$.
Substituting $Z$ and $g_{_Z}$ into the calculation of the vertex correction
we get
\be
\left( {|\delta\Lambda_{\vec k}|\over \Lambda_\circ} \right)_{Z=1}
\propto g^2 \quad \longrightarrow \quad
\left( {|\delta\Lambda_{\vec k}|\over \Lambda_\circ} \right)_{Z}
\propto Z^2 g_{_Z} = Z^2 {g^2 \over Z^2} = g^2
\ee
The factor $1/Z$ in (\ref{gZ}) therefore exactly cancels the effect of $Z^2$ in
vertex correction calculations and our calculated value remains unchanged.
Looking back at (\ref{osf}) but now remembering that $gZ$ is actually what 
one obtains from fitting to $T_c$, we find $\omega_{sf}=1.06 meV$ even
after the inclusion of the renormalization factor $Z$.  The mismatch of this 
value with the reported value of Ref. \cite{MP2} reflects the 
inconsistency of the theory rather than a mistake in our calculations.

\subsection{Sign of the Vertex Corrections}

The sign of the vertex correction is important because if it is positive,
the vertex corrections tend to increase the effect of the spin fluctuations 
on the electron
system resulting in a higher $T_c$. On the other hand if it is
negative, it decreases the spin-fermion coupling and reduces the $T_c$.
We saw that at the optimal doping, our calculation indicates a negative
sign for the first vertex correction. On the other hand, Ref.
\cite{CM,CMM} report a positive sign at the optimal doping but negative sign
near the antiferromagnetic instability. 
In this section 
we discuss what a change in the sign of the vertex correction between
strong and weak doping would entail physically 
- essentially it implies a phase transition of some sort for which there is
no experimental evidence that we know of.
We give our argument based on a fact with which everyone agrees: near the 
antiferromagnetic instability the sign of the vertex correction has to 
be negative to suppress the effect of spin-fermion coupling \cite{schrieffer1}.

If the sign of the vertex corrections at the optimal doping is negative, as 
we found, then the vertex corrections tend to decrease the dressed coupling.
Decreasing doping, increases the effect of spin fluctuations and therefore
increases the magnitude of the vertex corrections. This can be checked
by Eq. (\ref{vtx}). Decreasing doping would increase $\chi_Q$ ($\propto
\xi^2$) and thereby the vertex correction. Consequently, 
the dressed coupling would be decreased even further. 
This trend continues until the 
total vertex vanishes completely at the edge of the ordered phase. Thus 
the flow of the renormalized coupling when changing the doping is
continuous. 
On the other hand if the vertex corrections 
have positive sign at the optimal doping, decreasing the doping would
increase the dressed vertex. Continuation of this trend towards zero
doping would give a very large dressed coupling which is unphysical.

%
%

Interestingly it was speculated by Chubukov et al \cite{chubu3,CM}
that perhaps a weak phase transition (a Lifshitz transition
) might exist in the intermediate doping regime. 
The transition would change the big Fermi surface, compatible
with Luttinger's theorem, to small hole pockets centered at
$(\pi/2a,\pi/2a)$ and symmetry related points.
If this were really true, and
moreover followed from the Hubbard model, it might save the consistency of
the NAFL approach, in that it would make it possible for a sign change of
the vertex corrections, at some critical $g=g_c$.
To our knowledge, no experimental justification exists yet \cite{arpes}
and the evolution of the Fermi surface is not
widely believed. Without this evolution, the positive sign of the vertex
correction at optimal doping is not physical.
At the end we point out that the suppression of the spin-fermion 
coupling due to the vertex corrections, also gives a natural way to explain why
$T_c$ reduces at low dopings in spite of the increase in the spin susceptibility
\cite{CM} (although other reasons can also be thought of, e.g. reduction
of charged carriers, etc.).

\section{Concluding Remarks}

We re-emphasize here that these results do not
necessarily invalidate the internal consistency
of the Fermi liquid starting point, in NAFL theory. However they do show that
the theory cannot be trusted quantitatively, at least in the usual RPA form.
As is well known the RPA is not a ``conserving approximation", and for spin
fluctuation theories this makes it unreliable
(cf. ref.\cite{stamp2}, especially section 3). 
It is useful to compare the case of nearly ferromagnetic $^3$He liquid,
where vertex corrections are also quite large, and where use of the paramagnon
model yields values for $m^*/m$ which are off by a large factor 
\cite{20}. Thus if we use melting curve Landau parameters,
$Z_0 \sim 0.75$ and $F_1 \sim 15$, we infer a value for the Stoner factor
$S \sim 24$ which yields $m^*/m = {9\over 2} \ln S \sim 15$ , in the paramagnon
model. This is 
roughly 2.5 times the correct value of $\sim 6$ (note that the
first vertex correction is $\delta\Lambda/{\bar I} \sim \ln S \sim 3$
in this model), and no amount of self-consistent summing of
diagrams can cure this numerical problem.

Similar problems can clearly occur in the present NAFL model.
We believe this is the main reason for the
difficulty one encounters in the MP models, in determining a value for $g$
that (a) gives the correct superconducting $T_c$, and (b) is consistent with
 the observed spin susceptibility. 

To check the structure at higher order, we have also estimated 
the contributions from the graphs
containing 2 spin fluctuation lines (there are actually 7 distinct
 graphs at this
level), and found that some of them are also large for the values of $g$ used
above. Thus,
just as for the case of nearly ferromagnetic
$^3$He,  we see no reason to believe, 
{\it for the values of the parameters given in the
table}, that performing infinite graphical sums 
will lead to results which 
are numerically more reliable, even if they do converge to some smaller 
renormalized vertex-there will always be other diagrams with large values,
which will in general give uncontrolled contributions. 

It is interesting to compare our results with some other 
investigations. As we mentioned briefly, in the weak-coupling limit, 
Chubukov \cite{chubu2,CM} has calculated the leading vertex corrections to $g$, 
concentrating on the gapless case; in the case where there is a gap, 
he finds that the renormalized coupling is large 
(for a large Fermi surface considered in our investigation).
On the other hand Schrieffer \cite{schrieffer1} has argued that a correct
formulation of the theory, even 
in the weak-coupling limit, must take account of the short-range local 
antiferromagnetic order even in the normal state-if this done, he finds
that a weak-coupling calculation shows very strong
{\it suppression} of the vertex
when one is close to the antiferromagnetic transition. This theory seems
rather interesting-note that a related calculation by Vilk and Tremblay
\cite{tremblay,26}finds that the existence of such a 
short-range antiferromagnetic order in 2-d will cause a breakdown
of the Fermi-liquid starting point itself when the correlation length 
becomes larger than the single particle de Broglie wavelength! Thus 
the question of what is the correct theory itself is rather
confused, even in the weak-coupling regime. It is certainly not clear how any of
these arguments will work in the regime discussed in this paper, when $g$ is 
not small enough to control the magnitude of the vertex corrections. 

It is of course crucial that these higher-order corrections also be
included in any version of this theory that tries to reconcile different
experiments - as emphasized by Pines \cite{pines1}, the 
justification of the theory stands or falls on
its ability to do this 
{\it quantitatively}. It is possible that such a program
might succeed if one can show that the actual parameters $g$, $\omega_{sf}$,
and $\chi_Q$ are such that $|\delta\Lambda_{\vec k}|/\Lambda_\circ$ is considerably
less than one
(i.e., if one is genuinely in the weak-coupling regime).
 This would also be true of  
versions of the theory in which $\omega_{sf}$ depends on $g$,
whilst the spin gap becomes an independent parameter \cite{chubu2}; or
of the theory of
Schrieffer cited above \cite{schrieffer1}.
On the other hand if  $|\delta\Lambda_{\vec k}|/\Lambda_\circ > O(1)$, we see no
hope that such a scheme could succeed 
quantitatively (in, e.g., the calculation of $T_c$),
 since the vertex corrections
become large.


\part{Vortices in d-wave Superconductors}

\chapter{Introduction}

Although at a microscopic level, theories of high $T_c$ superconductors, 
especially for normal state properties, are not well established, 
superconducting properties of cuprates are better understood. 
Almost everybody now agrees
that the symmetry of the order parameter is d$_{x^2-y^2}$-wave, and
the existence of nodes in the superconducting gap is well established
\cite{Hardy,van}.
A highly anisotropic d$_{x^2-y^2}$-wave order parameter, with four nodes
in the superconducting gap is a unique property of cuprates ans also
some organic superconductors, 
with no analog in conventional superconductors. The effect of this
anomalous symmetry on different observable quantities has been a 
subject of investigation in recent years.

An early theoretical
investigation of the weak-field response of a $d_{x^2-y^2}$ superconductor
by Yip and Sauls \cite{Yip} predicted a direction dependent non-linear
Meissner effect, associated with the quasiclassical shift of the excitation
spectrum due to the superflow created by the screening currents. Maeda
{\em et al.} \cite{maeda} reported experimental evidence for such an effect
in Bi$_2$Sr$_2$CaCu$_2$O$_y$, but subsequent experiments \cite{chris}
failed to confirm their findings and the situation remains controversial.
A similar effect was also studied independently by Volovik \cite{volovik}.
In the mixed state, he predicted a
contribution to the residual density of states (DOS) proportional to the
inter-vortex distance $\sim\sqrt{H}$. Such a contribution was identified in
the specific heat measurements on YBa$_2$Cu$_3$O$_{7-\delta}$
(YBCO) by Moler {\em et al.}\cite{moler}
but later this interpretation was disputed by Ramirez
\cite{ramirez} who found evidence for a similar effect in a conventional
superconductor V$_3$Si and by others\cite{others}.
Kosztin and Leggett \cite{Leggett} predicted
that the nonlocal response at very low temperatures will lead to
a $T^2$ dependence of the penetration depth in clean samples
in contrast to the linear $T$-dependence obtained from the local theory for
$d$-wave materials.

Another important characteristic of high $T_c$ compounds is their short
coherence length $\xi_0$; responsible for strong fluctuation effects near $T_c$ 
\cite{kamal}. The large ratio  $\kappa= \lambda_0 / \xi_0 \gg 1$
($\lambda_0$ is the magnetic penetration depth) also makes them extremely
type-II superconductors.
Existence of a mixed state, characterized by a regular array of
magnetic flux lines penetrating the material, is perhaps one of the most
striking properties of type-II superconductors. The original pioneering work
of Abrikosov \cite{aaa}, based on the
solution of Ginsburg-Landau (GL) equations \cite{gl}
near the upper critical field $H_{c2}$, predicted a triangular flux lattice.
This prediction was subsequently verified by low field magnetic decoration
experiments on a variety of conventional superconductors. In stronger fields
neutron scattering
experiments revealed significant deviations from perfect triangular lattices
\cite{review1} which where attributed to anisotropies in the electronic
band structure and other effects. These were modeled by GL theories containing
additional higher order derivative terms reflecting the
material anisotropies \cite{review1}.

Based on the experience with conventional superconductors one would expect
even richer behavior of flux lattices in the copper-oxide superconductors.
In high-$T_c$ cuprates much of the experimental and
theoretical effort has been  focused on the sizable region of the phase
diagram just below $T_c(H)$ in which the vortex lattice properties are
completely dominated
by thermal fluctuations \cite{review3}. While understanding the physics of this
fluctuation dominated regime poses a very difficult statistical mechanics
problem, investigation of the equilibrium vortex lattice
structures at low temperatures may provide clues about the microscopic
mechanism in these materials. 

Neutron scattering \cite{Keimer}
and STM \cite{Maggio} experiments on YBa$_2$Cu$_3$O$_{7-\delta}$ 
compound, revealed
vortex lattices with centered rectangular symmetry and different
orientations with respect to the ionic lattice. 
These have been modeled by
phenomenological GL theories appropriate for anisotropic superconductors,
containing additional quartic derivative terms
\cite{Takanaka} or a mixed gradient coupling to an order parameter with
different symmetry \cite{Berlinsky,Franz,Ren,Heeb}. These works found
structures in qualitative agreement with experiment, but their inherent
shortcoming is a large number of unknown phenomenological parameters and
subsequent lack of predictive power. Also, the GL theory is only solvable
for a vortex lattice near $H_{c2}$, which is experimentally inaccessible in
cuprates away from $T_c$. The observed behavior of the vortex lattice may 
also be understood by incorporating penetration depth anisotropy and 
twin-boundary pinning without involving any effects associated with 
mixing symmetries or gap anisotropy \cite{Walker}. 

Muon-spin-rotation
($\mu$SR) experiments \cite{sonier-thesis,sonier1,sonier2,sonier3},
on the other hand, show an
unusual magnetic field dependence in their line-shapes for the magnetic field
distribution. This has been attributed to a field dependent penetration depth
- which is expected in the Meissner state because of quasiparticle accumulation
at gap nodes \cite{Yip}.  It was also modeled using an approach based on the
Bogoliubov- de Gennes (BdG) equations in a square lattice
tight-binding model \cite{wang}.

At intermediate fields $H_{c1}<H\ll H_{c2}$, properties of the flux
lattice are determined primarily by the superfluid response of the
condensate, i.e. by the relation between the supercurrent $\vec j$ and
the superfluid velocity $\vec v_s$. In conventional isotropic
strongly type-II superconductors this relation is to a good 
approximation that of simple proportionality,
\be
\vec j=-e\rho_s\vec v_s,
\label{j0}
\ee
where $\rho_s$ is a superfluid density. More generally, however, this
relation can be both {\em nonlocal} and {\em nonlinear}.
The concept of nonlocal response
dates back to the ideas of Pippard \cite{Tinkham} and is related to the
fact that the current response must be averaged over the finite size of the
Cooper pair given by the coherence length $\xi_0$. In strongly type-II
materials the magnetic field varies on a length scale given by the
London penetration depth $\lambda_0$, which is much larger than $\xi_0$ and
therefore nonlocality is typically unimportant unless there exist strong
anisotropies in the electronic
band structure \cite{kogan}. Nonlinear corrections arise
from the change of quasiparticle population due to
superflow which, to leading order,  modifies  the excitation spectrum
by a quasiclassical shift \cite{Tinkham}
\be
{\cal E}_k=E_k +{\vec v_f}\cdot{\vec v_s},
\label{spec}
\ee
where $E_k=\sqrt{\epsilon_k^2+\Delta_k^2}$ is the BCS energy.
Again, in clean, fully gapped conventional superconductors this
effect is typically negligible except when the current approaches the pair
breaking value. In the mixed state this happens only in the close vicinity
of the vortex cores which occupy a small fraction of the total sample
volume at fields well below $H_{c2}$. The situation changes dramatically
when the order parameter has nodes, such as in $d_{x^2-y^2}$ superconductors.
Nonlocal corrections to (\ref{j0}) become important for the response of
electrons with momenta on the Fermi surface close to the gap nodes even
for strongly type-II materials. This can be understood by realizing that
the coherence length, being inversely proportional to the gap \cite{Tinkham},
becomes very large close to the node and formally diverges at the nodal
point. Thus, quite generally, there exists a locus of points on the Fermi
surface where $\xi \gg\lambda_0$ and the response becomes highly nonlocal.
This effect was first discussed by Kosztin and Leggett \cite{Leggett}
for the Meissner state and by us \cite{Franz2} in the mixed state. Similarly,
the nonlinear corrections become important in a d-wave superconductor.
Eq.\ (\ref{spec}) indicates that finite areas of gapless excitations appear
near the node for arbitrarily small $v_s$. 

One of the simplest theories to study the magnetic behavior of the 
type-II superconductors is the London model \cite{London}. 
In London theory the Free energy is written only in 
terms of magnetic field and is therefore very easy for calculation. Unlike in
GL theory, the order parameter does not appear explicitly in London free 
energy. The effect of the symmetry of the order parameter or
other Fermi surface anisotropies, is therefore not
contained in the London model. Simple generalization of the model 
however can incorporate these effects. The generalized model contains
anisotropic higher order and higher derivative terms reflecting the 
symmetries of the order parameter and the anisotropic Fermi surface.

In the next chapter, we first briefly introduce Ginsburg-Landau
(GL) theory, and London theory emphasizing the solutions
for a vortex lattice. Starting from a GL model, we then derive the leading
fourfold anisotropic corrections to the London equation making the usual 
assumption that the free energy was an analytic functional of the order 
parameter and field.
The number of new parameters in this model is far smaller than in 
the GL approach (a reasonable model contains only one new parameter 
which controls the strength of the symmetry breaking term) and numerical 
simulations are considerably easier. This
provides a useful tool  to study vortex lattice structure, pinning
by twin boundaries and the magnetic field distribution measured in
$\mu$SR experiments. The model is suitable to study the intermediate
field region $H_{c1}\ll H\ll H_{c2}$ which is experimentally most relevant but
traditionally difficult to handle within the GL theory.  Furthermore, this
approach can be extended to $T=0$ where GL theory breaks down
and the supercurrent becomes singular.
With increasing magnetic field this
model predicts a transition from triangular to centered rectangular and
eventually a square vortex lattice. While no direct experimental evidence
exists in cuprates at present to confirm such a prediction, a similar
transition has been recently observed in a boro-carbide material ErNi$_2$B$_2$C
\cite{Yaron} and has been described by a similar London model \cite{kogan}.

In the last chapter of this thesis, we again derive a generalized London
free energy, but this time completely from a weak coupling microscopic 
model, including the nonlinear and nonlocal effects mentioned above.
We show that the dominant effect that determines the
vortex lattice geometry and the effective penetration depth as defined
in $\mu$SR experiments is the nonlocal corrections, while the
nonlinear corrections play a secondary role at low $T$. 
At high temperatures we obtain a nonlocal correction similar to the 
one suggested in chapter 5.
At low temperatures however, we find a novel singular behavior
directly related to the nodal structure of the gap which
completely changes the form of the London equation. This singular behavior
has profound implications for the structure of the vortex lattice
which, as a function of decreasing temperature, undergoes a series
of sharp structural transitions and attains a universal limit at $T=0$.
Our theory is now completely {\em parameter free}. 
The London free energy at low $T$ is non-analytic and its long
wavelength part is fully determined by the nodal structure of the gap
function. Such behavior is caused by the low-lying quasiparticle excitations
within the nodes and thus could never occur in conventional superconductors
with anisotropic band structures. 

Finally, in the last section of chapter 6, we talk about the experimental
justification of our low temperature nonlinear nonlocal theory. 
We especially emphasize $\mu$SR experiments.
Recent $\mu$SR data at high fields have justified our prediction for the Field
dependence of the penetration depth in vortex state.

\chapter{Phenomenological Model}

\section{Ginsburg-Landau Theory}

In 1950 Ginsburg and Landau \cite{gl} proposed a 
phenomenological theory to describe
the behavior of superconductors near their transition temperature. 
At a mean field level, the phase transition of superconductors is
second order (gauge field fluctuations however are believed to change
the order of the transition to a weak first order \cite{halperin2}).
A free energy density that can describe such a transition was proposed to be
\be
f={1\over 2m^*}|(-i\hbar \nabla - {e^*\over c} \vec A)\Psi|^2 +
\alpha |\Psi|^2 +{\beta\over 2} |\Psi|^4 + {H^2 \over 8\pi}
\label{glf}
\ee
where $e^*$ ($=2e$) and $m^*$ ($=2m$) are the charge and effective mass of a
Cooper pair, $\vec H$ and $\vec A$ are magnetic field and vector 
potential respectively. Superfluid density $n_s$ is 
related to the order parameter by $|\Psi|^2 = n_s / n$, where $n$ is the total
density of electrons.  Ginsburg-Landau (GL) equations can be obtained by
minimizing $f$ with respect to $\Psi$ and $\vec A$
\ba
&&\alpha \Psi + \beta |\Psi|^2\Psi + {1\over 2m}|(-i\hbar \nabla -
{e^*\over c} \vec A)^2\Psi=0  \label{gl1} \\ 
&&\vec j={c \over 4\pi} \nabla\times\vec B
={e^* \hbar \over 2im^*}(\Psi^* \nabla\Psi - \Psi\nabla\Psi^*)
-{e^{*2} \over m^*c}\Psi^*\Psi\vec A
\label{gl2}
\ea
where $\vec j$ is the current density.
In the absence of magnetic field (i.e. $\vec H=\vec A=0$) and for uniform $\Psi$,
the minimum of $f$ occurs at $|\Psi|=0$ when $\alpha > 0$, but at
$|\Psi|=\sqrt{-\alpha/\beta}$ when $\alpha < 0$. A continuous
transition therefore happens as $\alpha$ goes from positive to negative values.
Taking $\alpha \propto (T-T_c)$, one can study the critical behavior of
the superconductor and calculate all the critical exponents near $T_c$.
GL theory has been studied extensively in the 
literature in both Meissner and vortex states of superconductors. 
It is not our attempt in this thesis to discuss the solutions of GL equations  
for different problems.
Instead, we shall focus on deriving a London theory from 
GL theory, especially in the vortex state. Interested readers should refer
to Ref. \cite{dG,Tinkham} or any other standard 
text books in the subject, for more detailed information.

\section{London Theory in Vortex State}

Although we will use GL equations to derive the London equation, London theory was 
proposed long before GL, by F. and H. London in 1935 \cite{London}. 
The basic assumption in their theory is a local proportionality relation between
supercurrent $\vec j$ and vector potential $\vec A$
\be
\vec j = -{n_s e^2 \over mc} \vec A.
\label{jA}
\ee
This equation is usually taken as the definition of superfluid density $n_s$.
Eq. (\ref{jA}) is not a gauge invariant equation. However, to satisfy current
conservation, one has to fix the gauge to the London gauge
\be
\nabla \cdot \vec A = 0.
\ee
This way (\ref{jA}) and all the results following from it are gauge independent
and therefore physical. Eq. (\ref{jA}) plays a very important role in our
theory for the vortex lattice. 
Generalization of this equation will be used in this
chapter and also in the next chapter to develop a generalized London theory
to describe the vortex lattice properties of d-wave superconductors. 

Taking the curl of the both sides of (\ref{jA}) and
using the Maxwell's equation $\nabla \times \vec B = (4\pi / c) \vec j$ we
get
\be
\lambda^2 \nabla\times\nabla\times \vec B + \vec B=0
\label{ld0}
\ee
which is the famous London equation. $\lambda$ is called the {\it London 
penetration depth} and is given by
\be
\lambda = \sqrt{mc^2 \over 4\pi n_s e^2}
\ee
A free energy density that gives rise to (\ref{ld0}) via minimization
with respect to the magnetic field $\vec B$, is called the {\it London
free energy density}
\be
f_L = {1\over 8\pi} \left(\vec B^2 + \lambda^2 |\nabla\times \vec B|^2 \right)
\label{FL0}
\ee
We now obtain London equations (\ref{ld0}) from GL equations (\ref{gl1}) and
(\ref{gl2}).

\subsection{Derivation of the London Equation from GL Theory}

Let us begin our derivation of the London equation by writing the order parameter
as  $\Psi = \psi e^{i\phi}$, with $\psi$ and $\phi$ being real 
functions of position $\vec x$. Substituting $\Psi$ into (\ref{gl2}) we get
\be
\lambda^2\nabla\times \vec B + \vec A = {\Phi_0 \over 2\pi} \nabla \phi
\label{l0}
\ee
where $\lambda=\sqrt{m^*c^2 / 4\pi e^{*2}\psi^2}$ is the London
penetration depth and $\Phi_0=(hc / e^*)$ is the flux quantum.
In type II superconductors $\psi$ varies over a length scale $\xi$ much smaller than the
penetration depth $\lambda$. $\xi$ is usually called the coherence length and is of
the order of the extent of a cooper pair. In the Meissner state therefore, 
$\psi$ (and consequently 
$\lambda$) stays constant in the bulk of the superconductor away from the boundaries. 
The London equation (\ref{ld0}) can therefore be derived easily by taking the curl 
of Eq. (\ref{l0}) and keeping in mind that  $\nabla\times\nabla \phi =0$. 

In the vortex state on the other hand, $\psi$ is not a constant and actually vanishes
at the center of each vortex. Taking this into account gives rise to the effect of the
vortex core which is the subject of the next subsection.
Furthermore, in the derivation of (\ref{ld0}), we used
$\nabla\times\nabla \phi =0$ which is not true in the vortex state. 
The reason is that
around a vortex, $\phi$ has a topological winding number equal to 1. In other words, 
$\phi$ increases by $2\pi$ in every turn around a vortex core.  Integrating 
$(\nabla\times\nabla \phi) \cdot d\vec S$ over a small region around 
the vortex center we get
\be
\int (\nabla\times\nabla \phi) \cdot d\vec S = \oint \nabla \phi \cdot d\vec l
= \phi_f - \phi_i = 2\pi n 
\ee
where $n$ is the topological winding number; which is one in our case. 
This equation is satisfied for every infinitesimal region around the center of 
the vortex $\vec r_0$, only if 
$\nabla\times\nabla \phi = 2 \pi \delta^{(2)}(\vec r-\vec r_0)$.
The London equation in the presence of a vortex will therefore become 
(assuming constant $\psi$ again)
\be
\lambda^2 \nabla\times\nabla\times \vec B + \vec B= \Phi_0
\delta^{(2)}(\vec r-\vec r_0)
\label{l1}
\ee
To find a solution for a single vortex, it is easy to take the  magnetic 
field in
$z$-direction and write $\vec B=B(r) \hat{z}$. Eq. (\ref{l1}) then becomes
\be
- {\lambda^2 \over r}{d \over dr}(r {dB \over dr}) + B = \Phi_0
\delta^{(2)}(\vec r-\vec r_0)
\label{B0}
\ee
which is actually a Bessel's differential equation. 
The solution to this equation is
\be
B(r)={\Phi_0 \over 2\pi \lambda^2} K_0({r\over \lambda})
\ee
where $K_0$ is the zero-order modified Bessel function. Important to notice here 
is that $K_0$ diverges logarithmically as $r\rightarrow 0$; which is apparently
non-physical. The reason behind this non-physical behavior lies in the fact
that we did not incorporate the effect of the vortex core properly.
In other words, the order parameter $\psi$ always vanishes at the center
of the vortex. Therefore, it is not a good approximation to consider it to be
homogeneous everywhere in the lattice. We shall take into account 
this effect in more detail in the next subsection.

\subsection{The Effect of the Vortex Core}

Let us start from eq. (\ref{gl2}) again by writing it as
\be
\vec j(\vec r)=e^* n_s(\vec r) \vec v_s(\vec r)
\label{jL0}
\ee
where $n_s=\psi^2$ is the superfluid density and $\vec v_s$ is the 
superfluid velocity defined by
\be
\vec v_s = {\hbar \over m^*}(\nabla \phi - {e^*\over \hbar c} \vec A)
\label{vs0}
\ee
At the center of a vortex, $n_s$ vanishes. We can therefore write
\be
n_s(\vec r) = n_0 \eta(\vec r) 
\ee
where $\eta(\vec r)$ is a function with the property
\be
\begin{array}{cc}\eta(\vec r)=0 &\qquad {\rm at}\quad r=0 \\
\eta(\vec r)\rightarrow 1 & \hspace{1.5cm} {\rm for} \quad r/\xi \gg 1 \end{array}
\ee 
and $\xi$ is the coherence length of the superconductor. The exact form of
$\eta$ should come from a microscopic theory or a complete solution of GL
equation. Here, we do not derive $\eta$. Instead, we find the source function
for a few assumed forms for $\eta$. Before that, let us take the curl
of both sides of (\ref{jL0}), keeping in mind that 
\be 
\nabla \times (\eta \vec v_s) = {2\pi \hbar \over m^*}\eta(\delta^{(2)}(\vec r) - 
{1 \over \Phi_0}\vec B) + \nabla \eta \times \vec v_s =
{2\pi \hbar \over m^*\Phi_0}\vec B + \nabla \eta \times \vec v_s 
\ee
and also from (\ref{jL0})  
\be
\vec v_s = - \left({1 \over e^* n_0 \eta}\right) \vec j = 
-\left({c \over 4 \pi e^* n_0 \eta} \right) 
\nabla \times \vec B
\ee
The generalized London equation therefore becomes
\be
\lambda^2 \nabla \times \nabla \times \vec B + \vec B = \vec \rho(\vec r)
\label{ld1}
\ee
with a source term $\vec \rho(\vec r)$ defined by
\be
\vec \rho(\vec r)=(1-\eta)\vec B + \lambda^2 {\nabla \eta \over \eta} \times \nabla \times \vec B
\label{rho}
\ee
Equations (\ref{ld1}) and (\ref{rho}) can be viewed as a set of self-consistent
equations to be solved for $\vec B$ and $\vec \rho$. However, $\vec \rho(\vec r)$ can be 
simplified significantly in extremely
type II superconductors which have large $\kappa = \lambda / \xi$.
First notice that $\vec \rho$ is non-zero only in a region of size $\xi$
around the center of the vortex. In this region, both $\vec B$ and $\eta$
scale with $\xi$. Therefore the first term in the left hand side of (\ref{ld1})
and the second term on the right hand side of (\ref{rho}) are both 
$O(\lambda^2/\xi^2)B$ and therefore are the dominant terms in (\ref{ld1}).
Ignoring the subdominant terms, (\ref{ld1}) can be written as
\be
\nabla \times \vec u = {\nabla \eta \over \eta} \times \vec u 
\qquad \Longrightarrow \qquad \nabla \times \left({\vec u \over \eta}\right) =0
\label{u0}
\ee
where $\vec u = \nabla \times \vec B$. If we assume rotational symmetry and
let $\vec B = B(r) \hat{z}$ and therefore $\vec u = u(r) \hat{\theta}$ where
$\theta$ is the azimuthal angle, then (\ref{u0}) leads to 
\be
{1 \over r}{d \over dr}\left( r{u \over \eta}\right)=0
\qquad \Longrightarrow \qquad  u= C {\eta \over r}
\ee
$C$ is a constant that can be found, using the fact that 
$\int \vec \rho \cdot d\vec S = \Phi_0$, to be
$C= (\Phi_0 / 2\pi \lambda^2)$. Substituting all into (\ref{rho}), 
the source function now can be written
as $\vec \rho(r)=\rho(r)\hat{z}$, with
\be
\rho(r) = {\Phi_0 \over 2\pi r} {d\eta \over dr} 
\label{rho1}
\ee
Now let's calculate $\rho$ for different choices of $\eta$.\\\\
{\bf Example 1: Clem's Ansatz} \\
Let's assume that
\be
\eta(r)={r^2 \over R^2}  \qquad {\rm with} \qquad R^2= r^2 + \xi_v^2
\ee
where $\xi_v$ is some parameter of the order of the coherence length $\xi$.
Substituting into (\ref{rho1}), we find the source term to be
\be
\rho (r)= {\Phi_0 \over \pi} {\xi_v^2 \over R^4}
\ee
In order to calculate the magnetic field, it is easier to find the source term
in terms of the magnetic field, using (\ref{rho}). Ignoring the first term we get
\be
\rho (r)= - {2 \xi_v^2 \over R^3} {d B(r) \over dR}.
\ee 
Substituting back into (\ref{ld1}), with a little algebra, we get
\be
- {\lambda^2 \over R}{d \over dR}\left( R {dB \over dR} \right) + B = 0
\label{BR}
\ee
Comparing (\ref{BR}) with (\ref{B0}), the solution would be
\be
B(r)= {\Phi_0 \over 2\pi\lambda^2} K_0\left( {R \over \lambda} \right)
\label{Bcm}
\ee
This result was first obtained by Clem \cite{Clem}, directly by solving 
GL equation. It was later shown by Brandt \cite{Brandt0} that the 
magnetic field obtained this way in
a vortex lattice is consistent with the exact solution of the GL equations.
In Clem's approach, $\xi_v$ is a variational parameter to
be determined by minimizing the free energy. It turns out that in 
extremely type II superconductors $\xi_v = \sqrt{2} \xi$.
As is clear from (\ref{Bcm}), $B(r)$ is not divergent as $r \rightarrow 0$.
In fact, $B(0)= (\Phi_0/2\pi\lambda^2)K_0( \xi/\lambda)$.

Another way to find the magnetic field is by Fourier transforming (\ref{ld1})
to get
\be
\lambda^2 k^2 B(k) + B(k) = F(k)
\qquad {\rm or} \qquad
B(r)=\int {d^2k\over (2\pi)^2}{F(k)e^{i\vec k\cdot\vec r}\over 1+\lambda^2 k^2}
\label{Br}
\ee
where
\be
F(k) = \int d^2r e^{-i\vec k\cdot\vec r} \rho(r)
\ee
is the Fourier transform of the source function which sets an ultraviolet
cutoff in momentum integrations. 
In the present case, it is not difficult to show that
\be
F(k)= \Phi_0 \xi_v k K_1(\xi_v k)
\label{Fcm}
\ee
where $K_1$ is the first order modified Bessel function. Substituting (\ref{Fcm})
into (\ref{Br}), we get back to (\ref{Bcm}).\\\\
{\bf Example 2: Gaussian Source Function}\\
Now let us assume that
\be
\eta(r) = 1 - e^{-r^2/2\xi^2}
\ee
The resulting source function will then be
\be
\rho(r) = {\Phi_0 \over 2\pi r}{d\eta \over dr} = {\Phi_0 \over 2\pi \xi^2}
e^{-r^2/2\xi^2}
\label{gauss}
\ee
The corresponding cutoff function will also have a Gaussian form
\be
F(k)= \Phi_0 e^{-\xi^2 k^2 /2}
\label{Fk}
\ee
This form of momentum cutoff was first proposed by Brandt \cite{Brandt1}.
This is actually the form of cutoff we are going to use in our 
calculations throughout the rest of this thesis.
We will talk about the sensitivity of our solutions to this particular
choice of cutoff function, when appropriate later.

\subsection{Solution for a Vortex Lattice}

In a vortex lattice, taking the magnetic field in $z$-direction
($\vec B= B \hat{z}$), equation (\ref{ld1}) becomes
\be
- \lambda^2 \nabla^2 B + B = \sum_i \rho(\vec r - \vec r_i)
\label{ld2}
\ee
where $\vec r_i$ are lattice vectors indicating the position of vortices.
Taking the Fourier transform of both sides of (\ref{ld2}), we get
\be
B_k={F(k) \over 1+\lambda^2 k^2} \qquad \Longrightarrow \qquad
B(\vec r)= \bar B \sum_{\vec k} {e^{i\vec k\cdot\vec r}F(k) \over 1+\lambda^2 k^2}
\ee
where $\vec k$ are now reciprocal lattice wave vectors, $\bar B$ is the
average magnetic field given by $\bar B= \Phi_0 /\Omega$,
with $\Omega$ being the area of a unit cell. Unlike GL, the London
approach does not automatically determine the form of the lattice for us. 
Instead, one has to calculate the free energy for different configurations
of the lattice and find the one that minimizes the free energy. Although 
this sounds a formidable job, symmetries of the free energy help us in
our initial guess. In other words, the lattice that satisfies the symmetries
of the free energy the most, is the most likely lattice for the vortices to exist on.

\begin{figure}[h]
\epsfysize 6cm
\epsfbox[0 320 600 550]{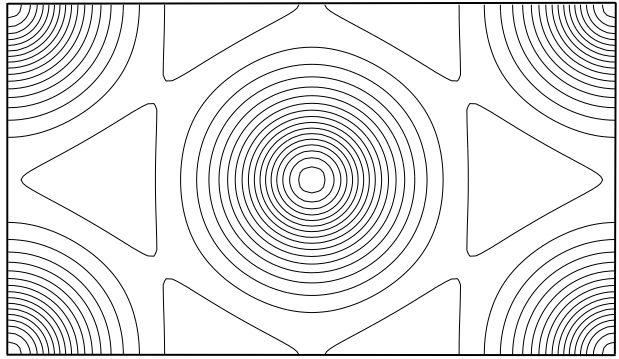}
\caption{Magnetic field distribution in a triangular lattice, using the
ordinary London equation with Gaussian cut off.}
\label{fgl0}
\end{figure}

Using (\ref{FL0}), the free energy per unit cell is given by 
\be
f_L = \int_{UC} d^2r [B^2 + (\nabla B)^2] = 
\sum_{\vec k}(1+\lambda^2 k^2)B_k^2
=\bar B^2\sum_{\vec k} {F(k)^2 \over 1+\lambda^2 k^2}
\ee
Substituting the Gaussian cutoff function (\ref{Fk}), we find
\ba
B(\vec r) &=& \bar B \sum_{\vec k} {e^{i\vec k\cdot\vec r} e^{-\xi^2 k^2/2}
\over 1+\lambda^2 k^2} \label{BL1} \\
f_L &=& \bar B^2\sum_{\vec k} {e^{-\xi^2 k^2} \over 1+\lambda^2 k^2}
\label{FL1}
\ea
These sums can be performed numerically quite easily.
The lattice configuration that minimizes the free energy (\ref{FL1}) 
is a triangular lattice. The contour
plot of the magnetic field in this lattice is shown in Fig \ref{fgl0}.

An important thing to notice is that the free energy (\ref{FL1}) is isotropic
and therefore does not determine the orientation of the lattice. In other
words, there is an infinite degeneracy in the orientation of the lattice. 
This degeneracy will be removed as soon as symmetry breaking terms are
added to the London equation. We will discuss this issue in more detail 
in the next two sections.

\section{Mixing of d-wave with s-wave}

So far we have established a London theory derived from GL theory, but we didn't 
mention anything about the symmetry of the order 
parameter. Since an s-wave order parameter is isotropic, whatever
we discussed in the previous sections is expected to hold for an s-wave  
superconductor. A d-wave order parameter on the other hand, is not isotropic.
Instead, it has fourfold anisotropy, and consequently, it is natural to expect 
this anisotropy to influence the structure of the vortex lattice. 
Surprisingly, the GL equations, as written in (\ref{gl1}) and (\ref{gl2}), 
have exactly
the same form for a d-wave order parameter as s-wave. All the
statements in the previous sections should therefore hold for a d-wave 
superconductor
as well. A way out of this dilemma, is to notice the fact that the 
GL free energy written in (\ref{glf}), is an approximation
and contains only the lowest 
order contributions. In other words, we have ignored all higher order
derivative terms. We have also overlooked the possibility of existence of
a second order parameter that in general can couple to the main (d-wave) 
order parameter.
It is conceivable that including these effects will introduce some fourfold 
anisotropic terms to the GL equations, 
reflecting the d-wave nature of the superconductor. In this section, we 
shall first consider the second possibility, namely including 
an s-wave order parameter coupled to the original 
d-wave order parameter. We shall see that the effect of this inclusion 
is to introduce higher order derivative terms in the
effective GL free energy written only in terms of the d-wave order parameter.
We shall then show how this additional high-derivative term will 
change the London free energy. This way we actually cover both possibilities
mentioned above. The material discussed in this section has been published in 
Ref. \cite{Affleck}.
   
A GL free energy density to describe mixing of d-wave and s-wave order 
parameters was first proposed by Joynt \cite{Joynt} as
\ba
f &=& \alpha_s|s|^2 + \alpha_d|d|^2 +
   \gamma_s|{\vec\Pi} s|^2 + \gamma_d|{\vec\Pi} d|^2  + f_4 + H^2/8\pi
\nonumber \\
   &&+\gamma_v\bigl[ (\Pi_y s)^*(\Pi_y d) - (\Pi_x s)^*(\Pi_x d) + {\rm c.c.}
\bigr].
\label{fgl}
\ea
where $\vec \Pi \equiv -i \nabla -e^*\vec A/\hbar c$ and $f_4$ contains the
quartic terms. $d$ and $s$ represent d-wave and s-wave order parameters 
respectively. In the bulk of the superconductor and in the absence
of magnetic field, we want only a d-wave order parameter to exist.
Therefore below $T_c$, we expect to have $\alpha_d<0$, but $\alpha_s>0$. 
In finite field $(H>H_{c1})$ on the other hand,
a small $s$-component  with a highly anisotropic spatial
distribution is nucleated in the vicinity of a vortex \cite{Berlinsky,Franz}.
This free energy has been extensively studied in the literature
\cite{Berlinsky,Franz,Ren,Heeb,Volovik}. It is well known that it gives rise to 
non-triangular equilibrium lattice structures \cite{Berlinsky,Franz}.
Our objective here, is not to discuss the solution to the GL equations resulting
from this free energy, but to focus on deriving the corrections
to the London equation that result from this free energy.
Our strategy will be to simplify the free energy (\ref{fgl}) by integrating
out the $s$-component in favor of higher order derivative terms in $d$.
In this process, some short length-scale information on the order parameter
is lost but  the magnetic field distribution is described accurately.

Using its Euler-Lagrange equation, $s$ can be expressed to the leading
order in $(1-T/T_c)$ as
\begin{equation}
s = (\gamma_v/\alpha_s)(\Pi_x^2-\Pi_y^2)d.
\end{equation}
Substituting this into $f$ gives the leading derivative terms in $d$
of the form:
\begin{equation}
f = \gamma_d[|\vec \Pi
d|^2-(\gamma_v^2/\gamma_d\alpha_s)|(\Pi_x^2-\Pi_y^2)d|^2] + \dots
\label{f4}
\end{equation}
Various additional corrections to the
free energy are obtained from integrating out $s$ more accurately,
taking into account the $\gamma_s|\vec \Pi s|^2$ term and  quartic
terms.  However these all involve higher powers of $\vec \Pi$ or other
terms that will not concern us. The coefficient of the second term has
dimensions of (length)$^2$. For future convenience, 
we will write it in the form $\epsilon\xi^2/3$,
where $\epsilon\equiv 3(\alpha_d\gamma_v^2/\alpha_s\gamma_d^2)$
is a dimensionless parameter which controls
the strength of the $s$-$d$ coupling and
$\xi\equiv\sqrt{\gamma_d/|\alpha_d|}$ is the GL coherence length.
We henceforth assume
$\epsilon \ll 1$.  As we will discuss in the next section, 
neutron scattering and STM
experiments probably support this assumption.
As mentioned before, a term of the form  $|(\Pi_x^2-\Pi_y^2)d|^2$ could
arise without invoking $s$-$d$ mixing. It can result from a systematic
derivation of higher order terms in the GL free energy starting with a
BCS-like model and taking into account the square symmetry of the Fermi
surface \cite{Hohenberg,Takanaka,Ichioka2,Feder}.

The free energy (\ref{f4}) is not bounded below,
exhibiting runaway behavior for rapidly varying $d$-fields.  This is in
fact cured by keeping additional higher derivative terms that also
arise from integrating out $s$.  
In fact, the approximation of Eq.
(\ref{f4}) will be sufficient for our purposes, yielding a local
minimum which we expect would become a global minimum upon including
the additional terms.

In the case of extremely type II superconductors - which we are considering 
here - the penetration depth $\lambda \gg\xi$.  We may therefore
assume that $|d(\vec r)|\approx d_0$, the zero field equilibrium
value, almost everywhere in the vortex lattice, except within a
distance of O($\xi$) of the cores.  This gives the London free energy,
\be
f_L = (1/8\pi)(\vec B)^2+\gamma_dd_0^2\{\vec v^2-(\epsilon
\xi^2/3)[(v_x^2-v_y^2)^2 
+(\partial_yv_y-\partial_xv_x)^2]\},\label{LFE}
\ee
written in terms of the superfluid velocity (we set $\hbar=m^*=1$),
\begin{equation}
\vec v \equiv \nabla \theta -(e^*/c)\vec A,
\end{equation}
where $\theta$ is the phase of $d$.

The corresponding London
equation, obtained by varying $f_L$ with respect to $\vec A$, is:
\be
{c\over4\pi}\nabla \times \vec B = \left({2e^*\over\hbar c}\right)
\gamma_dd_0^2\{\vec v -{\tiny{2\over 3}}\epsilon \xi^2[(\hat y v_y-\hat x
v_x)(v_y^2-v_x^2) 
-(\hat y \partial_y-\hat x
\partial_x)(\partial_yv_y-\partial_xv_x)]\}.\label{jL}
\ee
For many purposes it is very convenient  to
express $\vec v$ in terms of $\vec B$ and its derivatives, and then
substitute this expression for $\vec v$ back into $f_L$, giving an
explicit expression for $f_L$ as a functional of $\vec B$ only.  For
$\epsilon = 0$ this gives
\begin{equation}
\vec v^{(0)}=\nabla \times \vec B /B_0,
\label{v0}
\end{equation}
where $B_0\equiv \Phi_0/2\pi \lambda_0^2$  is a characteristic field 
of order $H_{c1}$,  and
\begin{equation}
f_L^0 = (1/8\pi )[\vec B^2+\lambda_0^2(\nabla \times \vec
B)^2].
\end{equation}
which is actually the ordinary London free energy (i.e. Eq. (\ref{FL0})).
Here the penetration depth, for $\epsilon = 0$ is
$\lambda_0^{-2}=8\pi \gamma_d(e^*d_0/\hbar c)^2$.  It is presumably not
possible to solve Eq. (\ref{jL}) in closed form for $\vec v$
as a function of $\vec B$ for $\epsilon\ne 0$.
However, this can be done readily in a
perturbative expansion in $\epsilon$.  The first order correction is:
\be
\vec v^{(1)}=
(2\epsilon \xi^2/3)\{(\hat y v_y^{(0)}-\hat x v_x^{(0)})[(v_y^{(0)})^2-
(v_x^{(0)})^2]
-(\hat y \partial_y-\hat x
\partial_x)(\partial_yv_y^{(0)}-\partial_xv_x^{(0)})\}
\ee
with $\vec v^{(0)}$ given by Eq. (\ref{v0}).  The
London free energy density, up to O($\epsilon $) is then:
\begin{equation}
f_L = f_L^0+
{\epsilon \lambda_0^2\xi^2\over 8\pi}[4(\partial_x\partial_yB)^2+
((\partial_xB)^2-(\partial_yB)^2)^2/B_0^2].
\label{fL}
\end{equation}
Note that we could have arrived at a similar conclusion by simply
writing down all terms allowed by symmetry in $f_L$, expanding in
number of derivatives and powers of $B$.  Square anisotropy is first
possible in the fourth derivative terms.  In principle, we should
 also include all isotropic terms to order $B^4$  and $\nabla
^4$.  However, assuming that these have small coefficients, they will
not be important.  As we shall see in the next chapter, similar
results can also be obtained from nonlocal effects due to the 
divergence of the coherence length along the node directions \cite{Franz2} 
and also from considering generation of quasi-particles near gap nodes 
\cite{Yip,Amin1}, in a range of temperature and field where
the supercurrent can be Taylor expanded in the superfluid
velocity.  More generally, the quadratic and quartic
terms in (\ref{fL})  have independent coefficients.

The corresponding London equation is obtained by varying $f_L$ with
respect to $\vec B(\vec r)$. For $B$ along the $z$ direction, one obtains
\begin{equation}
[1-\lambda_0^2\nabla^2+4\epsilon
\lambda_0^2\xi^2(\partial_x\partial_y)^2]B -\epsilon Q[B]=0,
\label{london}
\end{equation}
where
\be
Q[B]=2\lambda_0^2\xi^2 B_0^{-2}
[(\partial_x^2-\partial_y^2) B+\partial_xB\partial_x-\partial_y
B\partial_y] 
[(\partial_x B)^2-(\partial_y B)^2]
\ee
is the non-linear term arising from the last term in Eq. (\ref{fL}).

Our numerical calculation (next section) shows that, 
the effect of the non-linear term is negligible 
(Contrary to naive expectation, it doesn't
become more important with increasing applied field because the field
becomes nearly constant in the vortex lattice when the applied field is
large.) Thus to an excellent approximation one may neglect $Q[B]$ in the
London equation (\ref{london}).
To get a feeling for the effect of the extra nonlocal (four derivative)
term, consider a 
semi-infinite superconductor in its Meissner state with an interface
parallel to the $yz$ plane. The magnetic field in this case 
depends only on $x$; all the y-derivatives and
thereby the extra term vanish. The penetration depth will then remain
unchanged and equal to $\lambda_0$. On the other hand, if the interface is
$45^o$ rotated with respect to $x$ and $y$ axis ($a$ and $b$ directions), 
the magnetic field
will be a function of $(x+y)$. Substituting $B=B_0e^{-(x+y)/\sqrt{2} \lambda}$
into (\ref{london}), we get (neglecting the nonlinear part)
\begin{equation}
\lambda = \lambda_0\left[{1\over 2} + \sqrt{{1\over
4}-{\epsilon \xi^2\over \lambda_0^2}}\right]^{1/2},
\label{lmd0}
\end{equation}
Here we keep only the larger solution because it is the one that determines 
the decay rate of the field and therefore 
can be identified as the penetration depth. The penetration depth is
now longer along the crystal axes as is clear from (\ref{lmd0}).
 
To determine vortex lattice structure, we
insert source terms  $\sum_j\rho(\vec r-\vec r_j)$ at the vortex core
positions, $\vec r_j$, on the right hand side of Eq. (\ref{london}).
The source term we use is the Gaussian source term introduced
in eq. (\ref{gauss}). Taking Fourier transform, 
the magnetic field may be written explicitly as:
\be
B(\vec r) = \bar B\sum_{\vec k}{e^{i\vec k \cdot \vec r}e^{-k^2\xi^2/2}\over
1+\lambda_0^2k^2+4\epsilon \lambda_0^2\xi^2(k_xk_y)^2}.
\label{Bk0}
\ee
Here the sum is over all wave-vectors in the reciprocal lattice and $\bar
B$ is the average field.  The lattice constant is determined by the
condition that $\bar B\Omega=\Phi_0$, where $\Omega$ is the area of the
unit cell.  The lattice symmetry is then determined by minimizing the Gibbs
free energy ${\cal G}_L={\cal F}_L-H\bar B/4\pi$, where
\be
{\cal F}_L=\bar B^2 \sum_{\vec k}{e^{-k^2\xi^2}\over
1+\lambda_0^2k^2+4\epsilon \lambda_0^2\xi^2(k_xk_y)^2}
\label{Fk0}
\ee 

\section{Numerical Results and Experimental Realizations}

Ignoring the nonlinear term, one can perform the sums in (\ref{Bk0}) and
(\ref{Fk0}) numerically to find the magnetic field and lattice structure.
The correction due to the nonlinear term can then be found iteratively.
To do so, we start by writing (\ref{london}) in Fourier transformed 
form (with Gaussian source) as
\be
B^n_k = {e^{-k^2\xi^2/2} + \epsilon \widetilde{Q}[B^{n-1}_k] \over
1+\lambda_0^2k^2+4\epsilon \lambda_0^2\xi^2(k_xk_y)^2}
\label{Bkn}
\ee
where $\widetilde{Q}[B_k]$ is the Fourier transform of $Q[B]$ and the 
superscript $n$ represents the stage of iteration. The initial $B_k^0$
is found by neglecting the nonlinear term. At each
following step we calculate $\widetilde{Q}[B_k]$ using the known value
for $B_k$, substitute it into (\ref{Bkn}) to find the new $B_k$. We continue
this iteration until we get the desired accuracy. Having $B$, we then
can calculate the Gibbs free energy ${\cal G}_L={\cal F}_L-H\bar B/4\pi$
using (\ref{fL}). We calculate the Free energy this way for different
configurations of the vortex lattice. We find that 
a centered rectangular flux lattice, with
principal axes aligned with the ionic crystal lattice minimizes the
free energy. An angle $\beta$ between unit vectors (Fig. \ref{fgl1})
characterizes the dependence
of the lattice on $\epsilon$ and the
magnetic field. An example of such a centered rectangular lattice is shown
in Fig. \ref{fgl1}.
\begin{figure}[h]
\epsfysize 6cm
\epsfbox[-50 270 500 550]{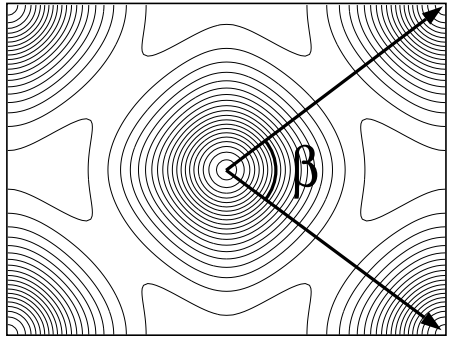}
\caption[Distribution of magnetic field in a vortex lattice for 
$\epsilon=0.3$ and $H=6.8$T]{
Distribution of magnetic field in a vortex lattice for $\epsilon=0.3$
and $H=6.8$T, leading to an angle of $\beta\simeq74^\circ$. We use
$\lambda_0=1400$\AA\  and $\kappa\equiv\lambda_0/\xi=68$.} 
\label{fgl1}
\end{figure}

In agreement with earlier results within GL \cite{Franz}
and Eilenberger \cite{Ichioka} formalisms, individual vortices are elongated
along the crystalline axes.  Figure \ref{fgl2}(a) shows the dependence of
Gibbs
free energy on $\beta$ for various values of $\epsilon$ at fixed applied
field $H=400B_0\simeq 6.8$T. For $\epsilon =0$ minimum occurs for
$\beta_{\rm MIN} = 60^\circ$,  corresponding to a hexagonal lattice. As
$\epsilon$ increases, $\beta_{\rm MIN}$ continuously increases and
for sufficiently large $\epsilon$, the flux lattice
becomes tetragonal with $\beta_{\rm MIN} = 90^\circ$.
For $\beta_{\rm MIN} \neq 90^0$,
there are always two solutions, related by a $90^\circ$ rotation,
in which the
long axis of the centered rectangle is aligned with either the $x$ or $y$
axis.  The degeneracy is much larger for $\epsilon = 0$, when the flux
lattice may have an arbitrary orientation relative to the ionic
crystal lattice.

\begin{figure}[h]
\epsfysize 8cm
\epsfbox[20 300 500 600]{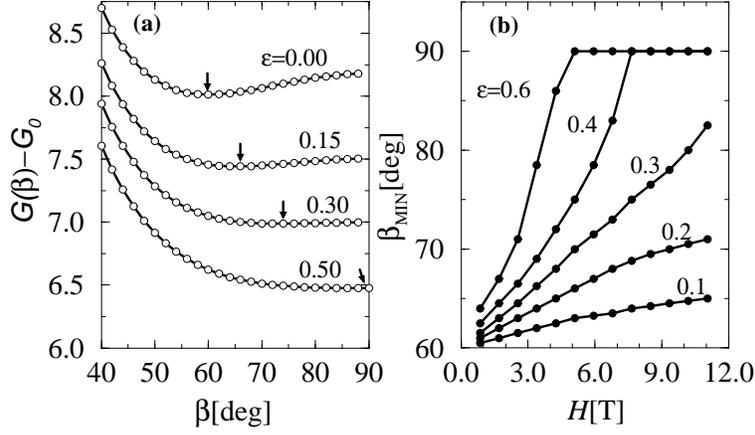}
\caption[a) Gibbs free energy as a function of $\beta$. 
b) Equilibrium angle $\beta_{\rm MIN}$ as a function of $H$.]{
a) Gibbs free energy as a function of $\beta$ for the same parameters
as Fig.  \ref{fgl1} and  various values of $\epsilon$. Arrows indicate
positions $\beta_{\rm MIN}$ of the minima and $G_0\equiv -H^2/8\pi$.
b) Equilibrium angle $\beta_{\rm MIN}$ as a function of $H$ for several
values of $\epsilon$.}
\label{fgl2}
\end{figure}

The dependence of
$\beta_{\rm MIN}$ on the applied field for various values of  $\epsilon$ is
displayed in Fig.  \ref{fgl2}(b). Clearly the anisotropic term becomes more
important at larger fields. Our perturbative elimination of $\vec v$ in
favor of $B$ breaks down when $\epsilon$ and $H$ are sufficiently large that
$\beta_{\rm MIN}$ differs significantly from $60^\circ$.  Furthermore, we
might expect higher order corrections to (\ref{LFE}) to be important in
this regime.   By fitting   Fig. \ref{fgl2}(b)
to experimental data on tetragonal materials such as
Tl$_2$Ba$_2$CuO$_{6+d}$ (once such data
become available) one can directly assess the magnitude of $\epsilon$, the
only unknown parameter in the model.

The analysis presented here,
can be easily extended to take into account effective
mass (i.e. penetration depth) anisotropy.  In a simple one-component
GL model, the derivative term is generalized to:
\begin{equation} f = \sum_{i=x,y,z}\gamma_i|\Pi_i d|^2.\end{equation}
We restrict our attention to fields along the $z$-axis.  The
anisotropy then can be removed by a rescaling of the $x$-coordinate and a
corresponding rescaling of the magnetic field.  The coherence length
and penetration depth anisotropies are considered the same:
$\xi_y/\xi_x=\lambda_x/\lambda_y$.  Making a simplifying
assumption that the higher derivative and mixed derivative terms in
${\cal F}$ are also simply modified by a rescaling by a common factor, it
then follows that the flux lattice shape is obtained by stretching
along the $x$-axis by the factor $\lambda_x/\lambda_y$.   We now obtain two
possible vortex lattices, both of centered rectangular symmetry, aligned with
the ionic lattice, with different angles, $\beta$. (Relaxing our simplifying
assumption may split the degeneracy between these two lattices.)
On the other hand, when $\epsilon = 0$, we may rotate the hexagonal
lattice by an arbitrary angle before stretching. This gives
an infinite set of oblique lattices with arbitrary orientation.

To compare theory with YBCO we should take into account twin
boundaries; which are the boundaries between different orientations
of crystal lattice. They may also tend to align the vortex lattice by pinning
vortices to the twin boundaries, at $\pm 45^0$ to the $x$-axis.  This
effect competes with alignment to the ionic lattice which we have been
discussing.  Only in the special case of a square vortex lattice does a
line of vortices occur at $\pm 45^0$.  If this is not the case, and if
pinning by twin boundaries is significant, then we should expect that
the vortex lattice will align with the ionic lattice far from twin
boundaries but will be deformed in the vicinity of a twin boundary in
an effort to align itself with the twin boundary. On the other hand,
for $\epsilon =0$, the vortex lattice would remain aligned with the twin
boundaries everywhere except within vortex lattice domain boundaries
which necessarily exist roughly midway between the twin boundaries.

Neutron scattering experiments on YBCO \cite{Keimer}
suggest that the vortex lattice is well-aligned with the twin boundaries and
is close to being centered rectangular (the ratio of lattice constants is
about 1.04) with $\beta \approx 73^0$, with weak dependence on $H$. This
corresponds to a rotation away from alignment with the ionic lattice by
$9^\circ$.  Four different orientational domains, related by reflection in the
$(1,1,0)$  axis and $90^0$ rotation were reported.
These results can be  rather well fitted \cite{Walker} by
the basic London model ($\epsilon =0$) with mass anisotropy.  
For $\lambda_x/\lambda_y=1.5$, a value roughly consistent with
infrared experiments, this lattice has about the right
shape. Taking into account the two crystallographic domains (related by
interchanging $\lambda_x$ and $\lambda_y$) there are altogether four
vortex lattice domains, as seen experimentally.  The experimental fact
that the vortex lattice appears to be well aligned with the twin
boundaries suggests that the tendency to align with the ionic lattice
 is small.   No evidence for a  bending of the vortex lattice (by
$9^0$) into alignment with the ionic lattice far from the twin boundaries
has so far been found.

STM imaging of the YBCO vortex lattice \cite{Maggio} also suggests that the
(highly disordered) lattice has approximately centered rectangular symmetry
with $\beta \approx 77^\circ$. However, no evidence for the $9^\circ$ tilt
into alignment with the twin boundaries was reported. Considering the 
observed anisotropy of the vortex cores it has been concluded that the mass
anisotropy alone cannot account for the measured $77^\circ$ angle of the
vortex lattice. It has been suggested that a mechanism related to the internal
symmetry of the order parameter (such as the one discussed in the present
thesis) needs to be invoked in order to reconcile these observations.

Low field Bitter decoration data on YBCO \cite{Dolan} show vortex
lattice geometry with a very small distortion from hexagonal, consistent with
a much smaller anisotropy $\lambda_x/\lambda_y=1.11-1.15$. One may be tempted
to attribute this apparent field dependence of $\beta$ to the effects
discussed above in connection with Fig.  \ref{fgl2}(b).
An alternative explanation is a poor
quality of samples used in the Bitter decoration experiments that may have
resulted in partial washing out of the $a$-$b$ plane anisotropy \cite{Timusk}.

\begin{figure}[h]
\epsfysize 8cm
\epsfbox[20 300 500 600]{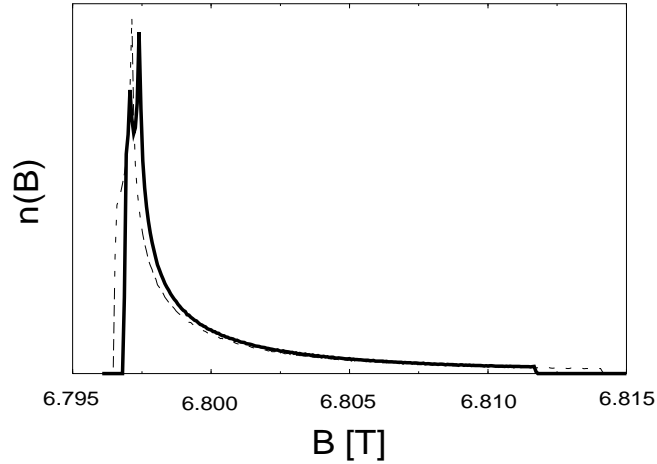}
\caption[Magnetic field distribution function.]
{Magnetic field distribution function; {\bf Solid line:} using the
same parameters as in Fig. \ref{fgl1}. {\bf Dashed line:} the same parameters
except for $\epsilon = 0$.}
\label{fgl3}
\end{figure}

$\mu$SR experiments measure the field distribution
$n(B)=(1/\Omega )\int \delta[B-B(\vec r)] dx dy$.  Our calculation
of this quantity is shown in
Fig.\ \ref{fgl3}.  For comparison we show our results with (solid line)
and without (dashed line) the extra nonlocal term. 
For $\beta \neq 60^\circ$, $B(\vec r)$ has
two inequivalent saddle points leading to two peaks in $n(B)$.
$n(B)$ is unaffected by effective mass anisotropy, as can be shown by the
rescaling transformation, mentioned above.  Existing $\mu SR$ experiments show
only a single peak \cite{Sonier} but this might be due to the large broadening
as a result of finite muon lifetime.

The weak field dependence of $\beta$, the alignment with twin boundaries in 
the neutron scattering experiments, and
the single peak in $n(B)$ suggest that $\epsilon$ is small in YBCO and
that the normal London model, together with twin boundary pinning, provides
a good fit to the data.  
STM and Bitter decoration data on the other hand, seem to favor finite 
$\epsilon$ and weak pining to twin boundaries.
Further experimental work, preferably on untwinned
YBCO or other tetragonal superconductors, will probably be
necessary to clarify the importance of square lattice anisotropy in
high-$T_c$ superconductors. 

Most of the experiments mentioned above are done at temperatures far below
$T_c$, i.e. beyond the validity range of GL theory. Thus our derivation of the
generalized London model from GL theory is not valid for most experimental 
situations. However, as we emphasized,
the corrections we introduce to the London free energy are the most general
analytic corrections possible by symmetry to the order of our concern. 
It is therefore reasonable that even at lower temperatures all the lowest 
order analytic corrections come with these forms. 
We will see in the next chapter that at very low 
temperatures, the lowest order corrections are actually non-analytic. This is 
because of the existence of the gap nodes; a unique feature of d-wave 
superconductors. Even those effects turn into the analytic forms discussed
here at higher temperatures.

\chapter{Microscopic Model}

\section{Gorkov Theory and Quasi-Classical Approximation}

Ginsburg Landau theory, discussed in the last chapter, is a theory that is
valid only near the transition temperature $T_c$. 
Most of the experiments however,
are done at temperatures far below $T_c$. Also, one of the
most essential characteristics of d-wave superconductors, i.e. the existence of
nodes in the superconducting gap, does not show a direct effect on the GL
equations. Thus a more sophisticated microscopic approach is necessary to
describe the vortex lattice properties of high $T_c$ 
superconductors. The microscopic theory we present here is based on
the theory developed by Gorkov \cite{Gorkov} in 1959. In his paper, Gorkov showed
how GL theory results from pairing theory at temperatures close to $T_c$.
Here, we don't give the detailed derivation of Gorkov's 
equations from the first principles. Our goal instead, is to derive relations
between supercurrent $\vec j$ and superfluid velocity $\vec v_s$ (or 
vector potential $\vec A$) starting from these equations. This relation is then
used to obtain a generalized London equation. For a detailed
derivation of the Gorkov equations, interested reader should refer to 
ref. \cite{dG,alex}. 

Let us start from an interaction Hamiltonian of the form
\be
H_{int}= \int d^3r d^3r^\prime \psi^\dagger_\uparrow({\vec r^\prime})
\psi_\uparrow({\vec r^\prime})\psi^\dagger_\downarrow({\vec r})
\psi_\downarrow({\vec r}) V({\vec r}-{\vec r^\prime})
\ee
The Gorkov equations for this Hamiltonian are 
\ba
\{ -{\partial \over \partial \tau} + {1 \over 2}[\nabla-{ie \over c}
{\vec A(\vec r)}]^2+\mu\} {\cal G}({\vec r}\tau,{\vec r^\prime}\tau^\prime) &&
\nn \\
+ \int \Delta({\vec r},{\vec r}^{\prime\prime}){\cal F}^\dagger
({\vec r}^{\prime\prime}\tau,{\vec r^\prime}\tau^\prime) d^3  r^{\prime\prime}
&=&\delta^{(3)}({\vec r}-{\vec r^\prime})\delta (\tau -\tau^\prime)
\label{g1} \\
\{ {\partial \over \partial \tau} + {1 \over 2}[\nabla + {ie \over c}
{\vec A(\vec r)}]^2+\mu\}{\cal F}^\dagger({\vec r}\tau,{\vec r^\prime}
\tau^\prime) && \nn \\
-\int \Delta^\ast({\vec r},{\vec r}^{\prime\prime}){\cal G}
({\vec r}^{\prime\prime}\tau,{\vec r^\prime}\tau^\prime) d^3  r^{\prime\prime}
&=&0
\label{g2}
\ea
where ${\cal G}$ and ${\cal F}$ are regular and anomalous parts of
the Nambu-Gorkov Green function respectively,
\begin{eqnarray}
{\cal G}({\vec r}\tau,{\vec r}^\prime\tau^\prime)\delta_{\sigma\sigma^\prime}&=&
-<{\rm T}_\tau \psi_{\sigma}({\vec r}\tau) \psi^\dagger_{\sigma^\prime}
({\vec r}^\prime\tau^\prime)> \label{g} \\
{\cal F}^\dagger({\vec r}\tau,{\vec r}^\prime\tau^\prime)&=&
-<{\rm T}_\tau \psi_\uparrow^\dagger({\vec r}\tau) \psi_\downarrow^\dagger
({\vec r}^\prime\tau^\prime)> 
\label{ag}
\end{eqnarray}
Here, and also throughout the rest of this chapter, we choose $\hbar=m=1$.
The order parameter $\Delta$ is related to the interaction potential $V$ by
\be
\Delta({\vec r}-{\vec r^\prime})=V({\vec r}-{\vec r^\prime})
<\psi_\uparrow({\vec r})\psi_\downarrow({\vec r}^\prime)>.
\label{delta1}
\ee
Equations (\ref{g1}) and (\ref{g2}) have to be solved for ${\cal G}$ and
${\cal F}$ self-consistently with $\Delta$.
The self-consistency equation follows from (\ref{ag}) and (\ref{delta1})
to be
\be
\Delta^*({\vec r}-{\vec r^\prime})=V({\vec r}-{\vec r^\prime})
{\cal F}^\dagger({\vec r} 0^+,{\vec r^\prime} 0)
\ee
Solving these equations self-consistently, in a general case at the presence of
Magnetic field is tedious if not impossible. However, substantial simplification
can be achieved using local approximation or gradient expansion. The method
we use here (with some minor differences), is a generalization of the 
gradient expansion method discussed in Ref. \cite{alex}. The generalization
is necessary to study d-wave superconductivity.

\subsection{Local Approximation}

Gorkov equations (\ref{g1}) and (\ref{g2}) take much simpler forms in Fourier space.
If the system is translationally invariant in time, ${\cal G}$ and ${\cal F}$ are only
functions of ($\tau-\tau^\prime$). We can therefore take their Fourier 
transform with respect to ($\tau-\tau^\prime$) and define
\begin{eqnarray}
&&{\cal G}_\omega({\vec r},{\vec r}^\prime)= \int {\cal G}({\vec r}\tau,
{\vec r}^\prime 0) e^{-i\omega_n\tau}d \tau \nonumber \\
&&{\cal F}^\dagger_\omega({\vec r},{\vec r}^\prime)= \int {\cal F}^\dagger({\vec r}\tau,
{\vec r}^\prime 0) e^{-i\omega_n\tau}d \tau
\end{eqnarray}
where $\omega_n=\pi T(2n-1)$ are fermionic Matsubara frequencies.
We also change the coordinates to ${\vec R=r^\prime}$ and 
${\vec \rho=r-r^\prime}$ and
take the Fourier transform with respect to ${\vec \rho}$ to get
\begin{eqnarray}
&&{\cal G}_\omega({\vec R},{\vec k})= \int {\cal G}_\omega({\vec R+\vec \rho,\vec R})
e^{-i{\vec k\cdot \vec \rho}} d^3\rho \\
&&{\cal F}^\dagger_\omega({\vec R},{\vec k})= \int {\cal F}^\dagger_\omega(\vec R+\vec \rho,
\vec R) e^{-i\vec k\cdot \vec \rho} d^3\rho \\
&&\Delta(\vec R,\vec k)= \int \Delta(\vec R+{\vec \rho \over 2},
\vec R-{\vec \rho \over 2}) e^{-i{\vec k \cdot \vec \rho}} d^3\rho
\end{eqnarray}
(\ref{g1}) therefore becomes
\be
\{i\omega_n + {1 \over 2}[i{\vec k} - {ie \over c}{\vec A(R}+i{\partial \over 
\partial {\vec k}})]^2 +\mu\}{\cal G}_\omega({\vec R},{\vec k})
+\int d^3 \rho \int d^3 r^{\prime\prime} e^{-i{\vec k\cdot \rho}}
\Delta({\vec R+\rho,r^{\prime\prime}}){\cal F}^\dagger_\omega
({\vec r^{\prime\prime},R })=1
\label{gf1}
\ee
The integral on the left hand side of (\ref{gf1}) can be written as
\begin{eqnarray}
&&\int d^3 \rho_1 \int d^3 \rho_2 e^{-i{\vec k\cdot (\vec \rho_1+\vec \rho_2)}}
\Delta({\vec R+\vec \rho_1 + \vec \rho_2,\vec R+\vec \rho_2}){\cal F}^\dagger_\omega
({\vec R+\vec \rho_2,\vec R}) \nonumber \\
&&=\lim_{\vec k^\prime \rightarrow \vec k} \int d^3 \rho_1 \int d^3 \rho_2 
e^{-i{\vec k^\prime\cdot \vec \rho_1}}e^{-i{\vec k\cdot \vec \rho_2}}
\Delta({(\vec R+\vec \rho_2+{\vec \rho_1 \over 2})+{\vec \rho_1 \over 2},
(\vec R+\vec \rho_2+{\vec \rho_1 \over 2})-{\vec \rho_1 \over 2}}){\cal F}^\dagger_\omega
({\vec R+\vec \rho_2,\vec R}) \nonumber \\
&&=\lim_{\vec k^\prime \rightarrow \vec k}\Delta ({\vec R}+
i{\partial \over\partial {\vec k}}+{i \over 2}{\partial \over\partial 
{\vec k}^\prime},{\vec k}^\prime){\cal F}^\dagger_\omega({\vec R, \vec k})
\end{eqnarray}
If ${\vec A(\vec R)}$ and $\Delta({\vec R,\vec k})$ are slowly varying 
functions of $\vec R$, then we can neglect all
${\partial \over\partial {\vec k}}$ and ${\partial \over\partial
{\vec k}^\prime}$ which are at most of the order of $\xi$.
We can also follow the same procedure for (\ref{g2}).
The resulting Gorkov equations in this limit will be
\ba
\{i\omega_n - {1 \over 2}[{\vec k} - {e \over c}{\vec A(\vec R)}]^2
 +\mu\}{\cal G}_\omega({\vec R},{\vec k})
+\Delta(\vec R,\vec k){\cal F}^\dagger_\omega(\vec R, \vec k)&=&1
\label{gk1} \\
\{-i\omega_n - {1 \over 2}[{\vec k} + {e \over c}{\vec A(\vec R)}]^2 
+\mu\}{\cal F}^\dagger_\omega({\vec R},{\vec k})
-\Delta^\ast({\vec R,\vec k}){\cal G}_\omega({\vec R,\vec k})&=&0
\label{gk2}
\ea
These are now algebraic equations instead of integral equations.
This approximation is usually referred to as {\it quasi-classical approximation}.
The reason is that the equations involve position $\vec R$ and momentum 
$\vec k$ at the same time, similar to classical equations. 
The self-consistency equation also takes the form
\be
\Delta^\ast({\vec R,\vec k})={1 \over \beta}\sum_n
\int d^3 k^\prime \widetilde{V}({\vec k-\vec k^\prime})
{\cal F}^\dagger_\omega({\vec R,\vec k^\prime})
\label{self}
\ee
where $\widetilde{V}({\vec k})$ is the Fourier transform of $V({\vec r})$.
The solutions to (\ref{gk1}) and (\ref{gk2}) are 
\ba
&&{\cal G}_\omega({\vec R,\vec k})={-i\omega_n - {1 \over 2}[{\vec k} + 
{e \over c}{\vec A(\vec R)}]^2 +\mu \over
\{i\omega_n - {1 \over 2}[{\vec k} - {e \over c}{\vec A(\vec R)}]^2
 +\mu\}\{-i\omega_n - {1 \over 2}[{\vec k} + {e \over c}{\vec A(\vec R)}]^2
 +\mu\}+\Delta^2({\vec k})} \nonumber \\
&&{\cal F}^\dagger_\omega({\vec R,\vec k})={\Delta^\ast({\vec k}) \over
\{i\omega_n - {1 \over 2}[{\vec k} - {e \over c}{\vec A(\vec R)}]^2
 +\mu\}\{-i\omega_n - {1 \over 2}[{\vec k} + {e \over c}{\vec A(\vec R)}]^2
 +\mu\}+\Delta^2({\vec k})} \nn
\ea
In the absence of magnetic field these solutions will be simplified to
\ba
&&{\cal G}_\omega({\vec k})={-i\omega_n-\epsilon_{\vec k} \over 
\omega^2+E_{\vec k}^2} \nonumber \\
&&{\cal F}^\dagger_\omega({\vec k})={\Delta^\ast({\vec k}) \over
\omega^2+E_{\vec k}^2}
\label{g&f}
\ea
with
\be
\epsilon_{\vec k}={1 \over 2}k^2-\mu 
\hspace{2cm} and \hspace{2cm}
E_{\vec k}=\sqrt{\epsilon_{\vec k}^2+\Delta^2({\vec k})}
\ee
The self-consistency equation (\ref{self}) also will be
\be
\Delta^\ast({\vec R,\vec k})=
\int d^3 k^\prime \widetilde{V}({\vec k-\vec k^\prime})
\Delta^\ast({\vec k^\prime}){1 \over \beta} \sum_n {1 \over
\omega^2+E_{\vec k^\prime}^2}
\label{self1}
\ee
Summing over Matsubara frequencies, the self-consistency equation becomes
\be
\Delta({\vec k})=-\int d^3 k^\prime \widetilde{V}({\vec k-k^\prime})
{\Delta({\vec k^\prime}) \over 2 E_{\vec k^\prime}}\tanh ({E_{\vec k^\prime} \over 2T})
\label{gap0}
\ee
which is the well known BCS gap equation.

In the presence of magnetic field, simplifications can be achieved if we  
introduce new 
Green's functions $\tilde{\cal G}$ 
and $\tilde{\cal F}$ by
\ba
\tilde{\cal G}(\vec r \tau,\vec r^\prime \tau^\prime)&=&
{\cal G}(\vec r \tau,\vec r^\prime \tau^\prime)
e^{-i\vec (\chi(\vec r)-\chi(\vec r^\prime))}
\nonumber \\
\tilde{\cal F}^\dagger({\vec r}\tau,{\vec r^\prime}\tau^\prime)&=&
{\cal F}^\dagger({\vec r}\tau,{\vec r^\prime}\tau^\prime)
e^{i  (\chi(\vec r)+\chi(\vec r^\prime))}
\ea
and also a new pairing potential
\be
\tilde{\Delta}({\vec r},{\vec r^\prime})=\Delta({\vec r},{\vec r^\prime})
e^{-i (\chi(\vec r)+\chi(\vec r^\prime))}
\ee
where $\chi({\vec r})$ is the phase of the operator $\psi({\vec r})$
at some fixed gauge.  
The Gorkov equations (\ref{gk1}) and (\ref{gk2}) for $\tilde{\cal G}$
and $\tilde{\cal F}$ become
\ba
\{i\omega_n - {1 \over 2}[{\vec k} + {\vec v}_s({\vec R})]^2
 +\mu\}\tilde{\cal G}_\omega({\vec R},{\vec k})
+\tilde{\Delta}(\vec R,\vec k)\tilde{\cal F}^\dagger_\omega(\vec R,\vec k)&=&1 \label{gi1} \\
\{-i\omega_n - {1 \over 2}[\vec k - \vec v_s(\vec R)]^2
 +\mu\}\tilde{\cal F}^\dagger_\omega({\vec R},{\vec k})
-\tilde{\Delta}^\ast({\vec R,\vec k})\tilde{\cal G}_\omega({\vec R,\vec k})&=&0
\label{gi2}
\ea
where ${\vec v}_s$ is the superfluid velocity defined by
\be
{\vec v}_s({\vec r})= \nabla \chi({\vec r})-{e \over c}{\vec A(\vec r)}
\ee
This is the same definition as in (\ref{vs0}) if we notice that $\chi=\phi/2$.
Shifting the chemical potential $\mu$ by $v_s^2/2$, we have
\be
{1 \over 2}({\vec k} \pm \vec v_s)^2-\mu =
\epsilon_{\vec k} \pm {\vec k\cdot \vec v}_s
\ee
Gorkov equations will then be simplified to
\be
\left( \begin{array}{cc} 
i\omega_n - \epsilon_{\vec k} - {\vec k\cdot \vec v}_s & \tilde{\Delta} \\
\tilde{\Delta}^\ast & i\omega_n + \epsilon_{\vec k} - {\vec k\cdot \vec v}_s 
\end{array} \right) 
\left( \begin{array}{c} \tilde{\cal G} \\ \tilde{\cal F} \end{array} \right)
=\left( \begin{array}{c} 1 \\ 0 \end{array} \right)
\ee
and their solutions will be
\begin{eqnarray}
&&\tilde{\cal G}_\omega({\vec R,\vec k})={i\omega_n - {\vec k\cdot \vec v}_s({\vec R})
+\epsilon_{\vec k} \over
[i\omega_n - {\vec k\cdot v}_s({\vec R})]^2-E_{\vec k}^2({\vec R})}\nonumber \\
&&\tilde{\cal F}^\dagger_\omega({\vec R,\vec k})={-\tilde{\Delta}^\ast({\vec R,\vec k}) \over
[i\omega_n - {\vec k\cdot \vec v}_s({\vec R})]^2-E_{\vec k}^2({\vec R})}
\end{eqnarray}
Notice that the only difference between these and (\ref{g&f}) is a Doppler 
shift (by $\vec k\cdot \vec v_s$) in the Matsubara frequencies.
We again sum over the Matsubara frequencies in the gap equation to get
\be
\tilde{\Delta}({\vec R,\vec k})= -\int d^3 k^\prime \tilde{V}({\vec k-\vec k^\prime})
{\tilde{\Delta}({\vec R,\vec k^\prime}) \over 2 E_{\vec k^\prime}({\vec R})}
\{ f[{\vec k^\prime\cdot v}_s({\vec R})-E_{\vec k^\prime}({\vec R})] -
f[{\vec k^\prime\cdot \vec v}_s({\vec R})+E_{\vec k^\prime}({\vec R})] \}
\label{gap}
\ee
where 
\be
f(E)={1 \over e^{\beta E} + 1}
\ee
is the Fermi distribution function and $\beta$ is the inverse temperature
$1/T$. Note that as ${\vec v}_s \rightarrow 0$, this reduces to (\ref{gap0}).

\subsection{Calculation of Current in a Superconductor}

The current density, in terms of the Green functions, can be written as
\begin{eqnarray}
{\vec j(r)}&=&{ie \over 2}\lim_{\vec r \rightarrow \vec r^\prime} \sum_\sigma
(\nabla_{\vec r^\prime}-\nabla_{\vec r})<\psi_\sigma^\dagger({\vec r^\prime})
\psi_\sigma({\vec r)}> - {e^2 \over c} {\vec A(\vec r)} \sum_\sigma
<\psi_\sigma^\dagger({\vec r})\psi_\sigma({\vec r)}> \nonumber\\
&=&{ie \over 2}\lim_{\vec r \rightarrow r^\prime} \sum_\sigma
(\nabla_{\vec r^\prime}-\nabla_{\vec r}){\cal G}({\vec r}0,{\vec r}^\prime 0^+)
-{e^2 \over c} {\vec A(\vec r)} \sum_\sigma {\cal G}({\vec r}0,{\vec r}0^+)\nonumber\\
&=& ie \lim_{\vec r \rightarrow \vec r^\prime} 
(\nabla_{\vec r^\prime}-\nabla_{\vec r})\tilde{\cal G}({\vec r}0,{\vec r}^\prime 0^+)
+2e {\vec v}_s({\vec r})\tilde{\cal G}({\vec r}0,{\vec r}0^+)
\end{eqnarray}
The current in this form is manifestly gauge invariant. Keeping in mind that the
superfluid density is $n_s({\vec r})=\sum_\sigma
<\psi_\sigma^\dagger({\vec r})\psi_\sigma({\vec r)}> = 
2\tilde{\cal G}({\vec r}0,{\vec r}0^+)$, the current can be written as
\be
{\vec j(\vec r)}=en_s\vec v_s(\vec r)+\vec j_{qp}({\vec r})
\label{jjs}
\ee
where 
\be
{\vec j}_{qp}({\vec r})=ie \lim_{\vec r \rightarrow r^\prime} 
(\nabla_{\vec r^\prime}-\nabla_{\vec r})\tilde{\cal G}({\vec r}0,
{\vec r}^\prime 0^+)
\ee
${\vec j}_{qp}$ turns out to be the current due to quasiparticle 
excitations in the superconductor.
Fourier transforming, using the same techniques discussed in the last subsection,
we get
\be
{\vec j}_{qp}({\vec R})=ie \int {d^3 k \over (2\pi)^3}
(\nabla_{\vec R}-2i{\vec k}){1 \over \beta}\sum_n
\tilde{\cal G}_\omega({\vec R,\vec k})
\ee
Again under quasi-classical approximation, one can neglect $\nabla_{\vec R}$
and write
\ba
{\vec j}_{qp}({\vec R})&=&{2e \over \beta}\sum_n
\int {d^3 k \over (2\pi)^3}{\vec k}\tilde{\cal G}_\omega({\vec R,\vec k}) \nonumber\\
&=&e \int {d^3 k \over (2\pi)^3}{\vec k}
\{(1+{\epsilon_{\vec k} \over E_{\vec k}})f[E_{\vec k}+{\vec k}\cdot 
{\vec v}_s({\vec R})] - (1-{\epsilon_{\vec k} \over E_{\vec k}})f[E_{\vec k}-
{\vec k}\cdot{\vec v}_s({\vec R})]\}
\ea
In the last step, we have performed a Matsubara sum.
Combining the two terms the current can be written as
\be
{\vec j}_{qp}({\vec R})=-2e \int {d^3 k \over (2\pi)^3}{\vec k}
f[E_{\vec k}-{\vec k}\cdot{\vec v}_s({\vec R})]
\ee
or
\be
{\vec j}_{qp}({\vec R})=-4eN_F \left< {\vec v}_F \int_0^\infty d \xi
f \left(\sqrt{\xi^2 + |\Delta|^2} - \vec v_F \cdot \vec v_s \right) \right>_{FS}
\label{jqp}
\ee
where $N_F$ is the density of states at the Fermi surface and
$\left< ... \right>_{FS}$ represents averaging over the Fermi surface.
Note that $\Delta$, in general, can depend on the point on the Fermi surface.
Equation (\ref{jqp}) has a natural interpretation. The argument of the Fermi
distribution function $f$ is the Doppler shifted quasiparticle energy at the 
particular point on the Fermi surface. Thus (\ref{jqp}) is actually 
a Fermi surface average of the current due to the quasiparticles excited
via thermal excitation plus the Doppler shift. This 
can be viewed as a normal fluid in the opposite direction of the superfluid
that tends to reduce the supercurrent density. At zero temperature, the
Fermi function is a step function which is zero unless its argument is negative.
Therefore (\ref{jqp}) is zero unless
\be
{\vec v}_F \cdot {\vec v}_s > |\Delta|
\label{vc}
\ee
Eq. (\ref{vc}) actually defines a critical value $v_c=|\Delta|_{\rm min}/v_F$ for 
the superfluid velocity $v_s$, 
beyond which the superconducting gap vanishes at some regions of the 
Fermi surface. Increasing $v_s$ beyond $v_c$ increases the number of 
quasiparticle excitations and eventually destroys the superconductivity.
Therefore, there is an interval of $v_s$ over which we can have 
superconductivity with a gap that vanishes in some region of the Fermi surface.
This is usually known as {\it gapless superconductivity}. For an s-wave
superconductor, (\ref{vc}) gives a finite critical value for 
the superfluid velocity.
For a d-wave superconductor (or any superconductor with nodes in the 
superconducting gap) on the other hand, the gap vanishes at the nodes even at
zero $v_s$. In other words, quasiparticles exist at node lines even for
infinitesimal $v_s$ and zero temperature. This phenomenon is highly
anisotropic because the nodes exist only at four symmetric points on the
Fermi surface. The effect 
of these nodal quasiparticles in the Meissner state was studied by Yip
and Sauls \cite{Yip} and others \cite{Valls}. Here, we focus on the effect in 
the vortex state; which has been studied in \cite{Franz2,Amin1}. 
We again derive a generalized London equation and solve
it numerically for a vortex lattice to calculate physical quantities.

\section{Generalized London Equation}

As we mentioned earlier, the fourfold anisotropic nature of the gap nodes
is expected to influence the magnetic properties of a d-wave superconductor. 
This influence will be via two distinct effects: a {\it nonlinear} effect
due to quasiparticle generation because of Doppler shift in the quasi-particle 
spectrum, and {\it nonlocal} effect due to divergence of the coherence length along
the nodal directions. In this section we study these effects in two separate
subsections. In what follows we assume that these two effects can be studied
independently and their effects are additive. In other words, we assume that
any inter-connection between these two effects produces higher order corrections
to our generalized London equation. In general the excitation of quasiparticles 
at the gap nodes can affect the coherence length and therefore the 
nonlocality of the system. Similarly nonlocalities can change the form 
of Eq. (\ref{jqp}). This effect which is neglected in our calculations 
has been studied extensively in the context of superfluid $^3$He \cite{v-w}.

\subsection{Nonlinear Corrections}

Let us first neglect any nonlocal effect and focus
on the quasiparticles generated at the gap nodes. As we mentioned at the end
of the last section,
excitation of quasiparticles at the gap nodes produces a current density
flowing in the  
direction opposite to the superfluid velocity -  sometimes called back-flow. 
The total current is given by (\ref{jjs}). We again define the 
superfluid velocity by
\be
\vec v_s={1\over 2}(\nabla \phi - {2 e\over c}\vec A) .
\label{vs}
\ee
with $\phi$ being the phase of the order parameter. In this definition,
$\vec v_s$ is in the direction of the superfluid, i.e. in the opposite direction
of motion of the Cooper pairs (which is used in \cite{Yip,Amin1} as the 
definition for the direction of $\vec v_s$).
The contribution of the quasiparticles generated at the
nodes to the total current is given by (\ref{jqp}) which we
rewrite it as 
\be
\vec j_{qp}=-4eN_F\int_{FS} ds \vec v_F(s) \int_0^\infty d\xi
f(\sqrt{\xi^2+\Delta(s)^2} - \vec v_F(s)\cdot\vec v_s)
\label{j1}
\ee
where $s$ parameterizes a point on the Fermi surface. $\Delta(s)$ is 
the superconducting gap which in general can have $s$-wave, $d$-wave, 
or other symmetries. At zero temperature (\ref{j1}) leads to
\be
\vec j_{qp}=-4eN_F\int ds \vec v_F(s) \theta (\vec v_F\cdot\vec v_s
-|\Delta|)\sqrt{(\vec v_F\cdot\vec v_s)^2-\Delta^2}
\label{j2}
\ee

\begin{figure}
\epsfysize 5cm
\epsfbox[-300 200 400 610]{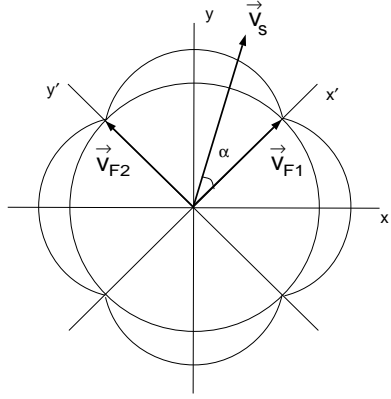}
\caption[Circular Fermi surface with a $d_{x^2-y^2}$ gap.]
{Circular Fermi surface with a $d_{x^2-y^2}$ gap. Quasiparticles
will be excited at nodes marked by $\vec v_{F1}$ and $\vec v_{F2}$ opposite to
$\vec v_s$.}
\label{fys1}
\end{figure}

At higher temperatures however, this gives the first term in the 
Sommerfeld expansion; which is a good approximation as long as 
$T<T^\star=T_c(H/H_0)$, where $H_0$ is of order of the
thermodynamic critical field, $H_c$ \cite{Yip}. The presence of
the $\theta$-function in (\ref{j2}), results in excitations
only at the nodes which are in the direction of $\vec v_s$.
Fig. \ref{fys1} illustrates a circular Fermi surface with a $d$-wave
gap. The quasiparticles are excited at the nodes marked by
$\vec v_{F1}$  and  $\vec v_{F2}$ in the direction of $\vec v_s$.
For a small enough $v_s$, the excitations stay very close to the
gap nodes. Therefore, one can linearize the gap function near the nodes
writing $\Delta(\theta)\simeq \gamma \Delta_0 \theta$,
with $\Delta_0$  the maximum gap and $\gamma$ defined by
\be
\gamma = {1\over \Delta_0}[{d\over d\theta}\Delta(\theta)]_{\rm node}
\label{gamma}
\ee
The most commonly used form for a d-wave gap is $\Delta(\theta)=\Delta_0
\cos (2\theta)$. In that case, (\ref{gamma}) leads to $\gamma=2$. 
The component of $\vec j_{qp}$ along the $x^\prime$-direction which is
diagonal to the $x$ and $y$ ($a$ and $b$) directions 
(as illustrated in Fig. \ref{fys1}) is then 
\begin{eqnarray}
j_{qpx^\prime}&=& -4eN_Fv_F\int_{-\theta_c}^{\theta_c}{d\theta \over 2\pi}
\sqrt{(v_Fv_s\cos\alpha)^2-|\gamma\Delta_0\theta|^2}\nonumber\\
&=& -en_s{v_s^2 \over v_0} \cos\alpha |\cos\alpha |= -en_s{v_{sx^\prime}
|v_{sx^\prime}|\over v_0}
\label{jx}
\end{eqnarray}
$\alpha$ is the angle between $v_s$ and $x^\prime$-axis, 
$n_s=N_F v_F^2$ is the superfluid
density,  $v_0=\gamma\Delta_0/v_F$ is some characteristic
velocity and $\theta_c$ is a cut off imposed by the $\theta$-function
in (\ref{j2}). Similarly the $y^\prime$-component is
\be
j_{qpy^\prime}= -en_s{v_{sy^\prime}|v_{sy^\prime}|\over v_0}
\label{jy}
\ee
and the total current thereby becomes
\be
\vec j=en_s[\vec v_s - (v_{sx^\prime}|v_{sx^\prime}|\hat x^\prime
+v_{sy^\prime}|v_{sy^\prime}|\hat y^\prime)/v_0].
\label{js2}
\ee
The nonanalytic nature of the effect is evident from this equation.  
It is now possible to write a free energy in such
a way  that (\ref{js2}) can be
obtained by minimization with respect to $\vec A$.
Keeping in mind that $\vec j=(c/4\pi)\nabla\times\vec B =
(c/4\pi)\nabla\times\nabla\times\vec A$ and $\partial/\partial \vec v_s
= -{c\over e}\partial/\partial \vec A$, the corresponding free energy 
density can be written as
\be
f=n_s\left[{1\over 2}v_s^2-{1\over 3v_0}(|v_{sx^\prime}|^3+
|v_{sy^\prime}|^3)\right]+{B^2\over 8\pi}
\label{fnl}
\ee

In general, it is possible to solve (\ref{js2}) for $\vec v_s$ in terms 
of $\vec j=(c/4\pi)\nabla\times\vec B$, substitute it into (\ref{fnl})
 and write down a London free energy 
only in terms  of $\vec B$ and its derivatives. However, instead of
solving (\ref{js2}) exactly, we find $\vec v_s$ perturbatively assuming
that the nonlinear part is much smaller than the linear part. This way we
get a polynomial correction to the London equation which is 
convenient for numerical purposes. To first order in perturbation theory
we have
\be
\vec v_s={c\over 4\pi e n_s} \left[\nabla\times\vec B 
+ {c\over 4\pi e n_s v_0}
\left((\nabla\times\vec B)_{x^\prime}|\nabla\times\vec B|_{x^\prime}\hat x^\prime
+(\nabla\times\vec B)_{y^\prime}|\nabla\times\vec B|_{y^\prime}
\hat y^\prime) \right) \right]
\ee
Substituting this into (\ref{fnl}) and keeping the lowest order terms, 
the London free energy density becomes
\be
f_L={1\over 8\pi}\Bigl[
B^2+\lambda_0^2(\nabla\times\vec B)^2+({2\pi\over 3\gamma})
{\xi_0\lambda_0^2\over B_0}(|(\nabla\times\vec B)_{x^\prime}|^3
+|(\nabla\times\vec B)_{y^\prime}|^3)\Bigr] 
\label{fL1}
\ee
where $\lambda_0=\sqrt{c^2/4\pi e^2n_s}$ is the zeroth order 
penetration depth, $\xi_0=v_F/\pi\Delta_0$
is the coherence length, $B_0\equiv \phi_0/2\pi\lambda_0^2$ is a characteristic
field of the order of $H_{c1}$ and $\phi_0=\pi c/e$ is the flux quantum. 
For magnetic fields in the $z$-direction, (\ref{fL1}) becomes
\be
f_L={1\over 8\pi}\Bigl[B^2+\lambda_0^2(\nabla B)^2+({2\pi\over 3\gamma})
{\xi_0\lambda_0^2\over B_0}(|\partial_{x^\prime}B|^3+|\partial_{y^\prime}B|^3)
\Bigr] = f^0_L + f_{\rm nl}
\label{fL2}
\ee
with $f^0_L$ and $f_{nl}$ representing the ordinary London free energy density 
and the leading nonlinear correction to the free energy density respectively.
The corresponding London equation is given by 
\be
-\lambda_0^2\nabla^2B+B-({2\pi\over \gamma}){\xi_0\lambda_0^2\over B_0}(\partial_
{x^\prime}^2B
|\partial_{x^\prime}B|+\partial_{y^\prime}^2B|\partial_{y^\prime}B|)=0
\label{London}
\ee
A similar London equation is also derived by Zutic and Valls
who investigated the effect in the Meissner state \cite{Valls}.

In order to find the magnetic field distribution 
in a vortex lattice, one has to insert a source term
$\Sigma_j \rho(\vec r - \vec r_j)$ on the right hand side of (\ref{London}) 
with $\vec r_j$ being the position of the vortices in the lattice.
The function $\rho(\vec r)$ takes into account the vanishing of the 
order parameter at the center of the vortex cores. 
Numerically, it is more convenient to work in  Fourier space rather than
real space. Fourier transforming (\ref{London}) with a proper source term 
on the right hand side yields
\be
B_{\vec k}+\lambda_0^2 k^2 B_{\vec k}-G_{\rm nl}(\vec k,B_{\vec k})=\bar B F(\vec k)
\label{ldn}
\ee
where $\vec k$ is a reciprocal lattice wave vector,  $G_{\rm nl}$ is the
Fourier transform of the nonlinear term and $\bar B$ is the average
magnetic field. The cut off function $F(\vec k)$ comes from the Fourier 
transformation of the source term and removes the divergences
by cutting off the momentum sums. We are going to use Gaussian cutoff    
(\ref{Fk}) in our numerical calculation in the next section.

\subsection{Nonlocal Corrections}

In our derivation of the corrections to the London equation in the previous 
subsection, we derived a local relation between current
density $\vec j$ and superfluid velocity $\vec v_s$ (or vector 
potential $\vec A$). This was a result of the {\it local approximation} we
used in our derivation. In fact, this relation is always
nonlocal over the length scale of the coherence length $\xi_0$; which is of the 
order of the finite spatial extent of the Cooper pair \cite{Tinkham}. 
Magnetic field in a superconductor varies in
the length scale given by London penetration depth $\lambda_0$ and therefore
nonlocal corrections to physical quantities, such as the effective
penetration depth, will be of order $\kappa^{-2}$, where 
$\kappa\equiv \lambda_0/\xi_0$ is the GL ratio. For strongly type 
II materials ($\kappa\gg 1$) such corrections are 
negligible. Since cuprate superconductors fall well within this class 
($\kappa$ is  of 50 for most) local 
electrodynamics is usually used. However, a closer examination suggests that 
this might not be justified in all situations, if, as it is widely believed,
these materials exhibit nodes in the gap. In such a case 
in place of the usual coherence length $\xi_0=v_F/\pi\Delta_0$
one is forced to define an angle dependent quantity, $\xi_0(\hat{p})=
v_F/\pi\Delta_{\hat p}$, which diverges along the nodes. Clearly, in the
vicinity of nodes the condition  $\lambda_0/\xi_0({\hat p})\gg 1$ is no 
longer satisfied and, in fact, the extreme nonlocal limit is achieved.
Nonlocal corrections therefore cannot be dismissed in unconventional
superconductors. The effect of these corrections on the 
temperature dependence of penetration depth at low $T$ was studies by Kosztin 
and Leggett \cite{Leggett}. Here, we discuss about the effect on the vortex 
lattice properties. What we follow here is based on a work published in Ref. 
\cite{Franz2}. From the above arguments, it is  clear that corrections 
due to the nonlocal effects at the gap nodes 
will be highly anisotropic and will in general break the rotational symmetry
of the flow field around the vortex, contributing an anisotropic
component to the inter-vortex interaction in the mixed state. 

Let us now find these nonlocal corrections to the London model
neglecting all the nonlinear effects discussed in the previous subsection.
In the next subsection, we will combine these two effects into a single London
equation to be used in numerical evaluations.
Nonlocal relation between ${\vec j}$ and ${\vec A}$ is conveniently written
in the Fourier space as \cite{Tinkham}
\begin{equation}
\vec j_{k}=-(c/4\pi)\hat Q(\vec k) A_{k}.
\label{n1}
\end{equation}
Here ${\hat Q(\vec k)}$ is the electromagnetic response tensor which can be 
computed, within the weak coupling theory, by generalizing the standard 
linear response treatment of Gorkov equations \cite{Lif} to an anisotropic gap. 
We find
\begin{equation}
Q_{ij}({\vec k})={4\pi T\over \lambda_0^2}\sum_{n>0}\left\langle
{\Delta_{\hat p}^2\hat v_{Fi}\hat v_{Fj} \over 
\sqrt{\omega_n^2+\Delta_{\hat p}^2}
(\omega_n^2+\Delta_{\hat p}^2+\gamma_{\vec k}^2)}\right\rangle,
\label{Q}
\end{equation}
where $\gamma_{\vec k}={\vec v}_F\cdot{\vec k}/2$, $\lambda_0=\sqrt{c^2/4\pi 
e^2v_F^2N_F}$ is the London penetration depth as before, $\omega_n$
are the Matsubara frequencies and the angular brackets again mean the Fermi 
surface averaging. Derivation of (\ref{Q}) involves a calculation of the 
perturbative correction to the Green's functions to linear order in $\vec A$ 
using Gorkov's equations. The result is then used to calculate the 
supercurrent $\vec j$ \cite{Lif}.
Eq. (\ref{Q}) is valid for arbitrary Fermi surface and gap 
function.  For isotropic gap one recovers an expression derived
by Kogan {\em et al.\ }from the Eilenberger theory \cite{Kogan2}.
One may simplify solving the London equation again by writing it 
in terms of magnetic field only.
Eliminating ${\vec j}$ from  (\ref{n1}) using the 
Amp\`{e}re's law ${\vec j}=(c/4\pi)\nabla\times{\vec B}$, one obtains
\begin{equation}
{\vec B_k}-{\vec k}\times[{\hat Q^{-1}(\vec k)}({\vec  k}\times{\vec B_k})]=0. 
\label{london1}
\end{equation}
To study vortex lattice using this formalism, it is necessary to insert a
cutoff function $F(k)$ on the right-hand side of the Eq. (\ref{london1}).
In our numerical calculations in the next section we again use Gaussian cutoff
function (\ref{Fk}).
For many purposes it is also convenient to write down the corresponding
London free energy, such that $\delta {\cal F}_L/\delta {\vec B_k}=0$ gives the above
London equation:
\begin{equation}
{\cal F}_L=\sum_{\vec k}[{\vec B_k}^2+ ({\vec k}\times{\vec B_k})
{\hat Q^{-1}(\vec k)}({\vec k}\times{\vec B_k})]/8\pi.
\label{free}
\end{equation}

At long wavelengths ${\hat Q(\vec k)}$ can be evaluated by expanding expression
(\ref{Q}) in powers of $\gamma_{\vec k}^2$. The zeroth order term 
\begin{equation}
Q_{ij}^{(0)}\equiv\delta_{ij}\lambda^{-2}
={4\pi T\over \lambda_0^2}\sum_{n>0}\left\langle
{\Delta_{\hat p}^2\hat v_{Fi}\hat v_{Fj} \over 
(\omega_n^2+\Delta_{\hat p}^2)^{3/2}}\right\rangle,
\label{Q0}
\end{equation}
clearly recovers the ordinary London free energy. Here,
$\lambda=\lambda(T)$ is the temperature dependent penetration depth which, 
at low temperatures, has the well known $T$-linear behavior 
\cite{Scalapino,Hardy}; for a $d_{x^2-y^2}$
superconductor with $\Delta_{\hat p}=\Delta_d(\hat p_x^2-\hat p_y^2)$
and a cylindrical Fermi surface (from now on we shall focus on this 
simple case) $\lambda^{-2}\approx\lambda_0^{-2}(1-2\ln2T/\Delta_d)$.
The leading nonlocal term is quadratic in $k$:
\begin{equation}
Q_{ij}^{(2)}
=-{4\pi T\over \lambda_0^2}\sum_{n>0}\left\langle
{\Delta_{\hat p}^2\hat v_{Fi}\hat v_{Fj} \over 
(\omega_n^2+\Delta_{\hat p}^2)^{5/2}}\gamma_{\vec k}^2 \right\rangle.
\label{Q2}
\end{equation}
The expression $Q_{ij}= \delta_{ij}\lambda^{-2}+Q_{ij}^{(2)}$ is
easily inverted to leading order in $k$ to get
$Q_{ij}^{-1}\approx\lambda^2[\delta_{ij}-\lambda^2 Q_{ij}^{(2)}]$. 
Substituting
this into (\ref{free}) and specializing to fields along the $z$-direction
we have
\begin{equation}
{\cal F}_L=\sum_{\vec k}B_{\vec k}^2[1+ \lambda^2k^2
+\lambda^2\xi^2(c_1k^4+c_2k_x^2k_y^2)]/8\pi.
\label{free-anal}
\end{equation}
Here $\xi=v_F/\pi\Delta_d$ and $\Delta_d$ is assumed to be a temperature
dependent solution to the appropriate gap equation. Dimensionless coefficients
$c_1$ and $c_2$ are given by
\renewcommand{\arraystretch}{1}
\begin{equation}
c_\mu
={\lambda^2\over\lambda_0^2}\pi^3\Delta_d^2T
\sum_{n>0}{1\over 2\pi}\int_0^{2\pi}d\theta
{\Delta_{\hat p}^2 w_\mu \over (\omega_n^2+\Delta_{\hat p}^2)^{5/2}},
\label{c12}
\end{equation}
where $w_1=\hat v_{Fx}^2 \hat v_{Fy}^2$, 
$w_2=(\hat v_{Fx}^2 -\hat v_{Fy}^2)^2-4\hat v_{Fx}^2 \hat v_{Fy}^2$ and 
the Fermi surface has been explicitly parameterized by the angle $\theta$
between ${\hat p}$ and $x$-axis; ${\hat v}_F=(\cos\theta,\sin\theta)$ and
$\Delta_{\hat p}=\Delta_d\cos2\theta$. 
Coefficients $c_1$ and $c_2$ depend on temperature through a dimensionless
parameter $t\equiv T/\Delta_d$. From (\ref{c12}) one can deduce their
leading behavior in the two limiting cases: for $t\ll 1$ we find
\begin{equation}
c_1={\pi^2\over 8}{\lambda^2\over\lambda_0^2}{1\over t}\quad , \quad c_2=-4c_1,
\label{clow}
\end{equation}
and for $t\gg 1$ (i.e., near $T_c$)
\begin{equation}
c_1=\alpha{\lambda^2\over\lambda_0^2}{1\over t^4}\quad , \quad c_2 =8c_1,
\label{chigh}
\end{equation}
where $\alpha=\zeta(5)(1-2^{-5})/8\pi^2=0.01272$. In the above $\lambda$ 
also depends on $t$, but this will be unimportant for the following 
qualitative discussion.

The free energy (\ref{free-anal}) formally coincides with (\ref{Fk0}),
deduced previously in the last chapter with
phenomenological considerations. The present model however, allows us
to evaluate the coefficient of the symmetry breaking term $c_2$ 
(or $\epsilon$ in (\ref{Fk0})).
At high temperatures, $c_2$ is found to be positive; in agreement with 
our previous result. As we showed in the last chapter, such 
a term leads to a centered rectangular vortex lattice
structure with principal axes oriented along $x$ or $y$ axes of the ionic
lattice (see Fig. \ref{fig1}a). 
The magnitude of distortion from a perfect triangular lattice is 
controlled by the magnitude of $c_2$ and grows with increasing magnetic field.
Eq. (\ref{chigh}) shows that at fixed field 
this distortion will initially grow with decreasing temperature. 
At low temperatures (\ref{clow}) predicts $c_2<0$. This will
lead to the same centered rectangular lattice but rotated by $45^\circ$ 
(Fig. \ref{fig1}b). Numerical
evaluation of (\ref{c12}) shows that $c_2$ passes through zero at 
$t^*\simeq0.19$. At this temperature the free energy (\ref{free-anal}) 
is isotropic and the lattice will be triangular at all fields. The 
change of sign of $c_2$ reflects
the competition between the two terms of different symmetry in $w_2$.
At $t\ll 1$ only the region around the node is important where 
$(\hat v_{Fx}^2 -\hat v_{Fy}^2)^2$ vanishes, while at $t\gg 1$ the average
involves the entire Fermi surface to which both terms in $w_2$ contribute.
This change of sign is a unique consequence of nodes in the gap function
and would not occur in conventional superconductors. 

\begin{figure}
\epsfysize 5cm
\epsfbox[-200 450 300 750]{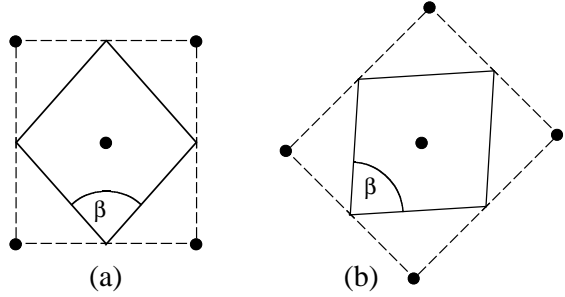}
\caption{Two high symmetry orientations of centered rectangular unit cell.}
\label{fig1}
\end{figure}

Another consequence of nodes
is the fact that, as can be seen from (\ref{clow}), both $c_1$ and
$c_2$ diverge as $1/t$ for $t\to 0$.  This divergence signals that the 
response tensor ${\hat Q(\vec k)}$ is a non-analytic function of ${\vec k}$
at $T=0$, and the expansion in powers of $\gamma_{\vec k}^2$ breaks down. 
Formally this is caused by the fact that at $T=0$, at the nodal point,
the expression (\ref{Q}) for $Q_{ij}({\vec k})$ contains a term
proportional to $1/\gamma_{\vec k}^2$. At $T=0$ the frequency sum in (\ref{Q})
becomes an integral which can be evaluated exactly with the result:
\begin{equation}
Q_{ij}({\vec k})={1\over \lambda_0^2}\left\langle
\hat v_{Fi}\hat v_{Fj}
{2{\rm arc\ sinh\ }y \over y\sqrt{1+y^2}} \right\rangle,
\label{Q-nono}
\end{equation}
where $y=\gamma_{\vec k}/\Delta_{\hat p}$. For small $k$ the dominant 
contribution to the angular average comes from the close vicinity of nodes
and can be evaluated by linearizing $\Delta_{\hat p}$ 
around the nodes. One finds that the
leading nonlocal contribution is {\em linear} in $k$ rather than quadratic.
For $Q_{ij}= \delta_{ij}\lambda_0^{-2}+Q_{ij}^{(1)}$, we have 
$Q_{xx}^{(1)}=Q_{yy}^{(1)}=-\mu(k_M\xi_0)$ and 
$Q_{xy}^{(1)}=Q_{yx}^{(1)}=-\mu(k_m\xi_0)
{\rm sgn}(\hat k_x\hat k_y)$, where $k_M=\max(|k_x|,|k_y|)$ 
and $k_m=\min(|k_x|,|k_y|)$. Prefactor 
$\mu=\pi^2/8\sqrt{2}=0.8723$ is exact in the sense that all corrections to
$Q_{ij}$ are $O(k^2)$. The resulting free energy at $T=0$ is 
\begin{equation}
{\cal F}_L=\sum_{\vec k}B_{\vec k}^2[1+ \lambda_0^2k^2
+\mu\lambda_0^2\xi_0 k_M
(k_M^2- k_m^2)]/8\pi.
\label{free-nono}
\end{equation}
The nonlocal term is clearly non-analytic in $k$. Its functional form is 
universal in the sense that it is independent of the Fermi surface structure
(as long as it has the same symmetry as the order parameter) and
the prefactor $\mu$ only depends on the angular slope of the gap function and
Fermi velocity at the node. 

The present calculation can be easily generalized to treat the effects of
Fermi surface anisotropy. As
mentioned above tetragonal anisotropy will not modify the $T\to 0$ universal
behavior but may lead to quantitative changes at higher temperatures.
Orthorhombic anisotropy, on the other hand, will modify even the $T\to 0$
limit. We expect that it will, to leading order, merely rescale the coordinate
axes, leading to the same structures as described above stretched by the
axes, leading to the same structures as described above stretched by the
appropriate factor \cite{Affleck}. It may further remove the degeneracy between
two equivalent lattices related by $90^\circ$ rotation. Also neglected in
our calculation is the effect of electronic disorder, which will remove
the non-analyticity of ${\hat Q(\vec k)}$ at longest wavelengths, just as
small finite temperature would.
Since the lattice structure is most sensitive to  ${\hat Q(\vec k)}$ at
finite $k\sim 1/l$ ($l$ is the vortex spacing), we expect our predictions
to be robust with respect to weak disorder.

\subsection{Combining Nonlinear and Nonlocal Corrections}

Having established theories for nonlinear and nonlocal effects independently,
we now would like to combine them in one London equation to be used in
numerical calculations. Nonlocal effect can be added to the generalized
London equation (\ref{ldn}), simply by replacing the second
term with its nonlocal counterpart
\be 
B_{\vec k}+{\cal L}_{ij}(\vec k)k_ik_j B_{\vec k}-G_{\rm nl}(\vec k,B_{\vec k})
=\bar B F(\vec k). 
\label{ldn2}
\ee 
Sums over $i$ and $j$ are implicit here. ${\cal L}_{ij}(\vec k)$ is related to
the electromagnetic response tensor $\hat Q(\vec k)$ by
\be
{\cal L}_{ij}(\vec k)=Q_{ij}(\vec k)/\det \hat Q(\vec k).
\label{L}
\ee
The electromagnetic response tensor $\hat Q(\vec k)$ is given by (\ref{Q})
for $T \ne 0$, and by (\ref{Q-nono}) at $T=0$. 

In the local case, we have 
${\cal L}_{ij}(\vec k)=\lambda_0^2\delta_{ij}$
which leads back to (\ref{ldn}). Taking $G_{\rm nl}$ to the
right hand side of (\ref{ldn2}), $B_{\vec k}$ can be obtained by
\be
B_{\vec k}={\bar B F(\vec k) + G_{\rm nl}(\vec k,B_{\vec k}) \over
1+{\cal L}_{ij}(\vec k)k_ik_j}.
\label{bk}
\ee
We will use this equation, in the next section, to find $B_{\vec k}$ iteratively 
for a specific lattice geometry. Having $B_{\vec k}$, the free energy can be 
easily calculated using
\be
{\cal F}={\cal F}_{\rm nl} + \sum_{\vec k}[1+{\cal L}_{ij}(\vec k)k_ik_j]B_{\vec k}^2 
\label{fe}
\ee
where ${\cal F}_{\rm nl}$ is the free energy due to the nonlinear term 
in Eq. (\ref{fL2})
\be
{\cal F}_{\rm nl} = ({2\pi\over 3\gamma}) {\xi_0\lambda_0^2\over B_0}
\int d^2 r \left( |\partial_{x^\prime}B|^3+|\partial_{y^\prime}B|^3 \right)
\ee

\section{Numerical Calculations}

Unlike Ginsburg-Landau, the London free energy cannot completely
determine the vortex lattice by a simple minimization. Instead,
one has to impose a set of source terms located at the position of
the vortices on an assumed lattice. The functional form of 
the source terms does not come from
the London theory and requires more fundamental treatments; as we 
discussed in Sec.(5.2.2). Information about the functional form of the source 
term is included in (\ref{bk}) via $F(k)$, and information about the positions 
of the vortices is included in the reciprocal lattice vectors $\vec k$. 
The free energy must then be minimized with respect to the positions of
the vortices in order to find the equilibrium lattice configuration.
In general, a 2D lattice can be determined by four parameters.
However, it turns out that a centered rectangular lattice is energetically 
more favorable than an oblique lattice. On the other hand,
the vortex lattice spacing is fixed
by the average magnetic field $\bar B$ ($\bar B \approx H$
away from $H_{c1}$). Thus we are left with two variational parameters, i.e.
the lattice orientation with respect to $a$ and $b$ directions and the
apex angle $\beta$ - the angle between the two basic vectors of the lattice.
We therefore find the vortex lattice geometry by minimizing ${\cal F}$ in 
with respect to the apex angle $\beta$ for different orientations of
the lattice.

In our numerical calculation, we first neglect the nonlinear effects and 
consider only the nonlocal corrections. At the end, we will add the effect of 
the nonlinear corrections to the results. We shall see that the nonlinear 
corrections are subdominant to the nonlocal ones. Therefore the results we 
obtain, ignoring the nonlinear effects, are actually very close to the reality.  
In our calculations, we first find $B_k$ using (\ref{bk}) but ignoring 
$G_{\rm nl}$ (i.e. ignoring the nonlinear corrections). Substituting (\ref{bk})
into (\ref{fe}), and also ignoring ${\cal F}_{\rm nl}$, 
we can calculate the free energy. 
At $T=0$, we can alternatively use (\ref{free-nono}) to calculate the free 
energy. 
\begin{figure}[h]
\epsfysize 8cm
\epsfbox[-30 100 400 430]{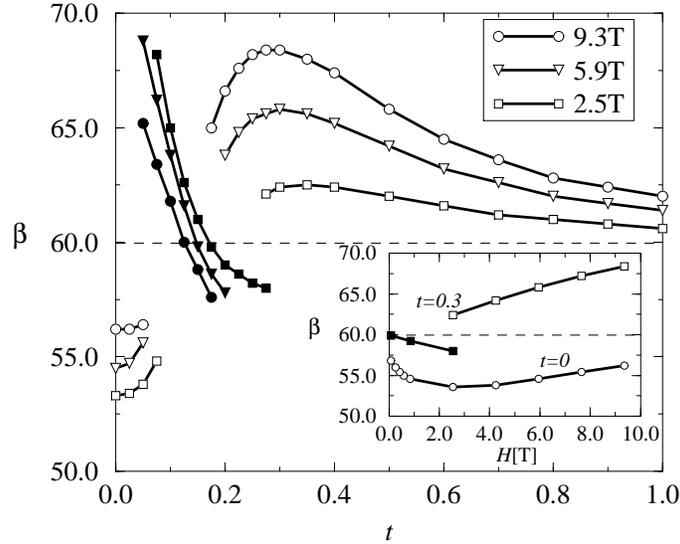}
\caption[Equilibrium angle $\beta$ as a function of reduced 
temperature $t=T/\Delta_d$ for various fields.]
{Equilibrium angle $\beta$ as a function of reduced temperature
$t=T/\Delta_d$ for various fields. Open symbols mark lattice with orientation
along $x$ or $y$ direction while solid symbols mark the lattice rotated
by $45^\circ$. We use $\lambda_0=1400$\AA  and $\kappa=68$. Inset: $\beta$ as
 a function of field at fixed $T$.}
\label{fig2}
\end{figure}
Numerical evaluation, shows that the free energy (\ref{free-nono}) gives rise to
a centered rectangular vortex lattice, aligned with $x$ or $y$ axes, but
now with the apex angle $\beta<60^\circ$; unlike (\ref{Fk0}) which gives a
$\beta>60^\circ$.
In order to map out the complete equilibrium $H$-$T$ phase diagram
we have carried out a
numerical computation of the vortex lattice structure using the full expression
for the response tensor ${\hat Q(k)}$, as given by Eqs.\ (\ref{Q}) and
(\ref{Q-nono}). We find that the free energy has two local minima
for centered rectangular lattices aligned with two high symmetry directions
(see Fig. \ref{fig1}), as expected from the tetragonal symmetry of the problem.
Which of the two becomes the global minimum depends on temperature and field.
The results are summarized in Fig. \ref{fig2}. For high temperatures, the exact
result agrees well with the one obtained from the long wavelength free
energy (\ref{free-anal}). The deformation of the lattice from perfect
triangular, grows with decreasing temperature, reaches maximum, and then
falls. Maximum distortion occurs around $t\simeq 0.3$, attaining
$\beta\simeq 70^\circ$ at 10T. Extrapolating this field dependence (see inset
to Fig. \ref{fig2}), the lattice should become square lattice around 
$H\approx 30$T, but
this field is outside the domain of validity of the London model.
At lower temperatures the distortion decreases but instead of going all the
way back to triangular at $t^*$, the lattice  undergoes a first order
phase transition to another centered rectangular lattice rotated by $45^\circ$
and with $\beta<60^\circ$. Further decrease of temperature causes the angle
to grow again. We note that the precise temperature at which it crosses
$60^\circ$ depends on field, but for all fields it is close to $t^*=0.19$, as
predicted by the long wavelength approximation.
At yet lower temperature we predict another first order transition to a
centered rectangular lattice along $x$ (or $y$) with $\beta<60^\circ$.
It has to be noted that the free energy difference between the two minima
is very small in the region where the
$45^\circ$ rotated lattice wins. It is thus
likely that real system, in which vortex pinning to various defects occurs,
will prefer to remain in the metastable state at these intermediate
temperatures. Instead of two consecutive first order transitions the experiment
would detect only a smooth crossover from a lattice with $\beta>60^\circ$
to the one with $\beta<60^\circ$. Alternatively domains with various
orientations may develop.

\begin{figure}
\epsfysize 7cm
\epsfbox[-200 300 400 750]{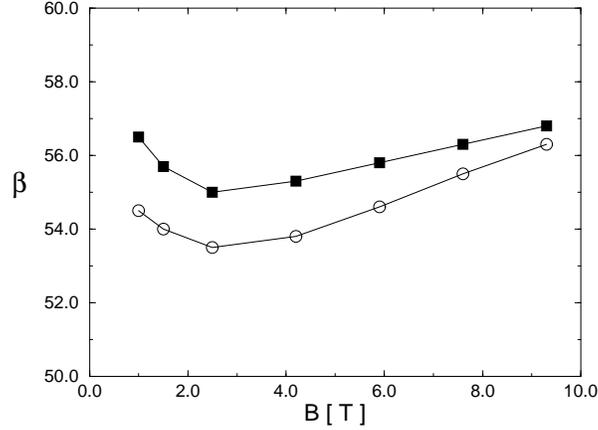}
\caption[Apex angle $\beta$ as a function of magnetic field $B$
at $T=0$.]
{Apex angle $\beta$ as a function of magnetic field $B$
at $T=0$. Circles represent the result of the calculation using
only the nonlocal corrections.  
Squares correspond to the calculations considering both nonlinear
and nonlocal corrections.}
\label{fys2}
\end{figure}

At this point, we can add the nonlinear corrections to our numerical 
calculations. The general strategy is to use (\ref{bk}) to 
calculate $B_{\vec k}$ iteratively. At each step $G_{\rm nl}$ has to be 
calculated numerically. The non-analytic form of $G_{\rm nl}$, prevents us
from using usual convolution integrals in our Fourier transformation. Instead,
we first calculate $G_{\rm nl}$ on a lattice in a unit cell of the
vortex lattice in position space.  Then we use
Fast Fourier Transformation (FFT) to calculate the Fourier transformed 
$G_{\rm nl}$ at the reciprocal lattice wave vectors $\vec k$. 
In our numerical calculation, we use
$\lambda_0=1400$\AA,  $\kappa=\lambda_0/\xi_0=68$ as before and also $\gamma=2$.

At $T=0$, the stable orientation of the lattice 
is again aligned with $a$ and $b$ axes. Fig. \ref{fys2}
shows the results of our numerical calculations for $\beta$ as a function of 
magnetic field. The upper curve (squares)  is the result of combined
calculation, considering both nonlocal and nonlinear
corrections, i.e. using (\ref{bk}) 
and (\ref{fe}). The lower curve (circles) on the other hand, corresponds 
to taking into account only the nonlocal corrections (this curve is the same 
as the $t=0$ curve in the inset of Fig. \ref{fig2}).
As is clear from Fig. \ref{fys2}, the difference between the two 
cases is small and about one or two degrees. Therefore, the phase diagram
given in Fig. \ref{fig2} retains its validity qualitatively, even after 
adding the nonlinear correction to the generalized London free energy.  

\begin{figure}
\epsfysize 7cm
\epsfbox[-200 300 400 750]{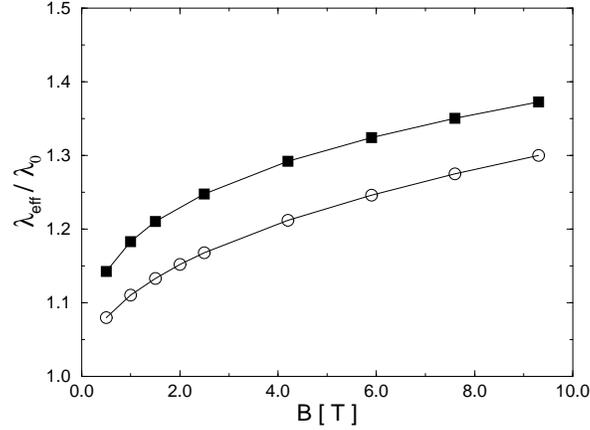}
\caption[The effective penetration depth as a function of the
magnetic field.]
{The effective penetration depth as a function of the
magnetic field. Circles represent the result of the calculations
considering only the nonlocal effects  whereas squares are the result
of combined calculation considering both nonlocal and nonlinear 
effects.}
\label{fys3}
\end{figure}

We also calculate the effective penetration depth $\lambda_{\rm eff}$
for different magnetic fields in almost the same way as it is calculated
from $\mu$SR data \cite{sonier-thesis}. In these experiments, the 
$\mu$SR precession signal obtained from experiment is fit to a signal
obtained by Fourier transforming a theoretical magnetic field distribution
function $n(B)$ defined by
\be
n(B^\prime)={1 \over \Omega}\int d^2r \delta[B^\prime - B({\vec r})]
\label{nb}
\ee
with $\Omega$ denoting the area of a unit cell. 
The magnetic field $B(\vec r)$ in 
(\ref{nb}) is calculated on a hexagonal vortex lattice with the same 
average magnetic field as the experimental field and using the 
ordinary London model with some cutoff function. 
The $\lambda$ which provides the best fit to data is considered as
$\lambda_{\rm eff}$. Here, we calculate $\lambda_{\rm eff}$ in a different
way (but similar in spirit), using the fact that in the ordinary London model,
for a general vortex lattice and for a large enough field,
\be
\overline{(B-\bar B)^2}\equiv \overline{\Delta B^2}=\bar B^2 
\sum_{\vec k \neq 0}{e^{-\xi^2 k^2} \over (1+\lambda^2 k^2)^2}
\simeq \lambda^{-4}\bar B^2 \sum_{\vec k \neq 0}
{e^{-\xi^2 k^2} \over k^4} = \Gamma(\xi) \lambda^{-4}
\ee
were $\Gamma(\xi)$ is a function that depends on the structure of the 
vortex lattice.
Associating all the field dependence of $\overline{\Delta B^2}$ 
in our calculation with the field dependence of an effective penetration depth 
$\lambda_{\rm eff}$, we can define $\lambda_{\rm eff}$ by
\be
{\lambda_{\rm eff} \over \lambda_0} = \left({\overline{\Delta B_0^2} \over
\overline{\Delta B^2}}\right)^{1 \over 4}
\label{lmd}
\ee
where $\overline{\Delta B_0^2}$ is the mean squared value of the magnetic
field $(B_0(\vec r)-\bar B)$ evaluated using the ordinary 
London model on a hexagonal
lattice with the same average field $\bar B$ and with the penetration 
depth $\lambda_0$. 

Fig. \ref{fys3} shows the result of our numerical calculation for
$\lambda_{\rm eff}$ using (\ref{lmd}). The lower curve corresponds to
the calculations including only the nonlocal correction. The upper curve  
corresponds to the result of the calculations using 
both nonlinear and nonlocal terms. 
The effect of the nonlinear term to the field dependence of $\lambda_{\rm eff}$ 
is almost nothing but an overall shift.
Fig. \ref{fys4} exhibits magnetic field distribution $n(B)$ 
at the average magnetic field $\bar B=5.9T$.
The solid line in Fig. \ref{fys4} represents the
magnetic field distribution calculated from the nonlinear-nonlocal London
equation. The double peak feature is a sign of existing two different saddle 
points in a unit cell of the lattice which is a result of having a 
$\beta \ne 60^\circ$. This line-shape is then compared with another line-shape 
(dashed line) obtained
from an ordinary London calculation but with a larger value of $\lambda_0$.
Similar line shapes with some additional broadening can also be produced 
from $\mu$SR data. The broadening is due to  lattice disorder,
interaction of muons with nuclear dipolar fields and finite lifetime of muons.
The resolution of the magnetic field as a result of this broadening is 
$\delta B \sim 10^{-3} T$. 
The difference between the solid line and dashed line in Fig. \ref{fys4}
as well as the double peak feature of the solid line is therefore
not observable by $\mu$SR experiments because of these broadening
effects. Thus as far as these line-shapes are concerned, 
it is difficult to distinguish
a nonlinear-nonlocal effect from a simple shift in the magnetic
penetration depth in the ordinary London model. 
Fig. \ref{fys5} compares the effect of
including both nonlinear and nonlocal corrections to the London equation
with the effect of including only the nonlocal term. Comparing
the two line-shapes, it is apparent that the effect 
of the nonlinear term is small compared to the nonlocal term as was 
emphasized before.

\begin{figure}[t]
\epsfysize 7cm
\epsfbox[-200 300 400 749]{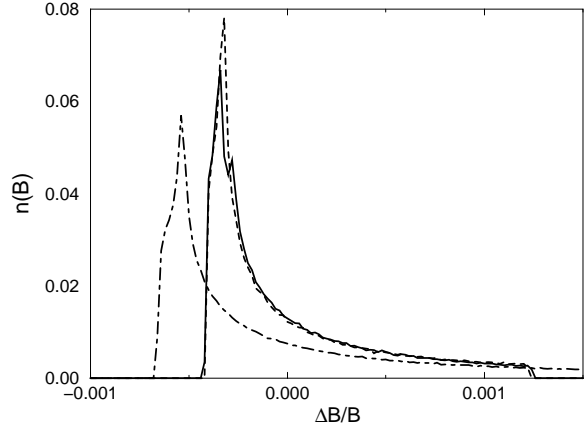}
\caption[Magnetic field distribution obtained from the nonlinear-nonlocal 
London equation as compared with the ordinary London calculation.]
{{\bf Solid line:}
Magnetic field distribution obtained from the nonlinear-nonlocal London
equation with $\bar B=5.9T$ and $\lambda_0=1400$ \AA. 
{\bf Dot-dashed line:} field distribution obtained from an ordinary
London equation on a hexagonal lattice with the same $\bar B$ and $\lambda_0$.
{\bf Dashed line:} Result of the same ordinary London calculation
but with $\lambda_0=1850$ \AA.}
\label{fys4}
\end{figure}

\section{Conclusion}

Ordinary London theory is not adequate to describe all the
different properties of a vortex lattice in high $T_c$ superconductors;
especially the properties resulting from the presence of the 
superconducting gap nodes or other
anisotropies on the Fermi surface. However, as we established
throughout this thesis, 
a generalized London model with appropriate higher order corrections which
take into account these anisotropic effects, can still
provide a fairly simple way to calculate different properties 
of a vortex lattice. The corrections we found to the London equation were 
classified into {\it nonlocal} and {\it nonlinear} ones. 
In both analytic and non-analytic
(low $T$) cases, nonlocal corrections play the dominant role in determining
the vortex lattice properties.

\begin{figure}[t]
\epsfysize 7cm
\epsfbox[-200 300 400 749]{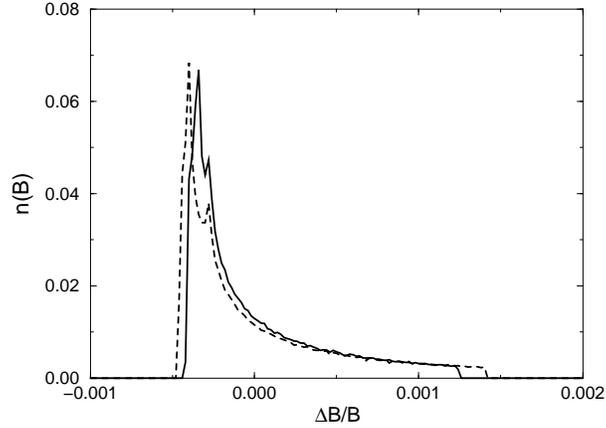}
\caption[Magnetic field distribution obtained from the nonlinear-nonlocal 
London equation and also from nonlocal London equation.]
{{\bf Solid line:} Magnetic field distribution obtained from the
nonlinear-nonlocal London equation. {\bf Dashed line:} Magnetic field
distribution obtained from a London equation including only the 
additional nonlocal term using the same parameters.}
\label{fys5}
\end{figure}

Equilibrium vortex lattice geometry exhibits novel field
and temperature dependence owing to the fourfold  
anisotropic effects expressed by the corrections to the London 
equation (Fig. \ref{fgl2}, Fig. \ref{fig2} and Fig. \ref{fys2}).
Numerical calculation of the lattice geometry is rather 
insensitive to the details of the vortex cores. The reason is that  
the details of the magnetic field inside the vortex cores
mainly affect the magnetic self-energies of the vortices. 
In magnetic fields far below $H_{c2}$ on the other hand, the vortex 
lattice geometry is mostly determined by the magnetic interaction 
energy between vortices which is not sensitive to the precise shape of the core.
Therefore, our replacement  of the vortex core by a simple Gaussian source 
term should be adequate for the vortex lattice structure calculations.

The effective penetration depth $\lambda_{\rm eff}$ also exhibits field
dependence at low temperatures as illustrated in Fig. \ref{fys3}.
Some important points need to be emphasized here:  

(i) The field dependence of $\lambda_{\rm eff}$ is not linear as is evident 
from Fig. \ref{fys3}. Variation of $\lambda_{\rm eff}$ with magnetic field is
faster at lower fields. At low magnetic fields,
the relative variation of the
effective penetration depth in our calculation is about 7\% for an
increase of 1T in the magnetic field which is close to 7.3\% variation
obtained from $\mu$SR data for an optimally doped and 9.5\% variation
for a detwinned underdoped YBCO single crystal  
using the same cutoff function as (\ref{Fk}) for fitting calculations
\cite{sonier-thesis,sonier2}. Most recent $\mu$SR results \cite{sonier-new} 
extrapolated to $T=0$, has provided a measurement of $\lambda_{\rm eff}$
as a function of magnetic field up to 7.5T. 
An excellent agreement between this result and our nonlinear-nonlocal theory
is presented in Fig. \ref{uSR}.

\begin{figure}
\epsfysize 7cm
\epsfbox[-200 300 400 750]{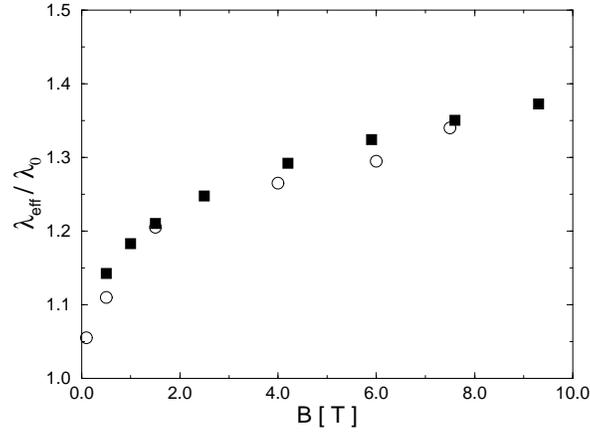}
\caption[Comparison between our nonlinear-nonlocal theory and a recent 
$\mu$SR result.]
{Comparison between our nonlinear-nonlocal theory and a recent 
$\mu$SR result at $T=0$. Squares represent our calculation of the effective 
penetration depth using nonlinear-nonlocal model (cf. Fig. \ref{fys3}).
Circles correspond to the recent $\mu$SR data \cite{sonier-new}. Our
$\lambda_0$ (= 1400 \AA) is a fitting parameter which might be different from the
measured value.}
\label{uSR}
\end{figure}

(ii) More importantly, 
this field dependence of $\lambda_{\rm eff}$ has a predominantly nonlocal
origin rather than a nonlinear one; contrary to what is generally believed.
The contribution of the nonlinear term to the total (minimized) 
free energy is almost
one order of magnitude smaller than the nonlocal term. What is more important
however, is the field dependence and $\beta$ dependence of these terms
not their orders of magnitude at
fixed $\beta$ and $B$. As we mentioned earlier, we consider $\beta$ as a variational
parameter to be fixed by minimizing the London free energy.
As can be inferred from Fig. \ref{fys2} and Fig. \ref{fys3},
the field dependence and $\beta$ dependence of the nonlinear 
term in the free energy is also smaller than the nonlocal term. 
It is worth noting that in the Meissner state, a linear nonlocal 
term can never produce field dependence in the penetration depth 
(as it is usually defined in that state) and
therefore a nonlinear term is necessary for such an effect \cite{Yip}.
In the vortex state on the other hand, a nonlocal term can result in 
a field dependent effective penetration depth, if we define it the
way it is defined in $\mu$SR experiments. We should emphasize here that
in this way of definition, $\lambda_{\rm eff}^{-2}$ does not directly 
give the superfluid density as it does in the Meissner state..

\begin{figure}[ht]
\epsfysize 7cm
\epsfbox[-200 300 400 750]{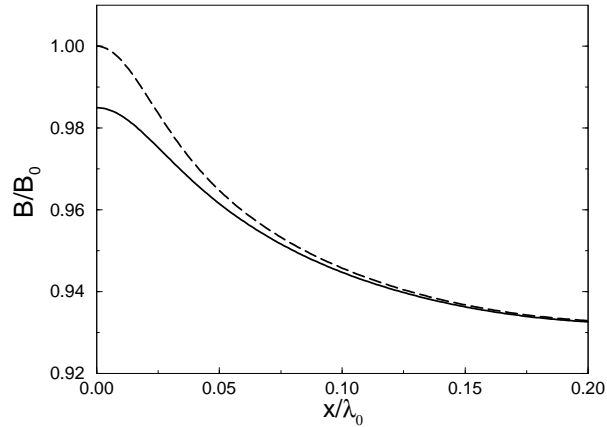}
\caption[Magnetic field as a function of the distance from the center of 
the vortex ($a$-direction) for an isolated vortex.]
{Magnetic field as a function of the distance from the center of 
the vortex ($a$-direction) for an isolated vortex. The solid line corresponds
to the nonlocal London equation whereas the dashed line represents an
ordinary London calculation with the same value of $\lambda_0$. 
$B_0=B(r=0)$ for the ordinary London vortex.}
\label{fys6}
\end{figure}

To understand this better, let's neglect the
nonlinear term and assume a linear but nonlocal London equation.
The total magnetic field, in this case, will be the superposition of the 
fields around
individual vortices. The magnetic field around an isolated vortex is given by
\be
B(\vec r)=\Phi_0\int {d^2k \over (2\pi)^2} {F(k)e^{i\vec k . \vec r} \over 
1+{\cal L}_{ij}(\vec k)k_ik_j}
\label{sglvx}
\ee
where $\Phi_0$ is the flux quantum and $F(k)$ is the cutoff function 
resulting from the source term. ${\cal L}_{ij}(\vec k)$ is defined
in (\ref{L}). The field decays with some constant (field independent) decay 
rate $\lambda$; which is the Meissner state penetration depth 
(see Fig. \ref{fys6}). Increasing the magnetic field does not affect the
profile of the magnetic field and therefore $\lambda$. Rather it squeezed 
the vortices together. For small values of 
$k$, ${\cal L}_{ij}(\vec k) \approx \lambda_o^2 \delta_{ij}$. Since small
$k$ corresponds to large $r$, one 
expects an isotropic field, similar to the local London case, far away 
from the vortex core. For large values of $k$ on the other hand, 
${\cal L}_{ij}(\vec k)$ has
strong $k$-dependence with four-fold anisotropy. Thus the closer to 
the vortex core, the more deviation form an isotropic ordinary London single 
vortex is expected, as is clearly shown in Fig. \ref{fys6}.
At low magnetic fields, the vortices are far apart and
their magnetic fields overlap in regions far away from their cores. The 
properties of the vortex lattice should then be similar to the ordinary 
london hexagonal lattice. As the magnetic field is increased, the vortices
come closer to each other. Although the profile of the magnetic field 
around each vortex remains unchanged, the overlap regions will be closer to
the vortex cores and will be more affected by the nonlocal term. Therefore,
it is conceivable that at large magnetic fields, the vortex lattice 
properties such as the magnetic field distribution will be affected by 
the nonlocal term in the London equation. The magnetic field near the 
vortex core is always reduced by the nonlocal term as is clear in 
Fig. \ref{fys6}. This is because ${\cal L}_{ij}(\vec k)-\lambda_0^2\delta_{ij}$ 
is a positive definite tensor for all $k$ and therefore the denominator
of the integrand in (\ref{sglvx}) is always larger 
than the corresponding ordinary London one.
As a result, the magnitude of $\overline{\Delta B^2}$ is smaller for 
the nonlocal case and therefore $\lambda_{\rm eff}$ defined by
(\ref{lmd}) tends to be larger. This explains why $\lambda_{\rm eff}/\lambda_0$
is always greater than one in Fig. \ref{fys3}. Fig. \ref{fys4} exhibits the
resemblance between a change in the magnetic field distribution due 
to the nonlocal term and due to 
a shift in the ordinary London penetration depth.
The slight difference between the solid and dashed lines in Fig. \ref{fys4} would
be unobservable in $\mu$SR experiments as a result of the broadening effects.
Since no field dependence due to the nonlocal term is expected in the Meissner 
state, $\lambda_{\rm eff}$ as defined here and also in $\mu$SR experiments
is expected to be conceptually different
from what is usually defined as penetration depth in the Meissner state,
although they are closely related.

(iii) The magnitude and field dependence of $\lambda_{\rm eff}$ is not
so sensitive to the apex angle $\beta$. In other words, a few degrees 
change in the variational parameter $\beta$ does not modify the 
magnetic field distribution 
as much as a variation in the average magnetic field does.  

(iv) Calculation of $\lambda_{\rm eff}$ is rather sensitive to the
form of the vortex source term. The importance of the source term in the
calculation of $\overline{\Delta B^2}$ has already been 
emphasized by Yaouanc {\em et al.}\cite{yaouanc}. 
In Ref.\cite{sonier-thesis}, $\lambda_{\rm eff}$ is obtained by fitting to 
the $\mu$SR data using both a Gaussian cutoff ((\ref{Fk})) and also the
cutoff function proposed by Hao {\em et al.}\cite{hao,yaouanc}. The 
difference between the two cases is significant and about 30\% for the
magnitude of $\lambda_{\rm eff}$ and even more (for detwinned sample)
for the relative variation with respect to
the magnetic field. This can explain
the importance of the source term in calculations of the effective 
penetration depth.

$\mu$SR experiments \cite{sonier-thesis,sonier3} on NbSe$_2$, which 
is believed to be a conventional
superconductor, also show a field dependence in the effective penetration 
depth; although it is much weaker than what is observed  in high $T_c$ 
compounds. Since there
is no node in the superconducting gap of these materials, the theory
presented in this thesis cannot explain this field dependence.  
However, since the size of the vortex core in these materials is large
and comparable to the vortex lattice spacing for the magnetic fields of
experimental interest, it is conceivable that a significant effect can come
from the cores as is pointed out in Ref. \cite{yaouanc}. 
Thus, a more careful consideration
of the vortex core might be necessary in order to have a better quantitative
explanation of the experimental results.



\end{document}